\RequirePackage{fix-cm}
\documentclass[twocolumn]{svjour3}          
\smartqed  
\usepackage{graphicx}
\usepackage{mathptmx}      
\usepackage{times}
\usepackage{epsfig}
\usepackage{graphicx}
\usepackage{amsmath}
\usepackage{amssymb}

\usepackage{multirow}
\usepackage{algorithm}
\usepackage{algorithmic}
\usepackage{subfigure}
\usepackage[misc]{ifsym}
\usepackage[colorlinks,
            linkcolor=blue,
            anchorcolor=blue,
            breaklinks,
            citecolor=blue]{hyperref}
 \usepackage{natbib}
\usepackage{etoolbox}
\makeatletter
\patchcmd{\NAT@citex}
  {\@citea\NAT@hyper@{%
     \NAT@nmfmt{\NAT@nm}%
     \hyper@natlinkbreak{\NAT@aysep\NAT@spacechar}{\@citeb\@extra@b@citeb}%
     \NAT@date}}
  {\@citea\NAT@nmfmt{\NAT@nm}%
   \NAT@aysep\NAT@spacechar\NAT@hyper@{\NAT@date}}{}{}

\patchcmd{\NAT@citex}
  {\@citea\NAT@hyper@{%
     \NAT@nmfmt{\NAT@nm}%
     \hyper@natlinkbreak{\NAT@spacechar\NAT@@open\if*#1*\else#1\NAT@spacechar\fi}%
       {\@citeb\@extra@b@citeb}%
     \NAT@date}}
  {\@citea\NAT@nmfmt{\NAT@nm}%
   \NAT@spacechar\NAT@@open\if*#1*\else#1\NAT@spacechar\fi\NAT@hyper@{\NAT@date}}
  {}{}
\makeatother

\begin{document}

\title{Hadamard Matrix Guided Online Hashing
}


\author{Mingbao Lin         \and
        Rongrong Ji         \and
        Hong Liu            \and
        Xiaoshuai Sun       \and
        Shen Chen           \and
        Qi Tian        
}


\institute{\hspace*{1.5em} Mingbao Lin$^1$ \at \hspace*{1.5em} lmbxmu@stu.xmu.edu.cn
           \and
           \Letter \hspace*{0.5em} Rongrong Ji$^{1,2}$ \at \hspace*{1.5em} rrji@xmu.edu.cn
           \and
           \hspace*{1.5em} Hong Liu$^1$ \at \hspace*{1.5em} lynnliu.xmu@gmail.com
           \and
           \hspace*{1.5em} Xiaoshuai Sun$^{1,2}$ \at \hspace*{1.5em} xiaoshuaisun.hit@gmail.com
           \and
           \hspace*{1.5em} Shen Chen$^1$ \at \hspace*{1.5em} chenshen@stu.xmu.edu.cn
           \and
           \hspace*{1.5em} Qi Tian$^3$   \at \hspace*{1.5em} tian.qi1@huawei.com
           \and
           $^1$\hspace*{1em} Media Analytics and Computing Laboratory, Department of \\
           \hspace*{1.4em} Artificial Intelligence, School of Informatics, Xiamen University,
           \\
           \hspace*{1.4em} China.  \at
           \and
           $^2$\hspace*{1em} Peng Cheng Laboratory, Shenzhen, China.  \at
           \and
           $^3$\hspace*{1em} Huawei Noah's Ark Lab, China. \at
}

\date{Received: date / Accepted: date}

\maketitle

\begin{abstract}
Online image hashing has attracted increasing research attention recently, which receives large-scale data in a streaming manner to update the hash functions on-the-fly.
Its key challenge lies in the difficulty of balancing the learning timeliness and model accuracy.
To this end, most works follow a supervised setting, \emph{i.e.}, using class labels to boost the hashing performance, which defects in two aspects:
First, strong constraints, \emph{e.g.}, orthogonal or similarity preserving, are used, which however are typically relaxed and lead to large accuracy drop.
Second, large amounts of training batches are required to learn the up-to-date hash functions, which largely increase the learning complexity.
To handle the above challenges, a novel supervised online hashing scheme termed \textbf{H}adamard \textbf{M}atrix Guided \textbf{O}nline \textbf{H}ashing (HMOH) is proposed in this paper.
Our key innovation lies in introducing Hadamard matrix, which is an orthogonal binary matrix built via Sylvester method.
In particular, to release the need of strong constraints, we regard each column of Hadamard matrix as the target code for each class label, which by nature satisfies several desired properties of hashing codes.
To accelerate the online training, LSH is first adopted to align the lengths of target code and to-be-learned binary code.
We then treat the learning of hash functions as a set of binary classification problems to fit the assigned target code.
Finally, extensive experiments demonstrate the superior accuracy and efficiency of the proposed method over various state-of-the-art methods. Codes are available at \url{https://github.com/lmbxmu/mycode}.

\keywords{Binary Code \and Online Hashing \and Hadamard Matrix \and Image Retrieval}

\end{abstract}

\section{Introduction}  \label{introduction}
Coming with the ever-increasing amount of visual big data, image hashing has attracted extensive research attention in the past decade \citep{weiss2009spectral,wang2010semi,liu2012supervised,gong2013iterative,liu2014discrete,shen2015supervised,gui2018fast,wang2018survey,liu2018dense,yang2018shared,deng2019unsupervised,deng2019two}.
Most existing works are designed to train hash functions one-off from a given collection of training data with/without supervised labels.
However, such a setting cannot handle the dynamic scenario where data are fed into the system in a streaming fashion.
Therefore, online hashing has been investigated recently \citep{huang2013online,onlinehashing,leng2015online,cakir2015adaptive,cakir2017online,fatih2017mihash,Chen2017FROSHFO,lin2019towards}, which receives streaming data online to update the hash functions instantly.
Online hashing merits in its superior efficiency in training and its timeliness in coping with the data variations.

The goal of online hashing is to update hash functions from the upcoming data batch while preserving the discriminability of binary codes for the past streaming data.
Existing works in online hashing can be categorized into either supervised methods or unsupervised methods.
For supervised methods, representative works include, but not limited to, OKH \citep{huang2013online,onlinehashing}, AdaptHash \citep{cakir2015adaptive}, OSH \citep{cakir2017online}, MIHash \citep{fatih2017mihash} and BSODH \citep{lin2019towards}.
For unsupervised methods, one can refer to SketchHash \citep{leng2015online} and FROSH \citep{Chen2017FROSHFO}.
In general, supervised online hashing methods typically achieves better results over unsupervised ones, which is mainly due to the use of labels to boost the hashing performance.

So far, online hashing retains as an open problem.
Its major challenge lies in the difficulty to make a tradeoff between model accuracy and learning efficiency.
To explain more explicitly, there exist two issues: First, existing online hashing methods rely on strong constraints to design robust hash functions, \emph{e.g.}, orthogonality \citep{cakir2015adaptive,leng2015online,Chen2017FROSHFO} and similarity preservation \citep{huang2013online,onlinehashing,fatih2017mihash,lin2019towards}, which however need to be relaxed in optimization and therefore lead to large accuracy drop.
Second, as validated in Sec.\,\ref{results}, existing online hashing methods require large amounts of training data to gain satisfactory results, which inevitably leads to low efficiency.
To handle the first issue, the work in \citep{cakir2017online} proposed to learn Error Correcting Output Codes (ECOC) to eliminate the heavy constraints in optimization.
However, the quality of ECOC remains inferior, which will lead to information loss as the streaming data grows.
Besides, the use of online boosting in \citep{babenko2009family} brings additional training burden.
In terms of the second issue, to our best knowledge, there is no work focusing on accelerating the online training, which remains as an open problem.

\begin{figure*}[!tb]
\begin{center}
\includegraphics[height=0.35\linewidth]{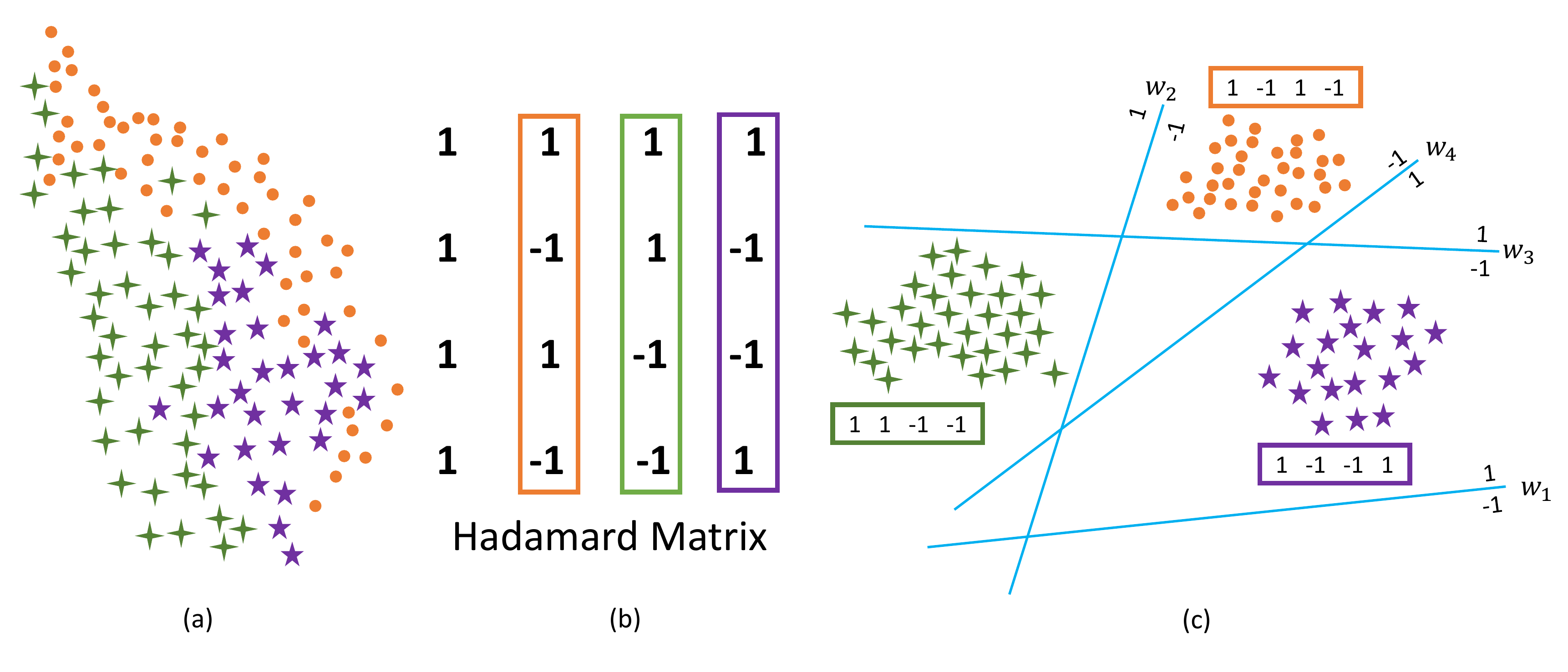}
\caption{The proposed Hadamard Matrix Guided Online Hashing framework.
Each time when a set of streaming data arrives (a),
the data points from the same class (denoted by one common shape and color) are assigned with a column (denoted by the same color) from a pre-generated Hadamard Matrix as the target code ($r^{*} = 4$ bits in this case) to be learned in the Hamming space (b).
And each row is regarded as a set of binary labels (-1 or +1).
The goal of our framework is to learn $r^{*}$ separate binary classifiers to predict each bit (c).
}
\label{graphic_example}
\end{center}
\vspace{-2em}
\end{figure*}

In this paper, we propose a simple yet effective online hashing method, termed \textbf{H}adamard \textbf{M}atrix Guided \textbf{O}nline \textbf{H}ashing (HMOH) to solve the aforementioned problems.
Our key innovation lies in the introduction of Hadamard matrix, each column of which serves as the target code to guide the learning of hash functions.
First, we propose to generate a Hadamard matrix via the Sylvester method \citep{sylvester1867lx}, which assigns individual column randomly to the streaming data with the same class label as their binary codes.
The Hadamard matrix by nature satisfies several desired properties of hashing, \emph{e.g.}, orthogonality and balancedness, which are beneficial to guiding the learning of hash.
Second, to align the size of Hadamard matrix with the to-be-learned binary codes, we further employ locality sensitive hashing (LSH) \citep{gionis1999similarity} to reduce the length of Hadamard codes, which has been proven to be effective in the following context.
Notably, both Hadamard matrix and LSH can be efficiently applied online, \emph{i.e.}, Hadamard matrix can be generated offline and LSH consisting of random projections is data-independent \citep{datar2004locality}.
Importantly, no extra training is needed, which differs our method from the existing online hashing \citep{cakir2017online} where the ECOC codebook is generated on-the-fly.
Third, the assigned binary codes are regarded as virtual category labels ($+1$ or $-1$), upon which the hash function is decomposed into a set of binary classification problems that can be well addressed by off-the-shelf online binary classification methods \citep{freund1999large,liu2015efficient,goh2001svm}.
Last, to preserve the information of the past streaming data while distilling the core knowledge of the current data batch, we further ensemble the learned models in previous rounds, which further boosts the retrieval performance.
Extensive experiments on four benchmarks, \emph{i.e.}, CIFAR-10, Places205, MNIST and NUS-WIDE, show that the proposed HCOH achieves better or competitive results to the state-of-the-art methods \citep{huang2013online,leng2015online,cakir2015adaptive,cakir2017online,fatih2017mihash,lin2019towards}.

The rest of this paper is organized as follows: In Sec.\,\ref{background}, the related works are discussed. The proposed HMOH and its optimization are presented in Sec.\,\ref{method}. Sec.\,\ref{experiment} reports our quantitative evaluations and analysis. Finally, we conclude this paper in Sec.\,\ref{conclusion}.

\section{Related Work} \label{background}
There are increasing endeavors of online hashing made in the recent years.
In generally, online hashing updates the hash functions sequentially and instantly along with the arriving data batch.
According to the different types, existing online hashing can be categorized into either supervised methods or unsupervised methods. The former includes, but not limited to, Online Kernel Hashing (OKH) \citep{huang2013online,onlinehashing}, Adaptive Hashing (AdaptHash) \citep{cakir2015adaptive}, Online Supervised Hashing (OSH) \citep{cakir2017online}, Online Hashing with Mutual Information (MIHash) \citep{fatih2017mihash} and Balanced Similarity for Online Discrete Hashing (BSODH) \citep{lin2019towards}. The latter includes, but not limited to, Online Sketching Hashing (SketchHash) \citep{leng2015online} and Faster Online Sketching Hashing (FROSH) \citep{Chen2017FROSHFO}.

%
Unsupervised methods consider the inherent properties among data, \emph{e.g.}, distribution and variance, to conduct online hashing.
The design of existing unsupervised online hashing is partially inspired from the idea of ``data sketching" \citep{liberty2013simple}, where a large dataset is summarized by a much smaller data batch that preserves the properties of interest.
For instance, Online Sketching Hashing (SketchHash) was proposed in \citep{leng2015online}, which maximizes the variance of every hashing bit among the sketched data and adopts an efficient variant of SVD decomposition to learn hash functions.
A faster version of SketchHash, termed FROSH, was proposed in \citep{Chen2017FROSHFO} to reduce the training time.
FROSH adopts the Subsampled Randomized Hadamard Transform (SRHT) \citep{lu2013faster} to speed up the data sketching process in SketchHash.

Supervised methods take advantage of label information to assist the learning of hash functions.
To our best knowledge, Online Kernel Hashing (OKH) \citep{huang2013online,onlinehashing} is the first of this kind.
OKH designs a prediction loss by using pairwise data and is optimized via a passive-aggressive strategy  \citep{crammer2006online}, based on which the updated hashing model is able to retain information learned in the previous rounds and adapt to the data in the current round.
Similar to OKH, Adaptive Hashing (AdaptHash) \citep{cakir2015adaptive} also assumes that the pairs of points arrive sequentially.
A hinge loss \citep{norouzi2011minimal} is defined to narrow the distance between similar pairs and to enlarge that between dissimilar ones.
Then Stochastic Gradient Descent (SGD) is deployed to update the hash functions.
In \citep{cakir2017online}, a two-step hashing framework was introduced, where binary Error Correcting Output Codes (ECOC) \citep{jiang2009efficient,kittler2001face,schapire1997using,zhao2013sparse} are first assigned to labeled data, and then hash functions are learned to fit the binary ECOC using Online Boosting \citep{babenko2009family}.
Cakir \emph{et al}. developed an Online Hashing with Mutual Information (MIHash) \citep{fatih2017mihash}.
Given an image, MIHash aims to separate the distributions between its neighbors and non-neighbors of Hamming distances.
To capture the separability, Mutual Information \citep{cover2012elements} is adopted as the learning objective and SGD is used to renovate the hash functions.
Balanced Similarity for Online Discrete Hashing (BSODH) \citep{lin2019towards} was recently proposed to enable learning hash model on streaming data, which investigates the correlation between new data and the existing dataset.
To deal with the data-imbalance issue (\emph{i.e.}, quantity inequality between similar and dissimilar data) in online learning, BSODH adopts a novel \emph{balanced similarity}, which also enables the use of discrete optimization in online learning for the first time.

In principle, supervised online hashing prevails over unsupervised methods by using the additional label information.
However, most existing supervised methods simply resort to learning robust binary codes under strong constraints like orthogonality and similarity preservation.
On one hand, a large volume of training data is required to obtain a competitive result, leading to poor efficiency.
On the other hand, the optimization process cannot be directly deployed with strong constraints.
The relaxation process further leads to low accuracy.
To sum up, the effectiveness and efficiency of existing online hashing cannot be simultaneously guaranteed, which is the main focus of this paper to address.
Note that our solution has a certain similarity to that of OSH \citep{cakir2017online}, which, as discussed in Sec.\,\ref{framework} and Sec.\,\ref{results}, fails in both effectiveness and efficiency.

A preliminary conference version of this work was presented in \citep{lin2018supervised}.
Besides more detailed analysis, this paper differs from our conference version in the following aspects:
$1$) Instead of simply using linear regression to learn the hash mapping, we further transfer the online retrieval problem into an online binary classification problem, which can be well solved by off-the-shelf algorithms and has achieved better results.
$2$) We extend the proposed method to multi-label benchmarks by proposing both ``majority principle" and ``balancedness principle".
$3$) We propose to ensemble the learned model in every round together, which is experimentally demonstrated to be more effective than using the updated model alone.
$4$) More extensive experiments are conducted to demonstrate the effectiveness and efficiency of the proposed method.

\section{The Proposed Method} \label{method}
In this section, we introduce the proposed HMOH method in details. The overall hashing framework is illustrated in Fig.\,\ref{graphic_example}.
A column from the Hadamard Matrix Fig.\,\ref{graphic_example}(b) is assigned to the newly arriving data from the same class as shown in Fig.\,\ref{graphic_example}(a).
The assigned code plays as the target code in the Hamming space.
The goal of the proposed method is to learn a set of binary classifiers to fit the target code as shown in Fig.\,\ref{graphic_example}(c).

\subsection{Problem Definition} \label{definition}
Suppose the dataset is formed by a set of $n$ vectors, $\mathbf{X} = \{\mathbf{x}_i\}_{i=1}^{n} \in \mathbb{R}^{d \times n}$, and accompanied by a set of class labels $ \mathbf{L} = \{l\}_{i=1}^n \in \mathbb{N}^n$.
The goal of hashing is to learn a set of hashing codes $\mathbf{B} = \{\mathbf{b}_i\}^n_{i=1} \in \{-1, +1\}^{r \times n}$ such that a desired neighborhood structure is preserved.
This is achieved by projecting the dataset $\mathbf{X}$ using a set of $r$ hash functions $H(\mathbf{X}) = \{h_i(\mathbf{X})\}_{i=1}^r$, \emph{i.e.},
\begin{equation}\label{hash_definition}
\mathbf{B} = H(\mathbf{X}) = sign(\mathbf{W}^T\mathbf{X}),
\end{equation}
where $\mathbf{W} = \{ \mathbf{w}_i \}_{i=1}^r \in \mathbb{R}^{d \times r}$ is the projection matrix and $\mathbf{w}_i$ is the $i$-th hash function.
The sign function $sign(x)$ returns $+1$ if the input variable $x > 0$, and $-1$ otherwise.
In the online setting, $\mathbf{X}$ comes in a streaming fashion and is not available once for all.
Hence, we denote $\mathbf{X}^t = \{ \mathbf{x}_i^t \}^{n_t}_{i=1} \in \mathbf{R}^{d \times n_t}$ as the input streaming data at $t$-stage,
denote $\mathbf{B}^t = \{ \mathbf{b}_i \}_{i=1}^{n_t} \in \{-1, +1\}^{r \times n_t}$ as the learned binary codes for $\mathbf{X}^t$,
and denote $\mathbf{L}^t = \{l_i^t\}_{i=1}^{n_t}$ as the corresponding label set, where $n_t$ is the size of streaming data at $t$-stage.
Correspondingly, the parameter $\mathbf{W}$ updated at $t$-stage is denoted as $\mathbf{W}^t$.
%

\subsection{Kernelization} \label{kernelization}
We use kernel trick to take advantages of linear models and meanwhile enable them to capture non-linear data patterns.
It has been theoretically and empirically proven to be able to tackle linearly inseparable data \citep{kulis2012kernelized,liu2012supervised,huang2013online,onlinehashing}.
We map data in the original space $\mathbb{R}^d$ to a feature space $\mathbb{R}^m$ through a kernel function based on anchor points.
Hence, we have a new representation of $\mathbf{x}_i$ that can be formulated by following:
\begin{equation}\label{kernel}
  z(\mathbf{x}_i) = [\kappa(\mathbf{x}_i, \mathbf{x}_{(1)}), \kappa(\mathbf{x}_i, \mathbf{x}_{(2)}), ..., \kappa(\mathbf{x}_i, \mathbf{x}_{(m)})]^T,
\end{equation}
where $\mathbf{x}_{(1)}, \mathbf{x}_{(2)}, ..., \mathbf{x}_{(m)}$ are $m$ anchors.
Without loss of generality, we simplify $z(\mathbf{x}^t_i)$ as $\mathbf{z}^t_i$ and simplify the kernelized representation of $\mathbf{X}^t$ as $\mathbf{Z}^t$.

To obtain these anchors, we follow the work in \citep{huang2013online,onlinehashing} to assume that $m$ data points can be available in the initial stage.
The learning process will not start until $m$ data points have been collected.
Then these $m$ data points are considered as $m$ anchors used in the kernel trick.
In terms of the kernel function, we use the Gaussian RBF kernel, \emph{i.e.}, $\kappa(\mathbf{x}, \mathbf{y}) = exp(-\| \mathbf{x} - \mathbf{y} \|^2 / 2{\eta}^2)$, where ${\eta}^2$ is known as the \emph{bandwidth} to be tuned in the learning process.

\subsection{The Proposed Framework}  \label{framework}
In this section, we introduce the framework of the proposed online hashing.
We first revisit the online hashing formulation based on the Error Correcting Output Codes (ECOC) \citep{cakir2017online}, which separates the learning process into two steps:
(1) When a new label is observed, the new target code, \emph{i.e.}, a codeword from the ECOC codebook, is assigned to it.
(2) All data that shares the same labels is proceeded to fit this codeword.
To that effect, \citep{cakir2017online} adopts the $0 - 1$ loss, which denotes whether the hash functions fit the assigned codewords.
The exponential loss with convexity is further used to replace $0 - 1$, and then SGD is applied to enable the optimization of hash functions.
To further improve the performance, a boosting scheme that considers previous mappings to update each hash function is used to handle the process of error-correlation.

\begin{figure}[!t]
\begin{center}
\includegraphics[height=0.4\linewidth]{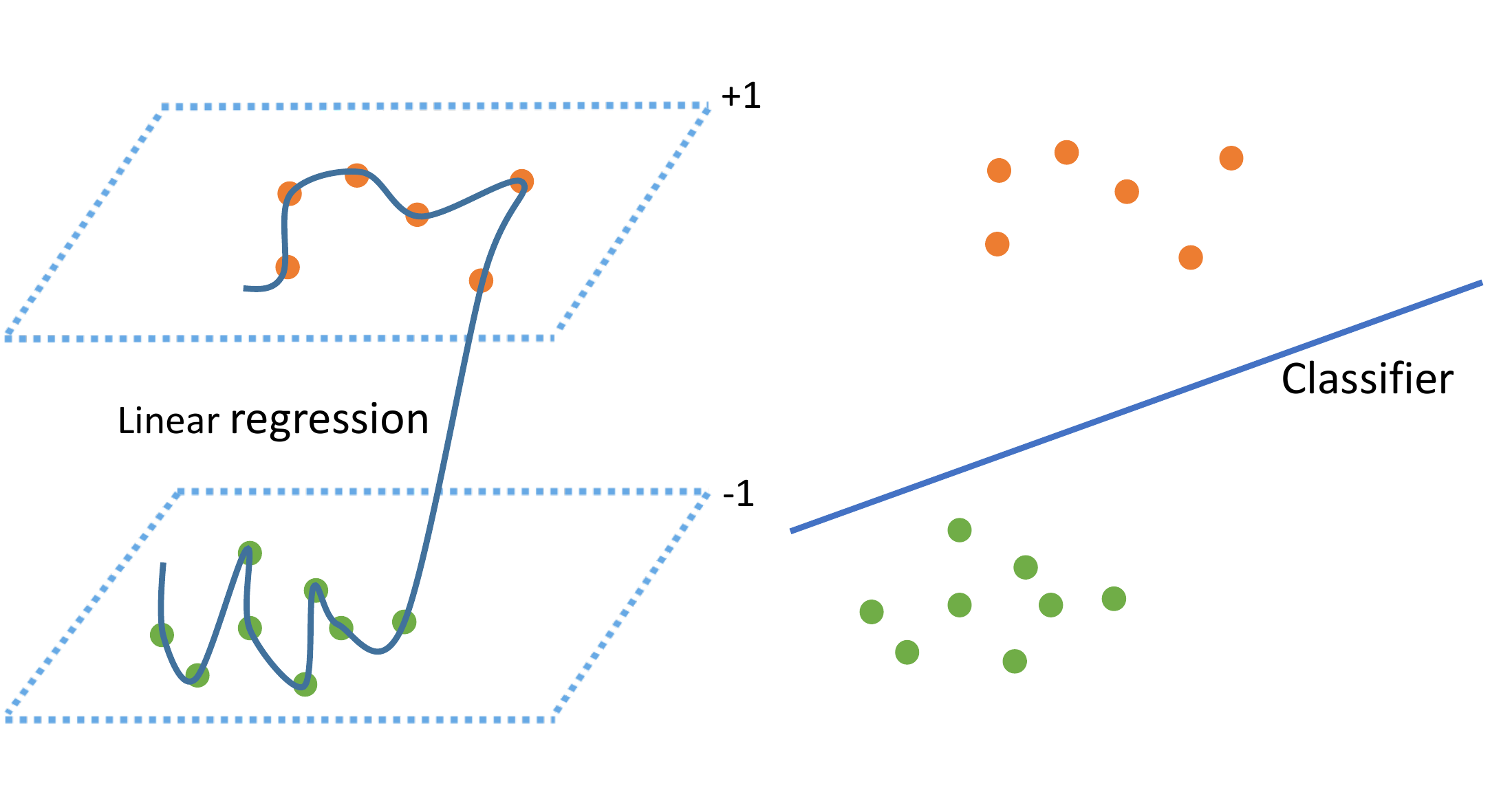}
\caption{\label{c_r} A comparison between linear regression and classifer.
Linear regression tries to fit binary codes with the best hyperplane which goes through the data points.
While classifer aims to find a hyperplane to split the data points.
Comparing with linear regression, the goal of classifer is much easier.
}
\end{center}
\vspace{-2em}
\end{figure}

However, there exist some issues in \citep{cakir2017online}.
First, the performance highly depends on the codebook construction, \emph{e.g.}, the distance between the target codes must be large enough to ensure error-correction.
However, the codebook quality is degenerated due to the random construction strategy in \citep{cakir2017online}.
Second, the use of exponential loss and boosting further increases the training time, which is a serious concern in online learning.
To sum up, the key points for a successful ECOC-based online hashing fall into \emph{a better ECOC codebook}, \emph{a loss function with less computation cost} and \emph{an efficient boosting algorithm}.
In terms of a better ECOC codebook, to our best knowledge, the basic idea of ECOC stems from the model of signal transmission in communication \citep{peterson1972error}.
Generally, the use of ECOC to guide the hashing learning contains two phases, \emph{i.e.}, ``encoding phase" and ``decoding phase".
As shown in Fig.\,\ref{graphic_example}(b),
in the encoding phase, the data points from the same class are assigned with one common column from the ECOC codebook $\mathbf{C} = \{ \mathbf{c}_i \}_{i=1}^{r^*} \in \{-1, +1\}^{r^* \times r^*}$.
In the decoding phase, the assigned column
$\mathbf{c}_{J(\mathbf{x}_i^t)}$ is regarded as the virtual multiple binary categories, where $J(\mathbf{x}_i^t)$ returns the class label of $\mathbf{x}_i^t$, \emph{i.e.}, $l^t_i$.
Therefore, in the case of $r^* = r$, \emph{i.e.}, the code length is the same with the size of virtual categories (the virtual categories can be seen as the target codes for hashing learning).
To this end, the preliminary work, \emph{i.e.}, HCOH \citep{lin2018supervised} simply considers the linear regression to fit the virtual categories as follows:
\begin{equation}   \label{previous}
\tilde{\phi}(\mathbf{x}_i^t;\mathbf{W}^{t-1}) =  \|H(\mathbf{x}_i^t) - \mathbf{c}_{J(\mathbf{x}_i^t)}\|_F^2,
\end{equation}
where $\|\cdot\|_{F}$ is the Frobenius norm of the matrix.
Nevertheless, there are some issues in such a learning approach:
To enable the optimization of non-convex sign($\cdot$) function in Eq.\,\ref{hash_definition}, HCOH has to relax the sign($\cdot$) function ranging in $\{-1, +1\}$, with tanh($\cdot$) function ranging in $(-1, +1)$.
On one hand, the relaxation process endures more quantization error.
On the other hand, the derivative of tanh($\cdot$) bears more computation burden.
Moreover, in the case of low hash bit, it is not appropriate to simply apply the Frobenius norm to fit the target code, due to its inferior performance as shown in \citep{lin2018supervised}.
To analyze the above issues, linear regression attempts to estimate the mapping from the input variables to numerical or continuous output variables.
However, the assigned binary codes indeed are in a discrete space (-1 or +1).
To fit data with the best hyperplane going through the data points is difficult as illustrated in the left part of Fig.\,\ref{c_r}.
One solution to solve these problems is to use classifier, which attempts to estimate the mapping function from the input variables to discrete or categorical output variables.
As shown in the right part of Fig.\,\ref{c_r}, different with linear regression, the classifier aims to find a hyperplane to split the data points, which is much easier.

%
On the contrary, in this paper, we consider the hash functions as a set of binary classifiers. And the virtual categories can be used as the corresponding class labels.
If $h_k(\mathbf{x}_i) = 1$, a given $\mathbf{x}_i$ belongs to the $k$-th virtual class, and vice versa.
Therefore, the online retrieval problem turns into training $r^*$ separate binary classifiers to predict each bit, which can be well addressed by off-the-shelf methods \citep{novikoff1963convergence,freund1999large,liu2015efficient,goh2001svm}.
To that effect, we consider the classical Kernelized Perceptron algorithm \citep{freund1999large}.
The perceptron based algorithms by nature can be seen as online methods since the binary classifiers are updated in a streaming fashion, which well satisfies the above requirements.
By simply removing the $sign(\cdot)$ in Eq.\,\ref{hash_definition}, we obtain the linear functions
$\hat{H}(\mathbf{X}^t) = \{\hat{h}_i(\mathbf{X}^t)\}_{i=1}^b$ as:
\begin{equation}\label{remove_sign}
  \hat{H}(\mathbf{X}^t) = {\mathbf{W}^{t-1}}^T\mathbf{X}^t,
\end{equation}

Given a kernelized training data point and its corresponding virtual categories ($\mathbf{z}_i^t, \mathbf{c}_{J(\mathbf{x}_i^t)}$), the loss for perceptron algorithm is as follows:
\begin{equation}\label{loss_perceptron}
    \phi(\mathbf{z}_i^t;\mathbf{W}^{t-1}) =  - {(\mathbf{c}_{J(\mathbf{x}_i^t)} \odot \mathbf{a}^t_i)}^T\hat{H}(\mathbf{z}_i^t),
\end{equation}
where $\odot$ stands for the Hadamard product (\emph{i.e.}, element-wise product) and $\mathbf{a}_i^t$ is a $0-1$ vector with the $k$-th element $\mathbf{a}_{ik}^t$ defined as $1$ if $\mathbf{x}_i^t$ is correctly classified as the virtual label $\mathbf{c}_{J(\mathbf{x}_i^t)k}$ by $\hat{h}_k(\mathbf{x}_i^t)$, and $0$ otherwise.
By considering all the data points $\mathbf{X}^t$ at the $t$-stage, the overall objective function can be re-written as:
\begin{align}\label{total_loss_perceptron}
    \Phi(\mathbf{Z}^t;\mathbf{W}^{t-1}) &= -\sum_{i=1}^{n_t}{(\mathbf{c}_{J(\mathbf{x}_i^t)} \odot \mathbf{a}^t_i)}^T\hat{H}(\mathbf{z}_i^t)   \nonumber \\
        & =  - tr\big({(\mathbf{c}_{J(\mathbf{X}^t)} \odot \mathbf{A}^t)}^T\hat{H}(\mathbf{Z}^t)\big),
\end{align}
where $\mathbf{A}^t = \{ \mathbf{a}_i^t \}_{i=1}^{n_t} \in \mathbb{R}^{r^* \times n^t}$ and $\mathbf{c}_{J(\mathbf{X}^t)} = \{ \mathbf{c}_{J(\mathbf{x}_i^t)} \}_{i=1}^{n_t} \in \mathbb{R}^{r^* \times n^t}$.

Above all, the Kernelized Perceptron algorithm merely considers the linear regression without any complex loss to replace the $sign(\cdot)$, which well satisfies the need for designing a good loss function with less computation cost.
Moreover, it also overcomes the inferior performance of the preliminary version \citep{lin2018supervised} in low hash bit, as demonstrated later in Sec.\,\ref{results}.

\subsection{Hadamard Matrix}  \label{hadamard_matrix}
Above all, the success of online hashing falls in encoding the ECOC matrix $\mathbf{C}$.
To analyze, an efficient hash code should satisfy that the variance of each bit is maximized and the bits are pairwise uncorrelated.
That is to say, in Fig.\,\ref{graphic_example}(b), half of the data in each row should be $+1$, and $-1$ for the other half \citep{wang2010semi}.
What's more, by designing columns in the ECOC matrix to have maximal Hamming distance from each other, we can get a method that is more resistant to individual bit-flipping errors (misclassification).
As above, a robust ECOC codebook $\mathbf{C}$ should satisfy:
1) Maximal Hamming distance between each row, which is for optimal hashing codes.
2) Maximal Hamming distance between each column, which ensures the resistance to misclassification.

To achieve these two goals, we consider the use of Hadamard Matrix \citep{horadam2012hadamard} as the backbone to construct the desired ECOC codebook.
In particular, on one hand, the Hadamard matrix is an $n$-order orthogonal matrix, \emph{i.e.}, both its row vectors and column vectors are pairwisely orthogonal, which by nature satisfies the principles of $1$) and $2$).
On the other hand, elements in the Hadamard matrix are either $+1$ or $-1$, \emph{i.e.},:
\begin{equation} \label{hadamard_definition}
\mathbf{H}\mathbf{H}^T = n\mathbf{I}_n, ~or ~ \mathbf{H}^T\mathbf{H}=n\mathbf{I}_n,
\end{equation}
where $\mathbf{I}_n$ is an $n-$order identity matrix.

Hence, Hadamard matrix can be used as an efficient ECOC codebook as shown in Fig.\,\ref{graphic_example}.

Though the existing theorems that describe the existence of Hadamard matrices of other orders \citep{paley1933orthogonal, williamson1944hadamard,goldberg1966hadamard,ockwig2005reticular},
we simply consider the $2^k$-order Hadamard matrices in this paper, which can achieve satisfactory performances as shown in Sec.\,\ref{experiment}.
To construct the $2^k$-order Hadamard matrices, the entry in the $i$-th row and the $j$-th column can be defined as:
\begin{equation}
H_{ij} = (-1)^{(i-1)\times(j-1)},
\end{equation}
or it can also be completed by a recursive algorithm as developed in \citep{sylvester1867lx}:
\begin{equation} \label{construction}
H_{2^k} = {
\left[ \begin{array}{cc}
H_{2^{k-1}} & H_{2^{k-1}}  \\
H_{2^{k-1}} & -H_{2^{k-1}}
\end{array}
\right ]}
~ \mathrm{and} ~
H_2 = {
\left[ \begin{array}{cc}
1 & 1  \\
1 & -1
\end{array}
\right ]}.
\end{equation}

Since the Hadamard matrix is limited to the $2^k$-order in this paper, each data point from the same class label are assigned with one common discriminative column from the Hadamard matrix, and the size of Hadamard matrix $r^*$ can be defined as follows:
\begin{equation}   \label{r_star}
\begin{aligned}
r^* = \min \{g|g=2^k, g\geq r, g \ge |\mathbf{L}|, k=1,2,3,...\},
\end{aligned}
\end{equation}
where $|\mathbf{L}|$ is the number of class labels in the dataset.
Therefore, based on the above discussion, we construct the square Hadamard matrix as $\mathbf{C}_{r^*}\in \{-1,1\}^{r^* \times r^*}$ as the ECOC codebook.
If data with new label is received, we randomly and non-repeatedly select a column representation to construct a virtual label vector for this data. Otherwise, the virtual label previously assigned to instances with the same label is given. Therefore, our scheme does not need to pre-define the category number of the dataset.

\subsection{Learning Formulation} \label{learning_formulation}

The derived formulation is based on the assumption of $r^* = r$ which may not be satisfied\footnote{Take the Places$205$ dataset as an example: There are in total $205$ categories. According to Eq.\,\ref{r_star}, $r^* = 256$ for the code length $r$ varying from $8$ to $128$.}.
To handle this problem, we further use the LSH to transform the virtual labels to obtain the same length of binary codes to the hash functions.
\begin{equation}\label{transformation}
  \tilde{\mathbf{c}}_{J(\mathbf{x}_i^t)} = sign(\tilde{\mathbf{W}}^T\mathbf{c}_{J(\mathbf{x}_i^t)}),
\end{equation}
where $\tilde{\mathbf{W}} = \{\tilde{\mathbf{w}_i}\}_{i=1}^r \in \mathbb{R}^{r^* \times r}$ with each $\tilde{\mathbf{w}}_i \in \mathbb{R}^{r^*}$ sampled from the standard Gaussian distribution, \emph{i.e.}, $\tilde{\mathbf{w}}_i \sim N(\mathbf{0}, \mathbf{\textit{I}})$ and $\mathbf{0}$, $\mathbf{\textit{I}}$ are all-zero vector and identity matrix, respectively
\footnote{When $r^* = r$, we set $\tilde{\mathbf{W}}$ as an identity matrix and the above equation still holds.}.
In the following, we theoretically demonstrate that $\tilde{\mathbf{c}}_{J(\mathbf{x}_i^t)}$ preserves the main property of $\mathbf{c}_{J(x_i^t)}.$
\textbf{Theorem $1$}: For any vector $\mathbf{w} = [w_1, w_2,..., w_{r^*}] \in \mathbb{R}^{r^*}$, each $w_i$ is $i.i.d.$ sampled from a Gaussian distribution with zero mean, \emph{i.e.}, $w_i \sim N(0, {{\sigma}}^2)$ where $\sigma$ is the variance.
The inner product between $\mathbf{w}$ and $\mathbf{c}$ satisfies
\begin{equation}\label{probability}
    P(\mathbf{w}^T\mathbf{c} > 0) = P(\mathbf{w}^T\mathbf{c} < 0).
\end{equation}

Before the proof of \textbf{Theorem $1$}, we first briefly give the following Proposition:

\textbf{Proposition $1$}: For any $X \sim N({\mu}_X, {\sigma}_X^2)$ and $Y \sim N({\mu}_Y, {\sigma}_Y^2)$, the following satisfies:
\begin{align}
    &X + Y \sim N({\mu}_X + {\mu}_Y, {\sigma}^2_X + {\sigma}^2_Y),\\
    &X - Y \sim N({\mu}_X - {\mu}_Y, {\sigma}^2_X - {\sigma}^2_Y),
\end{align}

Proof of \textbf{Theorem $1$}:
\begin{equation}  \label{proof}
\begin{aligned}
    \mathbf{w}^T\mathbf{c} = \sum_{i=1}^{r^*}w_ic_i = \sum_{i,c_i = +1}w_i - \sum_{i,c_i=-1}w_i.
\end{aligned}
\end{equation}

Based on \textbf{Proposition $1$} and $w_i \sim N(0, {\sigma}^2)$, we have
\begin{equation}
    \sum_{i,c_i = +1}\!\!\!w_i \sim N(0, \frac{r^*}{2}{\sigma}^2)~\mathrm{and}~
    (-\!\!\!\sum_{i,c_i=-1}\!\!\!w_i) \sim N(0, \frac{r^*}{2}{\sigma}^2),
\end{equation}

And then,
\begin{equation}\label{distribution}
  (\sum_{i,c_i = +1}w_i - \sum_{i,c_i=-1}w_i) \sim  N(0, r^*{\sigma}^2).
\end{equation}

The above inference verifies that the inner product between $\mathbf{w}$ and $\mathbf{c}$ obeys the Gaussian distribution with zero mean.
Therefore, $P(\mathbf{w}^T\mathbf{c} > 0) = P(\mathbf{w}^T\mathbf{c} < 0)$, which demonstrates the validity of \textbf{Theorem $1$}.

Further, we denote $P\big(sign(\tilde{\mathbf{w}}^T_j\mathbf{c}_{J(\mathbf{x}_i^t)}) = +1\big)$ as $P_{+1}$ and $P\big(sign(\tilde{\mathbf{w}}^T_j\mathbf{c}_{J(\mathbf{x}_i^t)}) = -1\big)$ as $P_{-1}$.
According to \textbf{Theorem $1$}, it is easy to derive that $P_{+1} = P_{-1} = 0.5$.
In terms of Eq.\,\ref{transformation}, we denote the number of $+1$ in the transformed virtual labels $\tilde{\mathbf{c}}_{J(\mathbf{x}_i^t)}$ as $M$.
It is comprehensible that $M \in \{0, 1, ..., r^*\}$ has a binomial distribution, which is written as:
\begin{equation}\label{binomial}
  M \sim B(r^*, P_{+1}).
\end{equation}

The probability function is given by
\begin{equation}\label{binomial_mass}
  P(M=m) = \binom{r^*}{m}P_{+1}^mP_{-1}^{r^* - m}.
\end{equation}

\textbf{Proposition 2}: For any binomial distribution $X \sim B(n, p)$, the probability of $P(X = k)$ reaches the maximal value when $k = k_{0}$, where
\begin{equation}\label{max}
  k_{0}=
    \begin{cases}
    (n+1)p \quad or \quad (n+1)p-1, & (n+1)p \in \mathbb{Z}, \\
    [(n+1)p], & \mathrm{otherwise},
\end{cases}
\end{equation}
where $[\cdot]$ denotes the integral function.
Hence, for Eq.\,\ref{binomial_mass}, $P(M=m)$ reaches its maximum when $m = [(r^* + 1)p_{+1}] = \frac{r^*}{2}$.
At this point, the number of $-1$ in the transformed virtual labels $\tilde{\mathbf{c}}_{J(\mathbf{x}_i^t)}$ is also $\frac{r^*}{2}$.

Therefore, with high probabilities, LSH can balancedly transform each column of the Hadamard matrix, which still gives an effective target code approximate to the requirement of $2)$.
Through similar analysis, we can also obtain that each row of the transformed Hadamard matrix shares similar property, which satisfies the requirement of $1)$.
Hence, applying LSH to transform the virtual categories can well preserve the discrepancy of Hadamard matrix.

Above all, we further reformulate Eq.\,\ref{total_loss_perceptron} by LSH-based random hashing as:
\begin{align}\label{total_loss_perceptron_lsh}
    \Phi(\mathbf{Z}^t;\mathbf{W}^{t-1}) &= -\sum_{i=1}^{n_t}{(\tilde{\mathbf{c}}_{J(\mathbf{x}_i^t)} \odot \mathbf{a}^t_i)}^T\hat{H}(\mathbf{z}_i^t)   \nonumber \\
        & =  - tr\Big((\tilde{\mathbf{c}}_{J(\mathbf{X}^t)} \odot \mathbf{A}^t)^T\hat{H}(\mathbf{Z}^t)\Big).
\end{align}

Putting Eq.\,\ref{remove_sign}, and Eq.\,\ref{total_loss_perceptron_lsh} together, we have the following overall objective function:
\begin{align}
    \Phi(\mathbf{Z}^t;\mathbf{W}^{t-1}) =  - tr\Big((\tilde{\mathbf{c}}_{J(\mathbf{X}^t)} \odot \mathbf{A}^t)^T\mathbf{W}^{{t-1}^T}\mathbf{Z}^t\Big).
\end{align}
To obtain $\mathbf{W}^t$, we adopt the classical SGD algorithm as follows:
\begin{equation}  \label{updation}
\mathbf{W}^t \leftarrow \mathbf{W}^{t-1} - {\lambda} \frac{\partial \Phi}{\partial \mathbf{W}^{t-1}},
\end{equation}
where ${\lambda}$ is the learning rate. And the partial derivative of $\Phi$ \emph{w.r.t.} $\mathbf{W}^{t-1}$ can be derived as:
\begin{equation} \label{derivative}
	\frac{\partial \Phi}{\partial \mathbf{W}^{t-1}} = -\mathbf{Z}^t(\tilde{\mathbf{c}}_{J(\mathbf{X}^t)}\odot\mathbf{A}^t)^T.
\end{equation}

\begin{figure}[!tb]
\begin{center}
\includegraphics[height=0.35\linewidth]{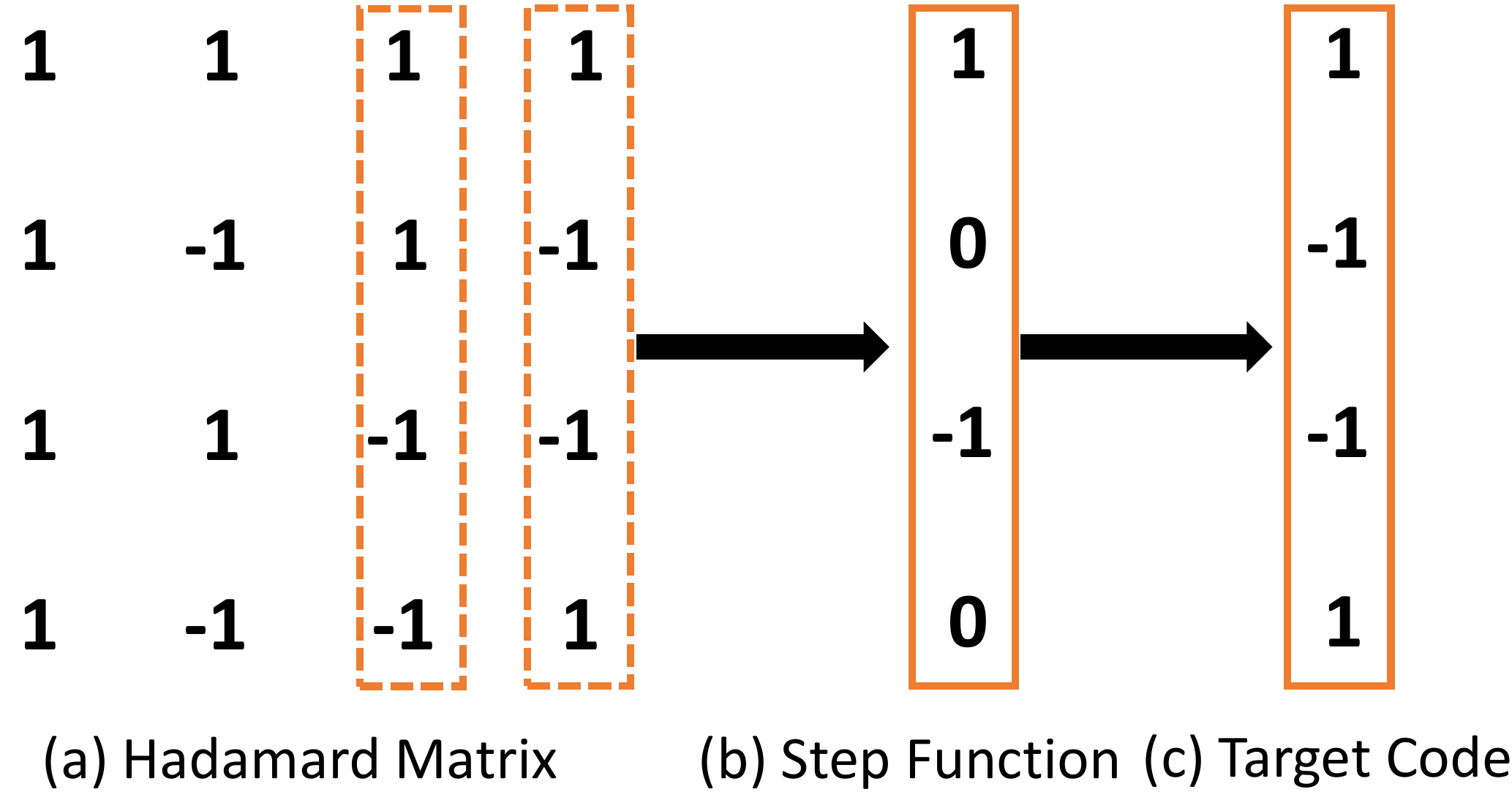}
\caption{A toy example of the multi-label case.
(a) More than one columns from the Hadamard Matrix will be selected as the target codes for multi-label data.
(b) To merge the multiple target codes into one binary vector while preserving the main property of Hadamard matrix, for each bit, we first vote for $+1$ or $-1$ based on the ``majority principle" and vote for $0$ if the number of $+1$ and $-1$ is equal.
(c) To balance the number of $+1$ and $-1$ as much as possible, the $0$ elements in (b) will be adjusted to $+1$ or $-1$, \emph{i.e.}, ``balanced principle".
}
\label{multi_label_instance}
\end{center}
\vspace{-2.8em}
\end{figure}

\subsection{Extended to Multi-label Case} \label{multi_label_case}
The framework elaborated above assigns one column of the Hadamard Matrix as the target code for data points from the same class, which however, may constrain the application in real-world scenarios since some images may be labeled with multiple classes.
To that effect, in this section, we further extend the proposed method to the multi-label case and demonstrate that the generated target code can also effectively approximate to the requirements of $1)$ and $2)$ in Sec.\,\ref{hadamard_matrix}, and thus can preserve the main property of the Hadamard matrix as in Sec.\,\ref{learning_formulation}.

In the multi-label case, we first rewrite the label set $\mathbf{L}^t = \{ l_i^t \}_{i=1}^{n_t}$ as $\mathbf{L}^t = \{ \mathbf{l}_i^t \}_{i=1}^{n_t}$ where $\mathbf{l}_i^t = \{ (l_i^t)_j \}_{j=1}^{o_t}$ and $(l_i^t)_j$ is the $j$-th class $\mathbf{x}_i^t$ belongs to and $o_t$ is the total categories of $\mathbf{x}_i^t$.
In this situation, there are $o_t$ target codes extracted from the Hadamard matrix since each class is randomly assigned with one column of the Hadamard matrix.
As illustrated in Fig.\,\ref{multi_label_instance},
to fuse these target codes into one vector, we propose to use the ``majority principle" and ``balancedness principle".
The initial target code for $\mathbf{x}_i^t$ can be defined as:
\begin{equation}\label{multi_label}
  \mathbf{\hat{c}}_{J(\mathbf{x}_i^t)} = step\big(\sum_{j}^{o_t} \mathbf{c}_{(l_i^t)_j} \big),
\end{equation}
where $\mathbf{c}_{(l_i^t)_j}$ is the $(l_i^t)_j$-th column of the Hadamard Matrix.
The step function $step(x)$ returns +1 if the input variable $x > 0$, 0 if $x = 0$, and $-1$, otherwise.
For simplicity, we denote the $t$-th bit of $\mathbf{\hat{c}}_{J(\mathbf{x}_i^t)}$ as $\hat{c}_t$.
The value of $\hat{c}_t$ falls in $\{+1, 0, -1\}$ and is voted based on the ``majority principle".
As shown in Fig.\,\ref{multi_label_instance}, $\hat{c}_t$ = +1 if the majority elements in the selected $t$-th row are $+1$;
$\hat{c}_t = -1$ if the majority elements in the selected $t$-th row are $1$;
$\hat{c}_t = 0$ if the numbers of $+1$ and $-1$ in the $t$-th row are equal.
Finally, the $0$ elements in $\mathbf{\hat{c}}_{J(\mathbf{x}_i^t)}$ are further re-assigned with $+1$ or $-1$ to balance the total number of $+1$ and $-1$ as far as possible, \emph{i.e.}, ``balancedness principle".
After all, we obtain the final target of $\mathbf{c}_{J(x_i^t)}$.

Below, we analyze that $\mathbf{c}_{J(\mathbf{x}_i^t)}$ in multi-label settings can preserve the main property of the Hadamard codebook.
For brevity, we first denote the $t$-th bit of $\mathbf{c}_{(l_i^t)_j}$ as $(c_j)_t$ and it is easy to obtain:
\begin{equation}\label{each_bit}
\begin{split}
    \hat{c}_t &= step\big(\sum_{j}^{o_t} (c_j)_t \big) \\
      &= step\big(\sum_{j}^{o_{t+}}(+1) + \sum_{j}^{o_{t-}}(-1)\big)   \\
      &= step(o_{t+} - o_{t-}),
\end{split}
\end{equation}
where $o_t = o_{t+} + o_{t-}$, $o_{t+}$ and $o_{t-}$ are the number of $+1$s and $-1$s, respectively.

\textbf{Theorem 2}: At the $t$-th stage, for any multi-label data point $\mathbf{x}_i^t$, it satisfies
\begin{equation}\label{theorem2}
  P(o_{t+} > o_{t-}) = P(o_{t+} < o_{t-}).
\end{equation}

Proof of \textbf{Theorem $2$}:

\emph{Case $1$}: $o_t$ is an odd number.
In this situation, it is easy to know that $o_{t+} \neq o_{t-}$.
When $o_{t+} \ge \frac{o_t + 1}{2}$, $o_{t+} > o_{t-}$, otherwise, $o_{t+} < o_{t-}$.
Hence, we have the following equations:
\begin{align}
    & P(o_{t+} > o_{t-}) + P(o_{t+} < o_{t-}) = 1.\label{sum1}\\
    & P(o_{t+} \ge \frac{o_t + 1}{2}) = P(o_{t+} > o_{t-}).\label{eq1}\\
    & P(o_{t-} \ge \frac{o_t + 1}{2}) = P(o_{t+} < o_{t-}).\label{eq2}
\end{align}

It's easy to obtain:
\begin{align}
  & P(o_{t+} \ge \frac{o_t + 1}{2}) = \frac{\sum_{i=\frac{o_t + 1}{2}}^{o_t}\binom{\frac{r^*}{2}}{i}\binom{\frac{r^*}{2}}{o_t-i}}{\binom{r^*}{o_t}}.\label{p1}\\
  & P(o_{t-} \ge \frac{o_t + 1}{2}) = \frac{\sum_{i=\frac{o_t + 1}{2}}^{o_t}\binom{\frac{r^*}{2}}{i}\binom{\frac{r^*}{2}}{o_t-i}}{\binom{r^*}{o_t}}.\label{p2}
\end{align}

Combining Eq.\,\ref{sum1}, Eq.\,\ref{eq1}, Eq.\,\ref{eq2}, Eq.\,\ref{p1} and Eq.\,\ref{p2}, we obtain:
\begin{equation}\label{odd}
  P(o_{t+} > o_{t-}) = P(o_{t+} < o_{t-}) = 0.5.
\end{equation}

\emph{Discussion 1}.
Based on Eq.\,\ref{each_bit} and Eq.\,\ref{odd}, we have that
the $t$-th bit of $\hat{\mathbf{c}}_{J(\mathbf{x}_i^t)}$, \emph{i.e.}, $\hat{c}_t$, can be either $+1$ with $50\%$ probability or $-1$ with $50\%$ probability.

\emph{Case $2$}: $o_t$ is an even number.

In this situation, the relationships between $o_{t+}$ and $o_{t-}$ are three aspects:
1. When $o_{t+} = \frac{o_t}{2}$, $o_{t+} = o_{t-}$.
2. When $o_{t+} > \frac{o_t}{2}$, $o_{t+} > o_{t-}$.
3. When $o_{t-} > \frac{o_t}{2}$, $o_{t-} > o_{t+}$.
Therefore, we have the following equations:
\begin{align}
     &P(o_{t+} = o_{t-}) + P(o_{t+} > o_{t-}) + P(o_{t+} < o_{t-}) = 1,\label{sum2}\\
     &P(o_{t+} > \frac{o_t}{2}) = P(o_{t+} > o_{t-}),\label{eq3}\\
     &P(o_{t-} > \frac{o_t}{2}) = P(o_{t+} < o_{t-}),\label{eq4}
\end{align}

The probabilities for them are listed as follows:
\begin{align}
    &P(o_{t+} = \frac{o_t}{2}) = \frac{\binom{\frac{r^*}{2}}{\frac{o_t}{2}}\binom{\frac{r^*}{2}}{\frac{o_t}{2}}}{\binom{r^*}{o_t}},\label{p3}\\
    &P(o_{t+} > \frac{o_t}{2}) = \frac{\sum_{i=\frac{o_t + 2}{2}}^{o_t}\binom{\frac{r^*}{2}}{i}\binom{\frac{r^*}{2}}{o_t-i}}{\binom{r^*}{o_t}},\label{p4}\\
    &P(o_{t-} > \frac{o_t}{2}) = \frac{\sum_{i=\frac{o_t + 2}{2}}^{o_t}\binom{\frac{r^*}{2}}{i}\binom{\frac{r^*}{2}}{o_t-i}}{\binom{r^*}{o_t}},\label{p5}
\end{align}

Combining Eq.\,\ref{sum2}, Eq.\,\ref{eq3}, Eq.\,\ref{eq4}, Eq.\,\ref{p3}, Eq.\,\ref{p4} and Eq.\,\ref{p5}, we have the following equations:
\begin{align}
    & P(o_{t+} = o_{t-}) = \frac{\binom{\frac{r^*}{2}}{\frac{o_t}{2}}\binom{\frac{r^*}{2}}{\frac{o_t}{2}}}{\binom{r^*}{o_t}}.\label{p6}\\
    & P(o_{t+} > o_{t-}) = P(o_{t+} < o_{t-}) = \frac{1 - P(o_{t+} = o_{t-})}{2}.\label{even}
\end{align}

Hence, together with Eq.\,\ref{odd} and Eq.\,\ref{even}, we have the demonstration of \textbf{Theorem $2$}.
\emph{Discussion 2}.
For Case $2$, $P(o_{t+} > o_{t-}) = P(o_{t+} < o_{t-}) < 0.5$.
However, when $P(o_{t+} = o_{t-})$, we adopt the ``balancedness principle" as demonstrated in Fig.\,\ref{multi_label_instance} to make sure the balance the number of $+1$ and $-1$.
Hence, the $t$-th bit of $\hat{\mathbf{c}}_{J(\mathbf{x}_i^t)}$, \emph{i.e.}, $\hat{c}_t$, is still either $+1$ with $50\%$ probability or $-1$ with $50\%$ probability.

Denote $P(\hat{c}_t = +1)$ as $\hat{P}_{+1}$ and $P(\hat{c}_t = -1)$ as $\hat{P}_{-1}$.
According to \emph{Discussion $1$} and \emph{Discussion $2$}, we have $\hat{P}_{+1} = \hat{P}_{-1} = 0.5$.
We denote the number of $+1$ in $\mathbf{c}_{J(\mathbf{x}_i^t)}$ as $\hat{M}$.
We can obtain that $\hat{M} \in \{0, 1, ..., r^*\}$ also has a binomial distribution.
According to \textbf{Proposition 2}, with a high probability, the proposed ``majority principle" and ``balancedness principle" can provide an effective target code for the multi-label data, which is approximate to the requirement of $2)$ in Sec.\,\ref{hadamard_matrix}.
Through similar analysis, it can prove that each bit from different data points sharing at least one different categories satisfies the requirement of $1)$.
Therefore, the proposed ``majority principle" and ``balancedness principle" can well extend the proposed method to the multi-label dataset.

\begin{figure}[!tb]
\begin{center}
\includegraphics[height=0.28\linewidth]{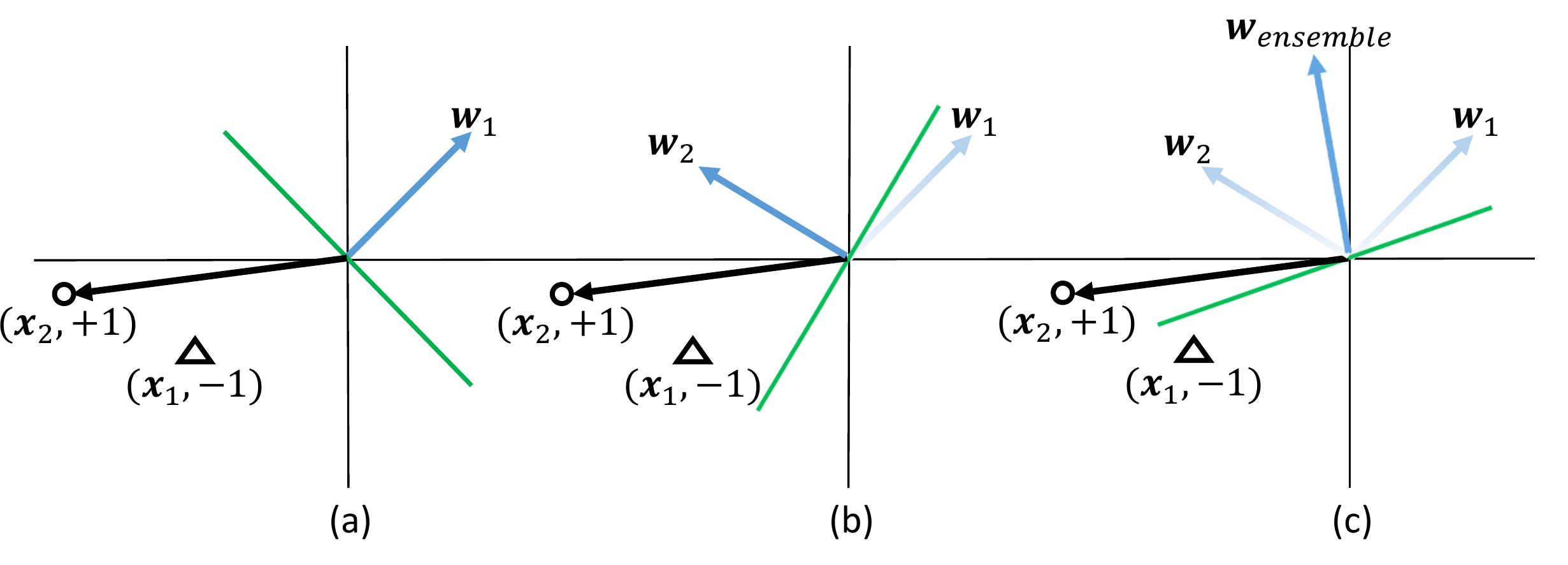}
\caption{A toy example of ensembling.
(a)  $\mathbf{w}_1$ is learned based on $\mathbf{x}_1$, and correctly classifies data point of $\mathbf{x}_1$ while misclassifying $\mathbf{x}_2$.
(b) $\mathbf{w}_2$ is learned based on $\mathbf{x}_2$, and correctly classifies data point of $\mathbf{x}_2$ while misclassifying $\mathbf{x}_1$.
(c) By taking both $\mathbf{w}_1$ and $\mathbf{w}_2$ into considerations, the final $\mathbf{w}_{ensemble}$ correctly classifies both $\mathbf{x}_1$ and $\mathbf{x}_2$.
}
\label{instance}
\end{center}
\vspace{-2.0em}
\end{figure}

\subsection{Ensemble Learning} \label{ensemble_learning}
When data comes sequentially, the Perceptron algorithm makes at most $(\frac{R}{\gamma})^2$ mistakes \citep{novikoff1963convergence}, where the margin $\gamma$ is defined as
$\gamma = min_{t \in [T]|\mathbf{x}^t\mathbf{w}^*|}$ and R is a constant such that ${\forall}_t \in [T], \parallel \mathbf{x}_t \parallel \le R$.
It guarantees a tight mistake bound for online classification.
But, the case in online retrieval is out of function.
Essentially, online classification simply considers prediction on the current streaming data.
However, for online retrieval, it has to preserve the information from the past dataset when learning from the current streaming data, since all data points are retrieved in the query stage.
Therefore, directly applying Perceptron algorithm to retrieval is far from enough.

To solve it, we consider the ensemble learning to learn a weighted combination of base models from the form of
\begin{equation} \label{ensemble_formulation}
	\mathbf{W}_{ensemble} = \sum_{t=1}^{T}{\pi}^t\mathbf{W}^t, \quad\quad\quad \emph{s.t.} \quad \sum_{t=1}^{T}{\pi}^t=1,
\end{equation}
where ${\pi}^t$ is the tunable parameter.
Empirically, we set ${\pi}^t = \frac{1}{T}$.
That is to say, each base model obtains an equal vote on the decision of ensembled model.

Fig.\,\ref{instance} shows a simple example of how ensemble strategy of Eq.\,\ref{ensemble_formulation} works, the quantitative results of which are shown in Sec.\,\ref{ablation_study}.
Generally, the Perceptron algorithm updates each time when a mistake occurs.
However, the updated model merely absorbs the misclassified data point to ensure its correctness.
As a consequence, the information from the past streaming data loses heavily.
Hence, the models updated at different stages are very independent, which is infeasible in a retrieval task.
The ensemble strategy to some extent integrates the independent models.
Besides, since the weighted parameter ${\pi}^t$ is fixed, there has no much time cost to acquire $\mathbf{W}_{ensemble}$ \footnote{Since it is just a matrix-addition operation at each stage.}.

We summarize the proposed Hadamard Matrix Guided Online Hashing (HMOH) in Alg. \ref{alg1}\footnote{$\tilde{W}$ is a random matrix that need not be optimized. When $r=r^*$, we set $\tilde{W}$ as an identity matrix}.

\begin{algorithm}[t]
\caption{ Hadamard Matrix Guided Online Hashing}
\renewcommand{\algorithmicrequire}{\textbf{Input:}}
\renewcommand{\algorithmicensure}{\textbf{Output:}}
\begin{algorithmic}[1]
\REQUIRE
Training data set $D$ with feature space $\mathbf{X}$ and label space $\mathbf{L}$, the number of hash bits $r$, the learning rate $\eta$, the total number of streaming data batches $L$.
\ENSURE
The hash codes $\mathbf{B}$ for training space $\mathbf{X}$ and the projection coefficient  matrix $\mathbf{W}$.\\
\STATE
Initialize $\mathbf{W}^0$ and $\mathbf{W}_{ensemble}$ as all-zero matrices.
\STATE
Set the value of $r^*$ by Eq.\,\ref{r_star}.
\STATE
Generate Hadamard matrix as stated in Sec.\,\ref{hadamard_matrix}.

\IF{$ r = r^* $}
\STATE
Set $\mathbf{\tilde{W}}$ as an identity matrix.
\ELSE
\STATE
Randomize $\mathbf{\tilde{W}}$ from a standard Gaussian distribution.
\ENDIF

\STATE Transform the virtual categories by Eq.\,\ref{transformation} or via Sec.\,\ref{multi_label_case}.

\FOR {$t=1 \to T$}
	\STATE Kernelize $\mathbf{X}^t$ by Eq.\,\ref{kernel}.
    \STATE Obtain $\mathbf{W}^t$ by Eq.\,\ref{updation} and Eq.\,\ref{derivative}.
    \STATE $\mathbf{W}_{ensemble} \leftarrow \mathbf{W}_{ensemble} + \mathbf{W}^t$.
\ENDFOR
\STATE Set $\mathbf{W} = \frac{1}{T}\mathbf{W}_{ensemble}$.
\STATE
Compute $\mathbf{B} = sign(\mathbf{W}^T\mathbf{X})$
\end{algorithmic}
\label{alg1}
\end{algorithm}

\subsection{Time Complexity}\label{time_complexity}
From Alg. \ref{alg1}, at each updating stage, the training time is spent on kernelization of $\mathbf{X}^t$ in line $11$, the updating  of $\mathbf{W}^t$ in line $12$ and the matrix addition for $\mathbf{W}_{ensemble}$ in line $13$.
In line $11$, the time cost is $\mathcal{O}(n_tmd)$.
Updating $\mathbf{W}^t$ in line $12$ takes $\mathcal{O}(mn_tr^*)$.
And it also takes $\mathcal{O}(mr^*)$ in line $13$.
Above all, the total time complexity for the proposed HMOH is $\mathcal{O}(mn_td + mn_tr^*)$.
What's more, as experimentally demonstrated in Sec.\,\ref{ablation_study}, the suitable value of $n_t$ is $1$ for the proposed method.
And we denote $s = \max(d, r^*)$.
Hence, without loss of generality, the overall time complexity can be further abbreviated as $\mathcal{O}(ms)$.
Hence, our method is scalable.

\section{Experiments} \label{experiment}
In this section, we evaluate our Hadamard matrix guided learning framework for online hashing generation.
To verify the performance of the proposed HMOH, we conduct large-scale image retrieval experiments with several state-of-the-art methods
\citep{huang2013online,leng2015online,cakir2015adaptive,cakir2017online,fatih2017mihash,lin2018supervised,lin2019towards} on four widely-used datasets, \emph{i.e.}, CIFAR-$10$\citep{krizhevsky2009learning}, Places$205$ \citep{zhou2014learning}, MNIST \citep{lecun1998gradient} and NUS-WIDE \citep{nus-wide-civr09}.

\subsection{Experimental Settings} \label{experimental_setting}
\textbf{Datasets.}
The \emph{CIFAR-$10$} contains $60,000$ images from $10$ classes with each class containing $6,000$ instances.
Each image is represented by a $4,096$-dim feature, which is extracted from the \emph{fc7} layer of the VGG-$16$ neural network \citep{simonyan2014very} pre-trained on ImageNet \citep{deng2009imagenet}.
Following the settings in \citep{fatih2017mihash,lin2018supervised,lin2019towards}, the whole dataset is split into a retrieval set with $59K$ images and a test set with $1K$ images.
Besides, we randomly sample $20K$ images from the retrieval set to form a training set to learn the hash functions.

The \emph{Places$205$} is a subset of the large-scale Places dataset \citep{zhou2014learning} for scene recognition.
It contains $2.5$ million images with each image belonging to one of the $205$ scene categories.
Feature of each image is first extracted from the \emph{fc7} layer of the AlexNet \citep{krizhevsky2012imagenet} and then represented as a $128$-dim feature by performing PCA.
To split the entire dataset, following \citep{lin2018supervised}, we randomly select $20$ instances from each category and the remaining is treated as the retrieval set.
Lastly, a random subset of $100K$ images from the retrieval set is used to update the hash functions.

The \emph{MNIST} dataset contains $70K$ handwritten digit images from $0$ to $9$.
Each image is represented by $784$-dim normalized original pixels.
According to the experimental settings in \citep{lin2019towards},
the dataset is divided into a test set with $100$ examples randomly sampled from each class and a retrieval set with all remaining examples.
$20K$ images from the retrieval set is sampled to form a training set.

The \emph{NUS-WIDE} is collected from Flickr, which contains $296,648$ images.
All images are manually annotated with at least one label from $81$ concepts.
following \citep{zhou2014latent,liu2018dense}, we preserve $186,577$ labeled images from the whole dataset according to the top $10$ frequent labels.
In this dataset, each image is represented as a $500$-dim bag-of-visual-words feature.
We choose $2,000$ images from this dataset as the query set, and the remaining as the retrieval set.
From the retrieval set, $40$K images are randomly sampled as the training set.

\textbf{Evaluation Protocols.}
We report the experimental results using mean Average Precision (denoted as \textit{mAP}), Precision within a Hamming ball of radius $2$ centered on each query (denoted as \textit{Precision@H$2$}), \textit{mAP vs. different sizes of training instances curves} and their corresponding areas under the \emph{m}AP curves (denoted as \textit{AUC}), Precision of the top K retrieved neighbors (denoted as \textit{Precision@K}) and their corresponding areas under the Precision@K curves (denoted as \textit{AUC}), and Precision-Recall Curves.
Notably, when reporting the \emph{m}AP performance on Places-205, following the works in \citep{fatih2017mihash,lin2018supervised,lin2019towards}, we only compute the top $1,000$ retrieved items (denoted as \textit{mAP@1,000}) due to its large scale and time consumption.
The above metrics are evaluated under hashing bits varying among $8$, $16$, $32$, $48$, $64$ and $128$.

\textbf{Baseline Methods.} We compare our method with representative state-of-the-art online hashing algorithms, including Online Kernel Hashing (OKH) \citep{huang2013online}, Online Sketching Hashing (SketchHash) \citep{leng2015online}, Adaptive Hashing (AdaptHash) \citep{cakir2015adaptive}, Online Supervised Hashing (OSH) \citep{cakir2017online}, Online Hashing with Mutual Information (MIHash) \citep{fatih2017mihash} and Balanced Similarity for Online Discrete Hashing (BSODH) \citep{lin2019towards}.
Besides, to demonstrate the advantages and improvements of the proposed HMOH, we also compare it with the previous version, \emph{i.e.}, HCOH \citep{lin2018supervised}.
The public MATLAB codes of these methods are available. Our model is also implemented with MATLAB. All the experiments are performed on a server with a $3.60$GHz Intel Core I$7$ $4790$ CPU and $16$G RAM, and the experimental results are averaged over three runs.

\begin{table}[]
\centering
\caption{Parameter configurations on the four benchmarks.}
\label{setting}
\begin{tabular}{c|c|c|c|c}
\hline
Method     & CIFAR-$10$              &Places$205$        &MNIST         &NUS-WIDE\\
\hline
\hline
Kernel        &$\times$                &$\surd$           &$\surd$      &$\times$\\
\hline
$\sigma$      &$\times$                &6                 &10           &$\times$\\
\hline
$m$             &$\times$                &800               &300          &$\times$\\
\hline
$\lambda$     &0.5                     &0.01              &0.1          &0.1\\
\hline
$n_t$           &1                       &1                 &1            &1 \\
\hline
\end{tabular}
\vspace{-1.5em}
\end{table}

\begin{table*}[]
\centering
\caption{\textit{m}AP and Precision@H$2$ Comparisons on CIFAR-$10$ with $8$, $16$, $32$, $48$, $64$ and $128$ bits. %
The best result is labeled with boldface and the second best is with an underline.}
\label{map_precision_cifar}
\begin{tabular}{c|cccccc|cccccc}
\hline
\multirow{2}{*}{Method} & \multicolumn{6}{c|}{\textit{m}AP}                   & \multicolumn{6}{c}{Precision@H$2$}          \\
\cline{2-13}
                        & 8-bit & 16-bit & 32-bit &48-bit & 64-bit & 128-bit & 8-bit & 16-bit & 32-bit &48-bit & 64-bit
                        &128-bit \\
\hline
OKH                     &0.100  &0.134   &0.223   &0.252  &0.268  &0.350    &0.100  &0.175   &0.100   &0.452  &0.175
                        &0.372  \\
\hline
SketchHash              &0.248  &0.301   &0.302   &0.327  &0.326  &0.351   &0.256  &0.431   &0.385   &0.059   &0.004
                        & 0.001     \\
\hline
AdaptHash               &0.116  &0.138   &0.216   &0.297  &0.305  &0.293    &0.114  &0.254   &0.185   &0.093  &0.166
                        &0.164      \\
\hline
OSH                     &0.123  &0.126   &0.129   &0.131  &0.127  &0.125    &0.120  &0.123   &0.137   &0.117  &0.083
                        &0.038 \\
\hline
MIHash                  &0.512  &0.640   &0.675   &0.668  &0.667  &0.664    &0.170  &0.673   &0.657   &0.604  &0.500
                        &0.413   \\
\hline
BSODH                   &\underline{0.564}  &0.604   &\underline{0.689} &0.656  &0.709  &0.711    &0.305  &0.582   &0.691 &\underline{0.697}  &\underline{0.690}
                        &\underline{0.602}  \\
\hline
\hline
HCOH                    &0.536  &\underline{0.698}   &0.688 &\underline{0.707} &\underline{0.724}  &\underline{0.734}
                        &\underline{0.333}  &\underline{0.723}   &\underline{0.731} &0.694   &0.633  &0.471 \\
\hline
HMOH                    &\textbf{0.600}  &\textbf{0.732}   &\textbf{0.723} &\textbf{0.734} &\textbf{0.737}  &\textbf{0.749}
                        &\textbf{0.348}  &\textbf{0.756}   &\textbf{0.743} &\textbf{0.729}  &\textbf{0.710}  &\textbf{0.734} \\
\hline
\end{tabular}
\end{table*}
\begin{figure*}[!t]
\begin{center}
\includegraphics[height=0.48\linewidth]{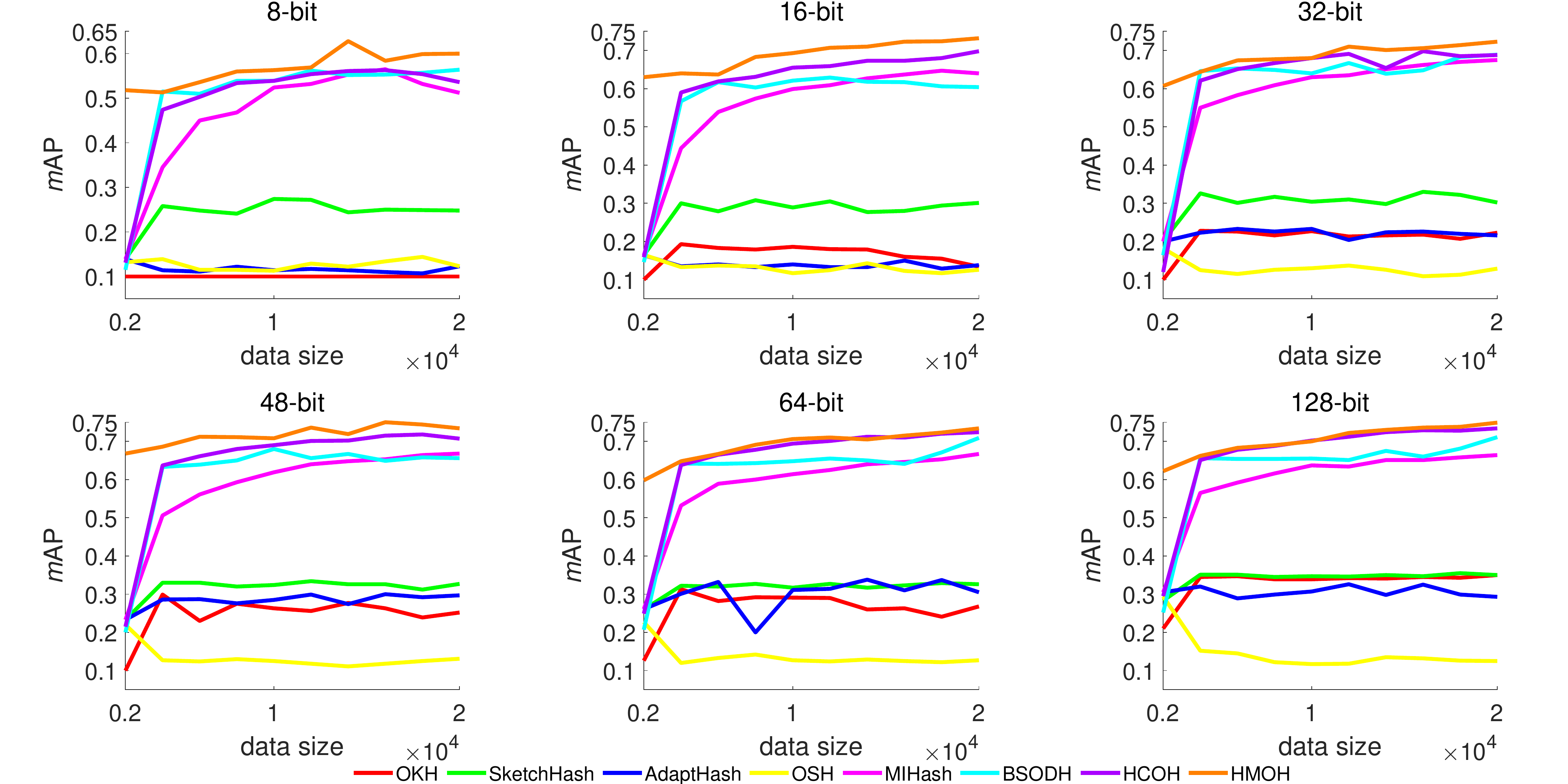}
\caption{\label{map_instance_cifar} \textit{m}AP performance with respect to different sizes of training instances on CIFAR-$10$.}
\end{center}
\vspace{-1em}
\end{figure*}
\begin{figure}[!t]
\begin{center}
\includegraphics[height=0.53\linewidth]{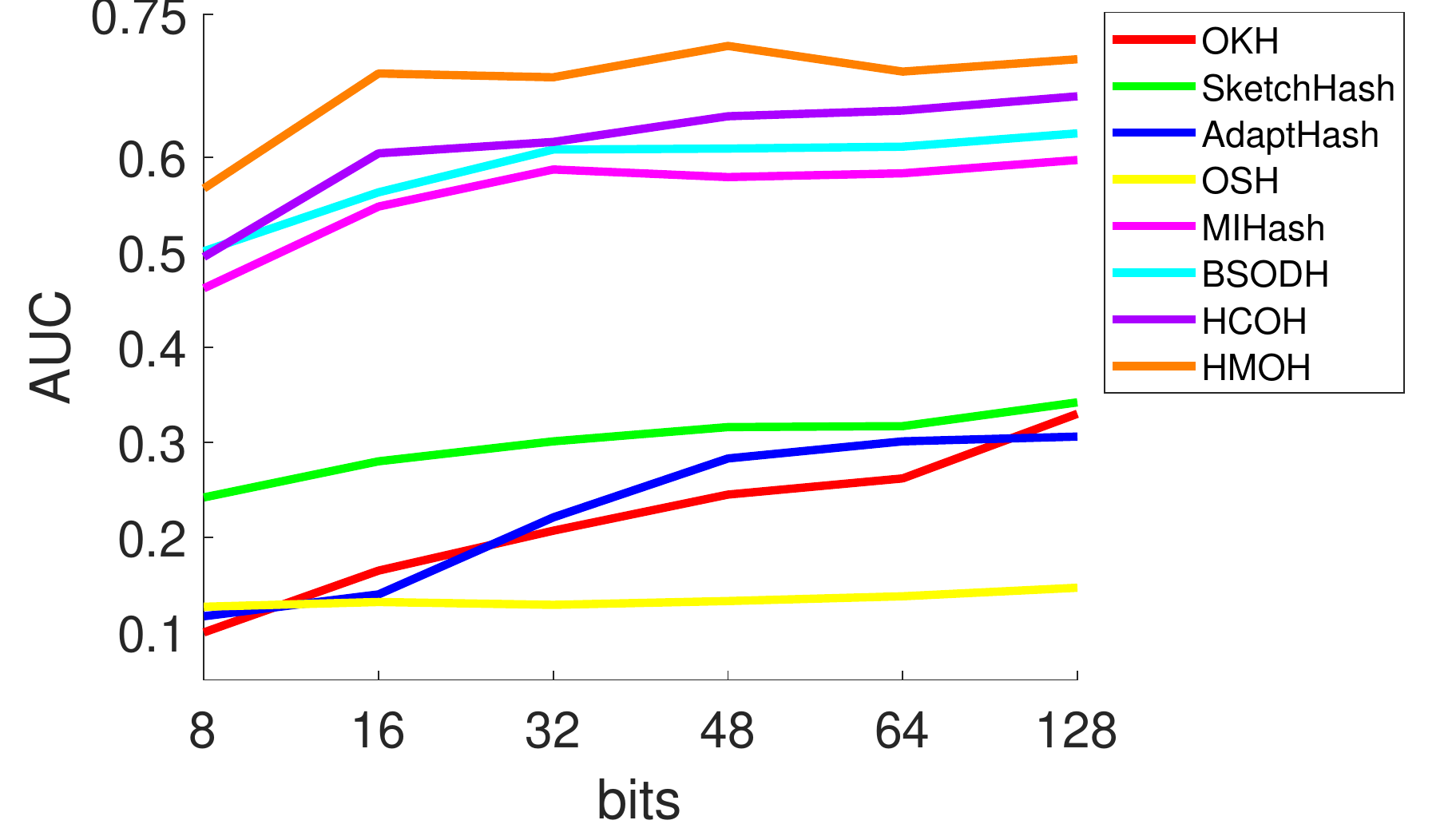}
\caption{\label{map_auc_cifar} AUC curves for mAP on CIFAR-$10$.}
\end{center}
\vspace{-2.5em}
\end{figure}
\begin{figure*}[!t]
\begin{center}
\includegraphics[height=0.48\linewidth]{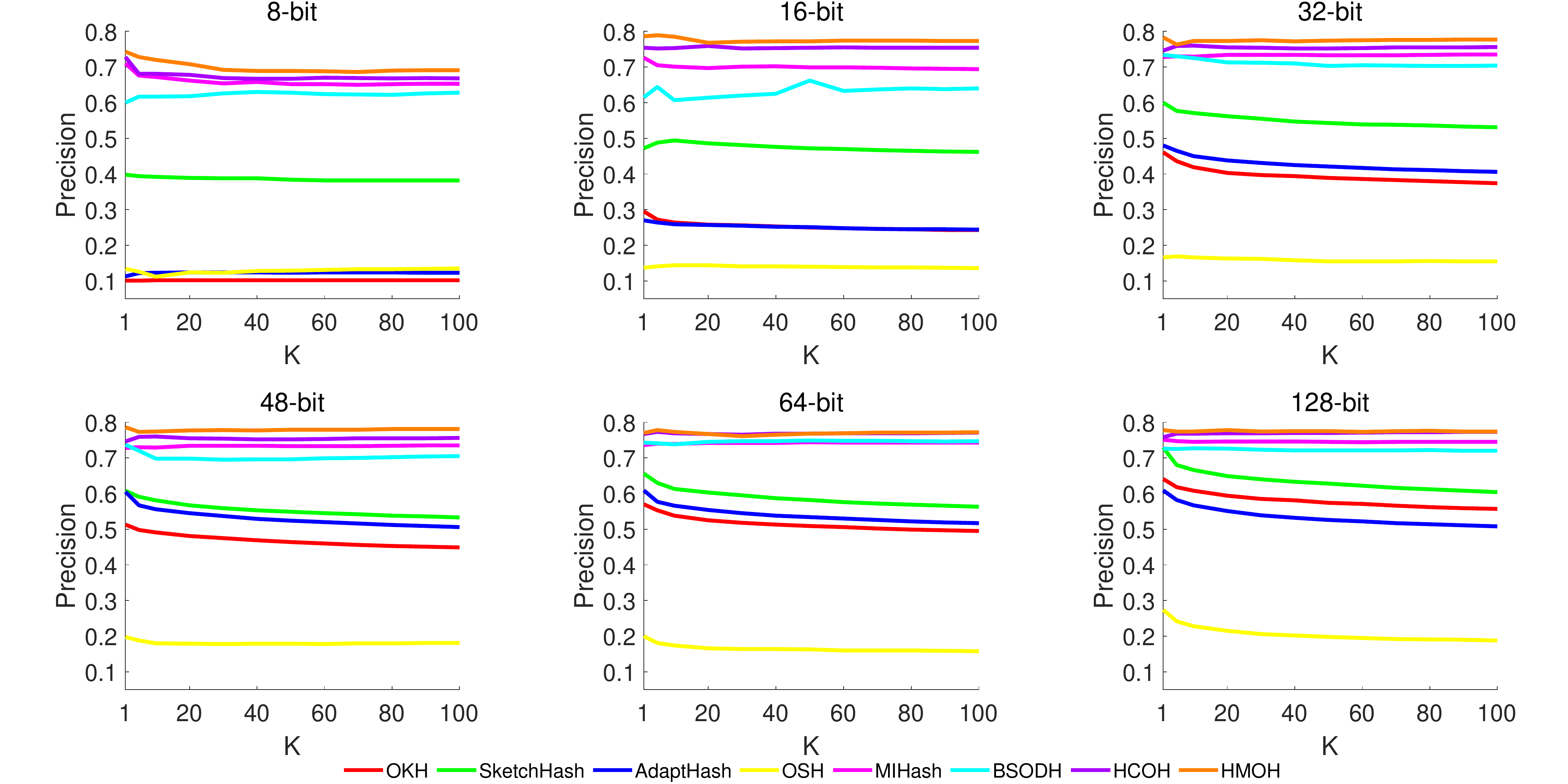}
\caption{\label{precision_cifar} Precision@K curves of compared algorithms on CIFAR-$10$.}
\end{center}
\vspace{-1em}
\end{figure*}
\begin{figure}[!t]
\begin{center}
\includegraphics[height=0.53\linewidth]{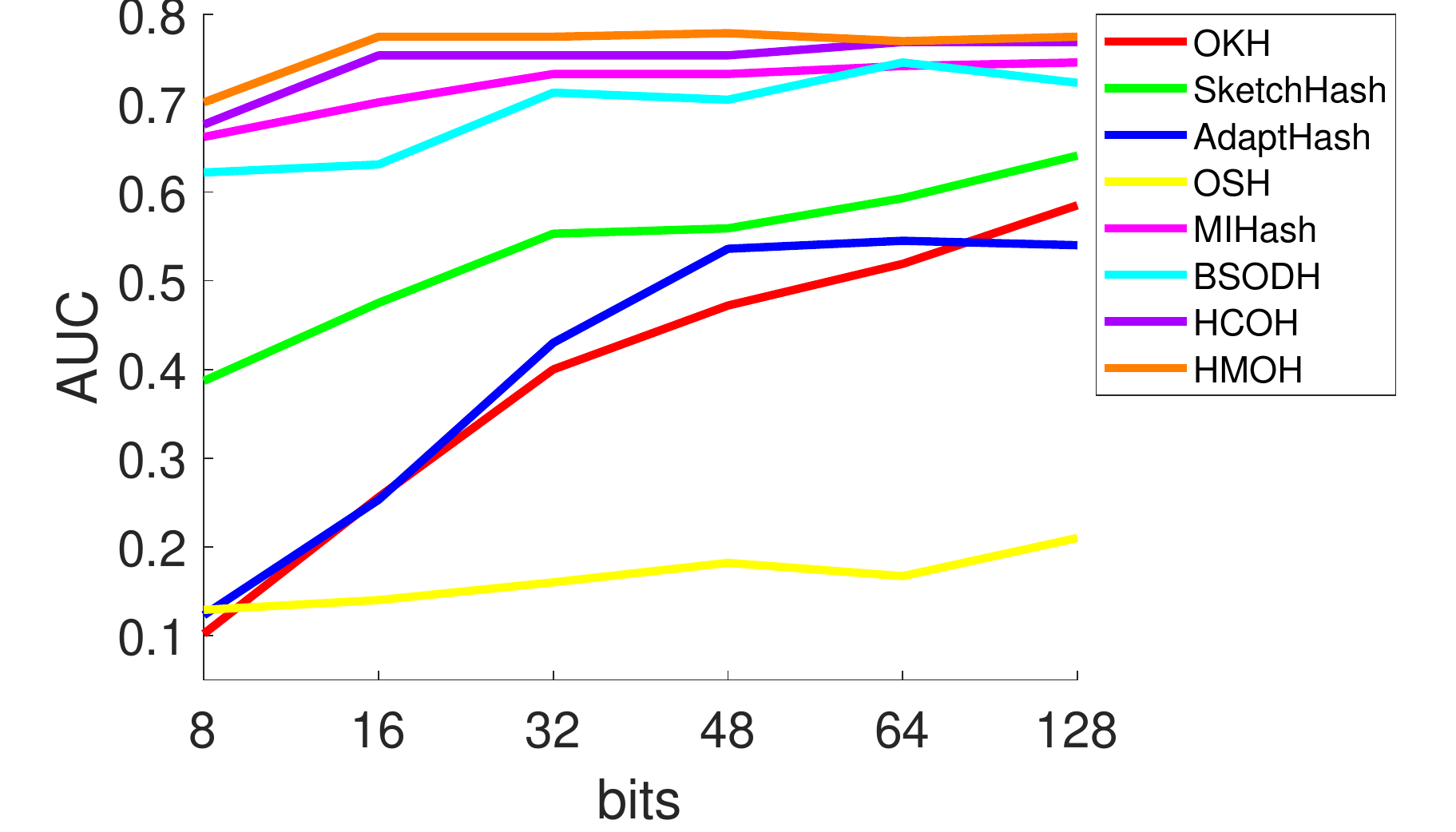}
\caption{\label{precision_auc_cifar} AUC curves for Precision@K on CIFAR-$10$.}
\end{center}
\vspace{-2.5em}
\end{figure}

\textbf{Parametric Settings.} We describe the parameters to be tuned during the experiments.
Since we share the same dataset configurations on CFIAR-10, Places205 and MNIST, we directly adopt the parameters as described in \citep{lin2018supervised,lin2019towards}, which have been carefully validated for each method.
For NUS-WIDE, we also conduct careful ablation studies for all methods and set the optimal values for all the hyper-parameters.
The following describes the parameter configurations for all compared baselines in details.
\begin{itemize}
\item \textbf{OKH}: The tuple ($C$, $\alpha$) is set as (0.001, 0.3), (0.0001, 0.7), (0.001,0.3) and (0.001, 0.5) on CIFAR-10, Places205, MNIST and NUS-WIDE, respectively.
\item \textbf{SketchHash}: The tuple $(sketch size, batch size)$ is set as (200, 50), (100, 50), (200, 50) and (200, 50) on CIFAR-10, Places205, MNIST and NUS-WIDE, respectively.
\item \textbf{AdaptHash}: The tuple $(\alpha, \lambda, \eta)$ is set as (0.9, 0.01, 0.1), (0.9,0.01,0.1), (0.8, 0.01, 0.2) and (1, 0.01, 0.5) on CIFAR-10, Places205, MNIST and NUS-WIDE, respectively.
\item \textbf{OSH}: On all datasets, $\eta$ is set as $0.1$ and the ECOC codebook $C$ is populated the same way as in \citep{cakir2017online}.
\item \textbf{MIHash}: The tuple $(\theta, \mathcal{R}, A)$ is set as (0, 1000, 10), (0, 5000, 10), (0, 1000, 10) and (0, 2000, 1) on CIFAR-10, Places205, MNIST and NUS-WIDE, respectively.
\item \textbf{BSODH}: The tuple $(\lambda, \sigma, {\eta}_s, {\eta}_d)$ is set as (0.6, 0.5, 1.2, 0.2), (0.3, 0.5, 1.0, 0.0), (0.9, 0.8, 1.2, 0.2) and (0.3, 0.1, 0.4, 1.2) on CIFAR-10, Places205, MNIST and NUS-WIDE, respectively.
\item \textbf{HCOH}: The tuple $(n_t, \eta)$ is set as (1, 0.2), (1, 0.1), (1, 0.2) and (1, 0.2) on CIFAR-10, Places$205$, MNIST and NUS-WIDE, respectively.
\end{itemize}

Specific descriptions of these parameters for each method can be found in \citep{huang2013online,leng2015online,cakir2015adaptive,cakir2017online,fatih2017mihash,lin2019towards,lin2018supervised}, respectively.
As for the proposed HMOH, we list the parameter configurations on the four benchmarks in Tab. \ref{setting}.
Notably, experiments of the proposed method without kernelization on CIFAR-$10$ and NUS-WIDE show better results. Hence, kernel trick is not applied in the cases of CIFAR-$10$ and NUS-WIDE.
Detailed analysis is conducted in Sec.\,\ref{ablation_study}.

Emphatically, for SketchHash \citep{leng2015online}, it has two limitations:
First, the training size has to be larger than the code length.
Second, the code length has to be smaller than the dimension of input features.
Therefore, we show its experimental results with hashing bit being $8$, $16$, $32$, $48$ to follow the works in \citep{lin2018supervised,lin2019towards} by setting the training size as $50$.
To evaluate the hashing bit of $64$ and $128$, the training size is set as $150$.
In the longer code length (\emph{e.g.}, 256-bit and 512-bit), the experiments can not be conducted on some benchmarks due to the second limitation.
All the experiments are run over three times and we report the averaged values in this paper.

\begin{figure*}[!t]
\begin{center}
\includegraphics[height=0.48\linewidth]{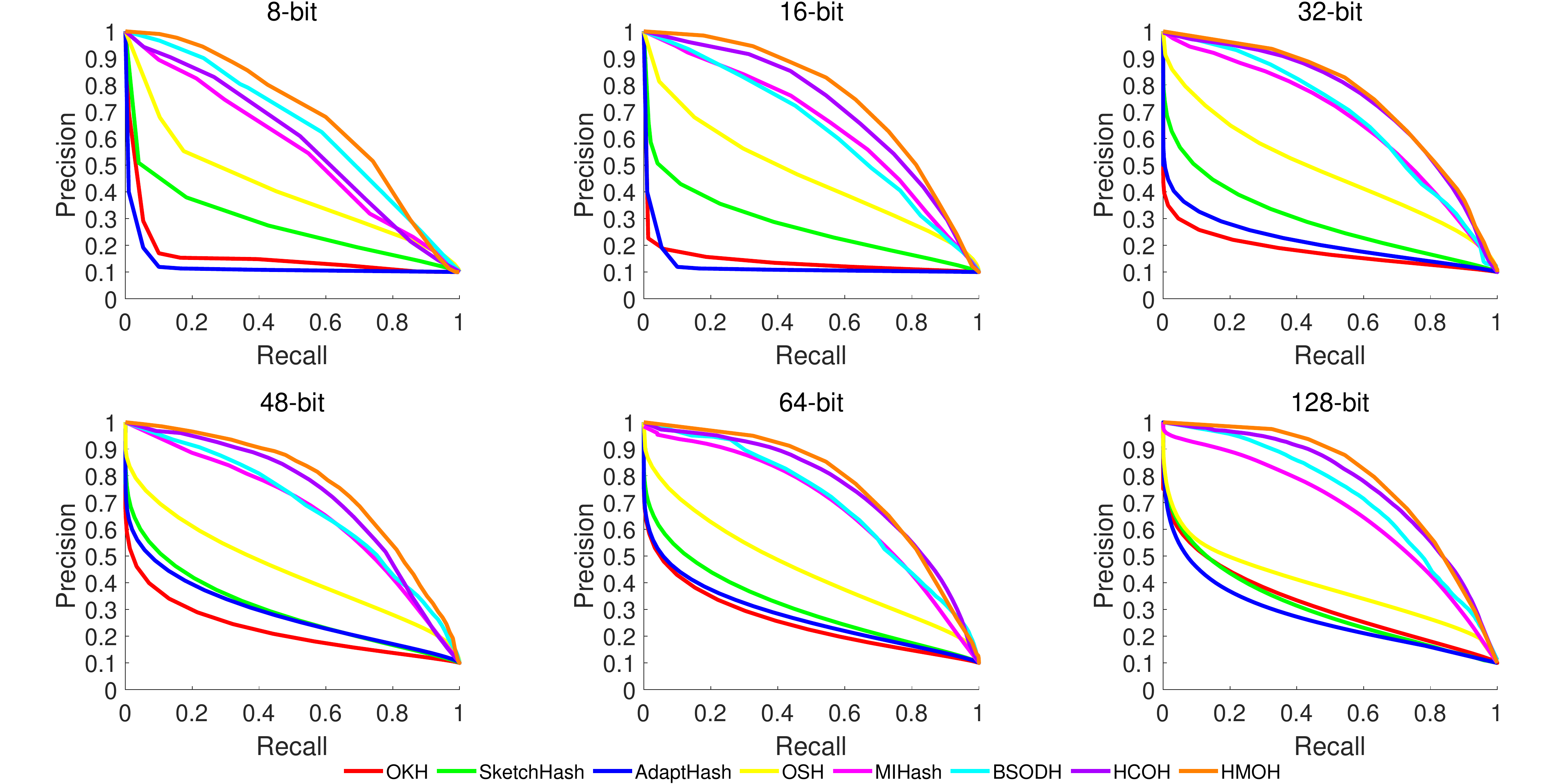}
\caption{\label{pr_cifar} Precision-Recall curves of compared algorithms on CIFAR-$10$.}
\end{center}
\vspace{-2em}
\end{figure*}

\subsection{Results and Discussions \label{results}}

\subsubsection{Results on CIFAR-$10$}
The \emph{m}AP and Precision@H$2$ values of the proposed HMOH and seven baseline methods on CIFAR-$10$ dataset are reported in Tab. \ref{map_precision_cifar}.
The \emph{m}AP vs. different sizes of training instances and their corresponding AUC curves are plotted in Fig.\,\ref{map_instance_cifar} and Fig.\,\ref{map_auc_cifar}, respectively.
The Precision@K curves and their corresponding AUC curves are shown in Fig.\,\ref{precision_cifar} and Fig.\,\ref{precision_auc_cifar}, respectively.
Finally, Fig.\,\ref{pr_cifar} depicts the Precision-Recall curves.

In terms of \emph{m}AP, from Tab. \ref{map_precision_cifar}, we can observe that the proposed HMOH obtains the best results in all cases and performs much better than the baselines in some cases, which well demonstrates its effectiveness.
Detailedly, compared with the best baseline, \emph{i.e.}, BSODH or MIHash, the proposed HMOH outperforms them by an average percentage of $7.478\%$.
Meanwhile, compared with the previous version of the proposed method, \emph{i.e.}, HCOH, HMOH in our paper obtains an average increase of $4.926\%$.
Regarding the Precision@H$2$, it can be observed that the proposed method still achieves superior retrieval results by a margin.
Quantitively, compared with BSODH or MIHash, the proposed HMOH achieves satisfactory performance with an average gain of $10.440\%$.
Similarly, HMOH obtains a substantial improvement of $13.971\%$ over the previous version of HCOH.
Noticeably, we observe that when the hashing bit grows up to $128$, most of other methods suffer a great deal of performance loss (\emph{e.g.}, MIHash: $0.500 \rightarrow 0.413$, BSODH: $0.690 \rightarrow 0.602$ and HCOH: $0.633 \rightarrow 0.471$). However, the proposed HMOH still shows an increasingly high Precision@H$2$ result ($0.710 \rightarrow 0.734$), which demonstrates its robustness.

Next, we further look into the \emph{m}AP over time for all the online hashing methods as depicted in Fig.\,\ref{map_instance_cifar} and their corresponding AUC results in Fig.\,\ref{map_auc_cifar}.
Based on Fig.\,\ref{map_instance_cifar}, we have the following two observations.
First, most cases of all hashing bits, the proposed HMOH yields the best \emph{m}AP results compared with other methods over time. This can be reflected in their AUC results in Fig.\,\ref{map_auc_cifar}.
In detail, the proposed HMOH surpasses the best baseline, \emph{i.e.}, BSODH by an average $15.170\%$ gain and outperforms the previous version of HCOH by an average increase of $10.532\%$.
The second observation is that Fig.\,\ref{map_instance_cifar} also implicates a stable generalization ability of the proposed HMOH.
That is, HMOH achieves satisfactory performance with only a small batch of training instances.
Especially, taking the case of code length being $48$ as an example,
when the size of training data is $2K$, the proposed HMOH gets an \emph{m}AP of $0.668$ compared with other state-of-the-art baselines, \emph{e.g.}, $0.233$ \emph{m}AP for MIHash, $0.200$ \emph{m}AP for BSODH and $0.215$ \emph{m}AP for HCOH.
To achieve similar performance, it takes $20K$ training instances for MIHash and BSODH, $10K$ training instances for our previous version of HCOH, which is inefficient.

Moreover, experimental results for Precision@K and their AUC curves are reported in Fig.\,\ref{precision_cifar} and Fig.\,\ref{precision_auc_cifar}, respectively.
We can find in Fig.\,\ref{precision_cifar} that in low code length ($\le 48$), HMOH transcends other methods by a clear margin.
While the proposed HMOH shows similar results with its previous version, \emph{i.e.}, HCOH, in large code length ($\ge 64$), it still holds the first position for all hashing bits.
Quantitively speaking, as far as their AUC performance in Fig.\,\ref{precision_auc_cifar}, the proposed HMOH consistently outperforms the best baseline by an average of $6.019\%$ AUC gain.
And compared with HCOH, our proposed HMOH still outperforms by an average increase of $2.249\%$.
Hence, the proposed HMOH shows a great improvement over the previous version and its effectiveness over other methods.

For further analysis, we plot the Precision-Recall curves in Fig.\,\ref{pr_cifar}.
From Fig.\,\ref{pr_cifar}, we can observe similar results to those in Tab.\,\ref{map_precision_cifar} (\emph{m}AP), Fig.\,\ref{map_instance_cifar} and Fig.\,\ref{precision_cifar}.
In most cases, the proposed HMOH and its previous version HCOH consistently outperform all other methods.
In the case of $8$-bit, BSODH ranks the second.
No matter what, the proposed HMOH generally performs the best from short code length to long code length.

\begin{table*}[]
\centering
\caption{\textit{m}AP@1,000 and Precision@H$2$ Comparisons on Places$205$ with $8$, $16$, $32$, $48$, $64$ and $128$ bits. The best result is labeled with boldface and the second best is with an underline.}
\label{map_precision_places}
\begin{tabular}{c|cccccc|cccccc}
\hline
\multirow{2}{*}{Method} & \multicolumn{6}{c|}{\textit{m}AP@1,000}                   & \multicolumn{5}{c}{Precision@H$2$}          \\
\cline{2-13}
                        & 8-bit & 16-bit & 32-bit &48-bit &64-bit & 128-bit & 8-bit & 16-bit & 32-bit &48-bit & 64-bit
                        & 128-bit \\
\hline
OKH                     &0.018  &0.033   &0.122   &0.048  &0.114  &0.258    &0.007  &0.010   &0.026   &0.017  &0.217
                        &0.075     \\
\hline
SketchHash              &0.052  &0.120   &0.202   &0.242  &0.274  &0.314    &\underline{0.017}  &0.066   &0.220   &0.176  &0.274
                        &0.016   \\
\hline
AdaptHash               &0.028  &0.097   &0.195   &0.223  &0.222  &0.229    &0.009  &0.051   &0.012   &0.185  &0.021
                        &0.022   \\
\hline
OSH                     &0.018  &0.021   &0.022   &0.032  &0.043  &0.164    &0.007  &0.009   &0.012   &0.023  &0.030
                        &0.059  \\
\hline
MIHash                  &\underline{0.094}&\underline{0.191}&0.244  &\underline{0.288} &0.308  &0.332    &\textbf{0.022}  &\underline{0.112}   &0.204  &0.242  &0.202
                        &0.069 \\
\hline
BSODH                   &0.035  &0.174   &0.250   &0.273  &0.308  &0.337    &0.009  &0.101   &0.241   &\underline{0.246} &\underline{0.212}
                        &\underline{0.101}  \\
\hline
\hline
HCOH                    &0.049  &0.173   &\underline{0.259} &0.280  &\underline{0.321}&\underline{0.347} &0.012  &0.082 &\underline{0.252} &0.179   &0.114
                        &0.036 \\
\hline
HMOH                    &\textbf{0.102}&\textbf{0.232}&\textbf{0.305} &\textbf{0.314} &\textbf{0.335}&\textbf{0.349}
                        &0.014&\textbf{0.137}&\textbf{0.296} &\textbf{0.262}  &\textbf{0.223}&\textbf{0.137}\\
\hline
\end{tabular}
\end{table*}

\begin{figure*}[!t]
\begin{center}
\includegraphics[height=0.48\linewidth]{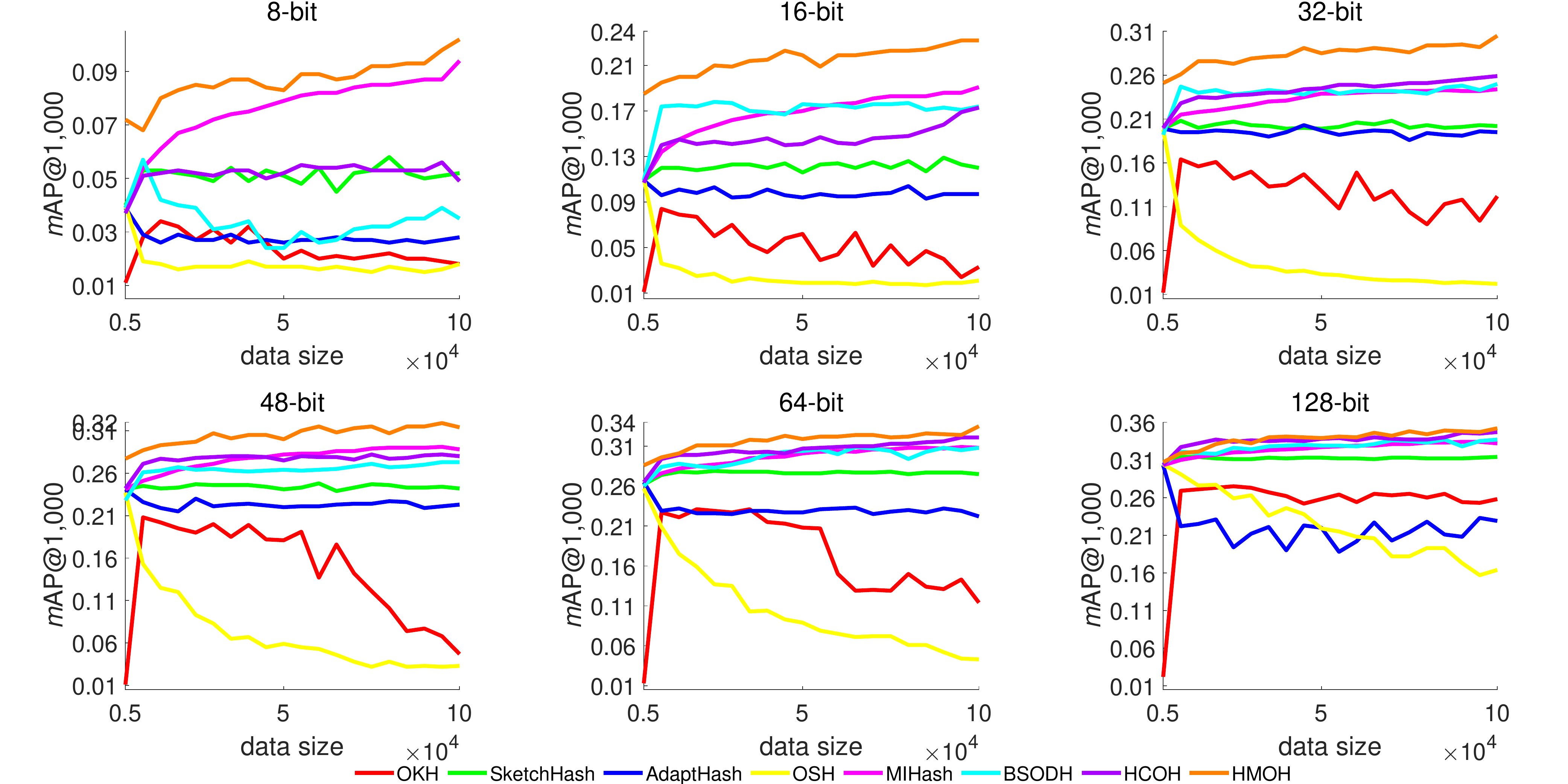}
\caption{\label{map_instance_places} \textit{m}AP performance with respect to different sizes of training instances on Places$205$.}
\end{center}
\vspace{-1em}
\end{figure*}
\begin{figure}[!t]
\begin{center}
\includegraphics[height=0.53\linewidth]{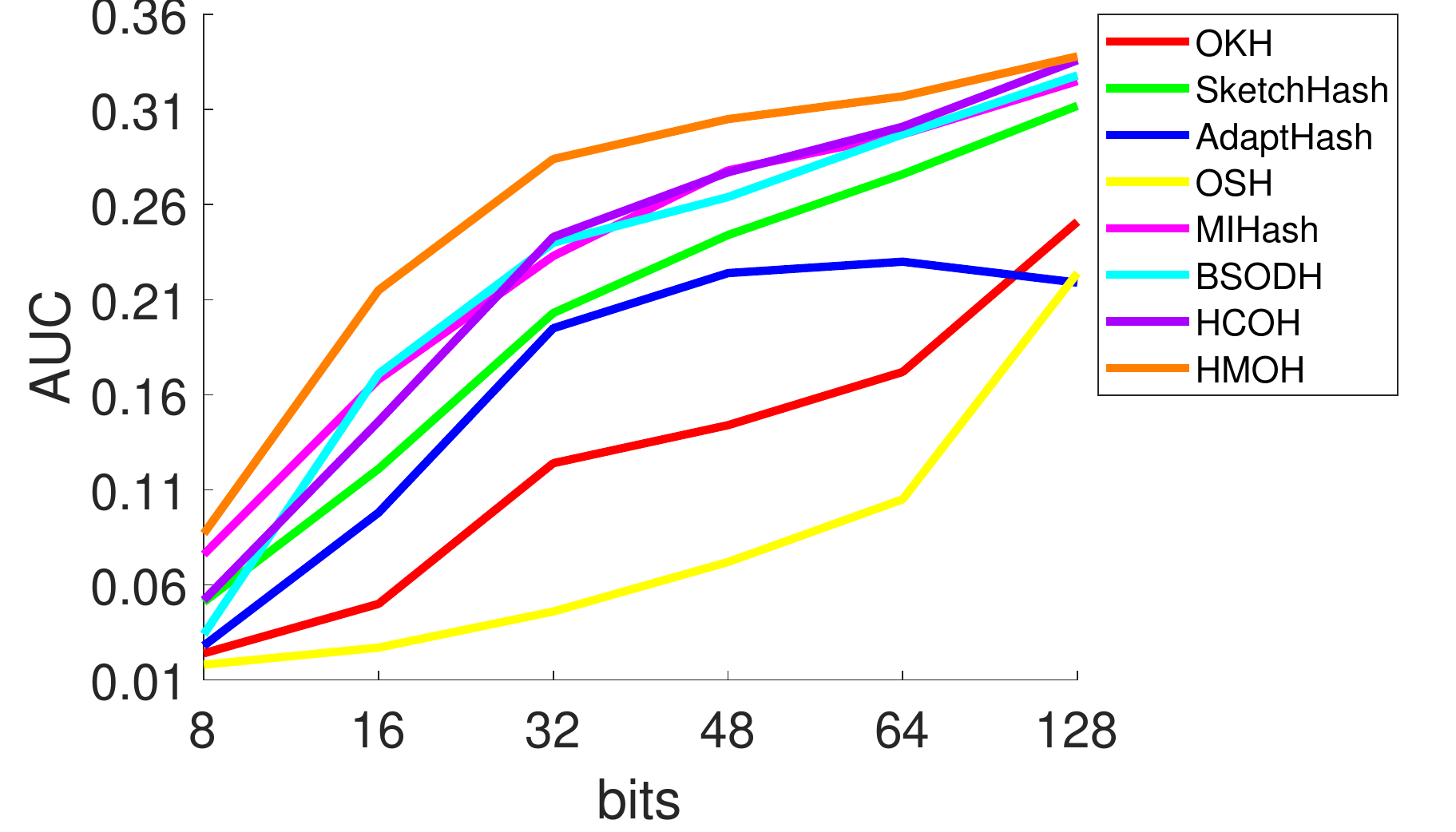}
\caption{\label{map_auc_places} AUC curves for mAP on Places$205$.}
\end{center}
\vspace{-2.5em}
\end{figure}
\begin{figure*}[!t]
\begin{center}
\includegraphics[height=0.48\linewidth]{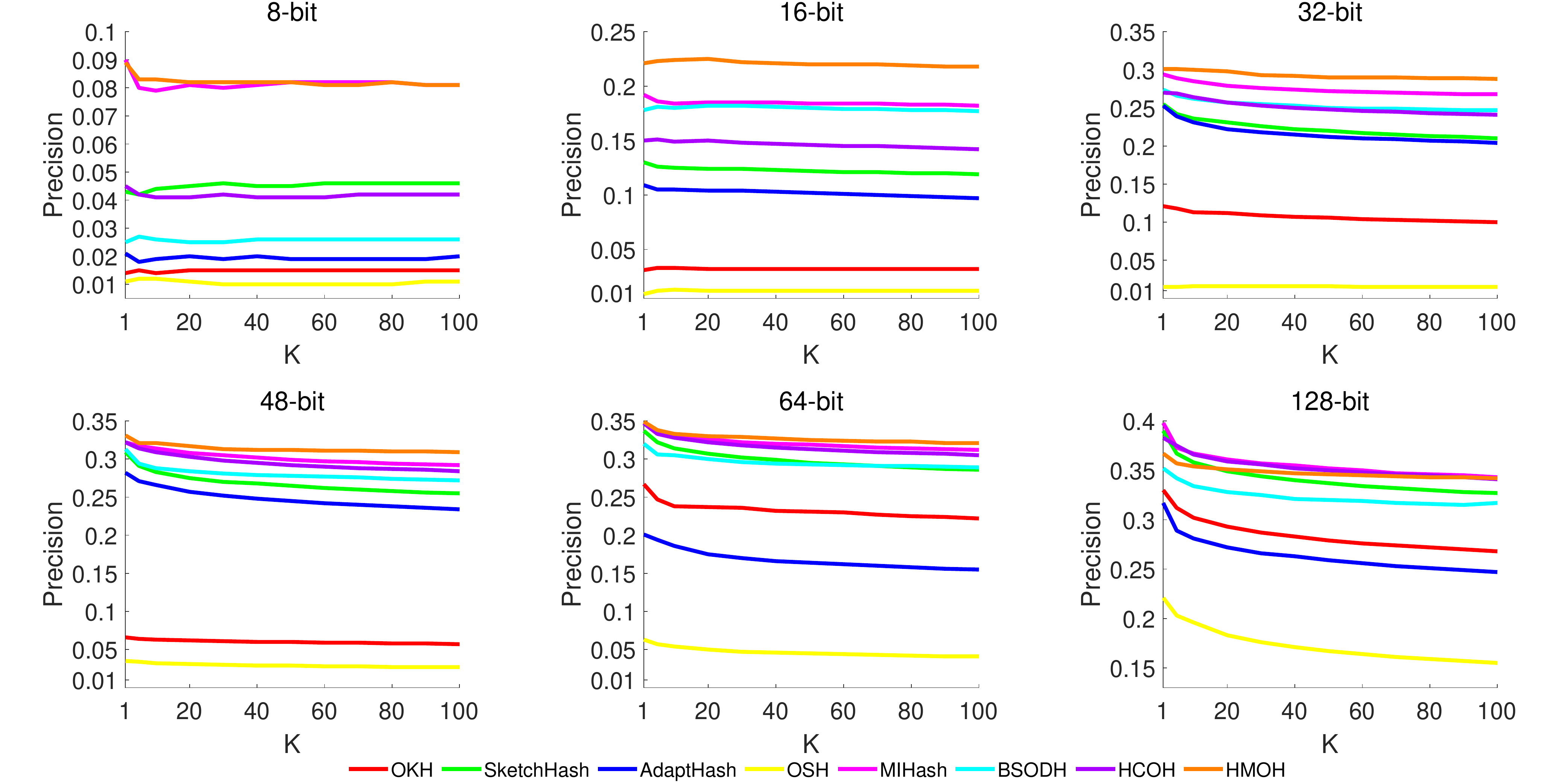}
\caption{\label{precision_places} Precision@K curves of compared algorithms on Places$205$.}
\end{center}
\vspace{-1em}
\end{figure*}
\begin{figure}[!t]
\begin{center}
\includegraphics[height=0.53\linewidth]{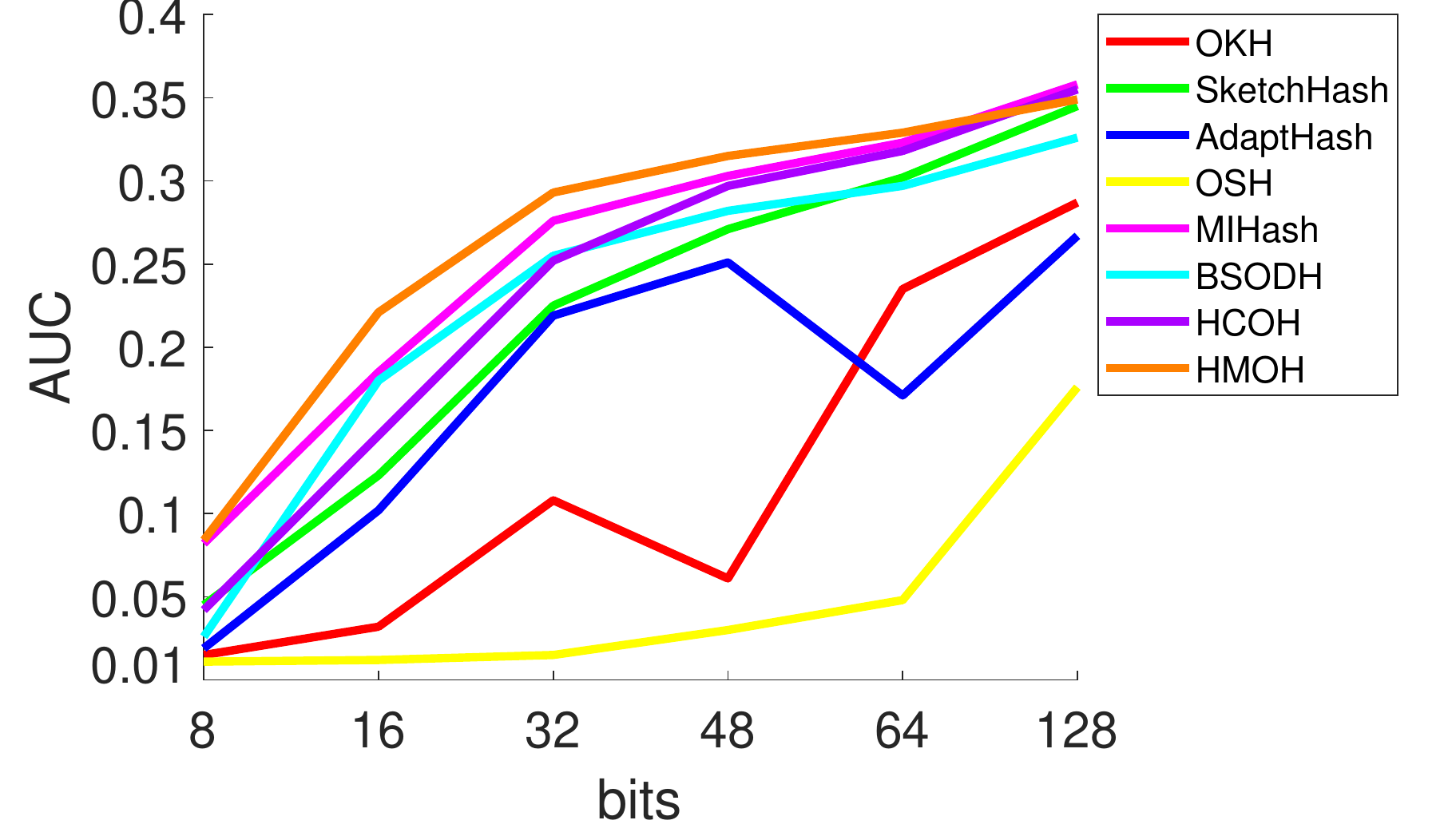}
\caption{\label{precision_auc_places} AUC curves for Precision@K on Places$205$.}
\end{center}
\vspace{-2.5em}
\end{figure}
\begin{figure*}[!t]
\begin{center}
\includegraphics[height=0.48\linewidth]{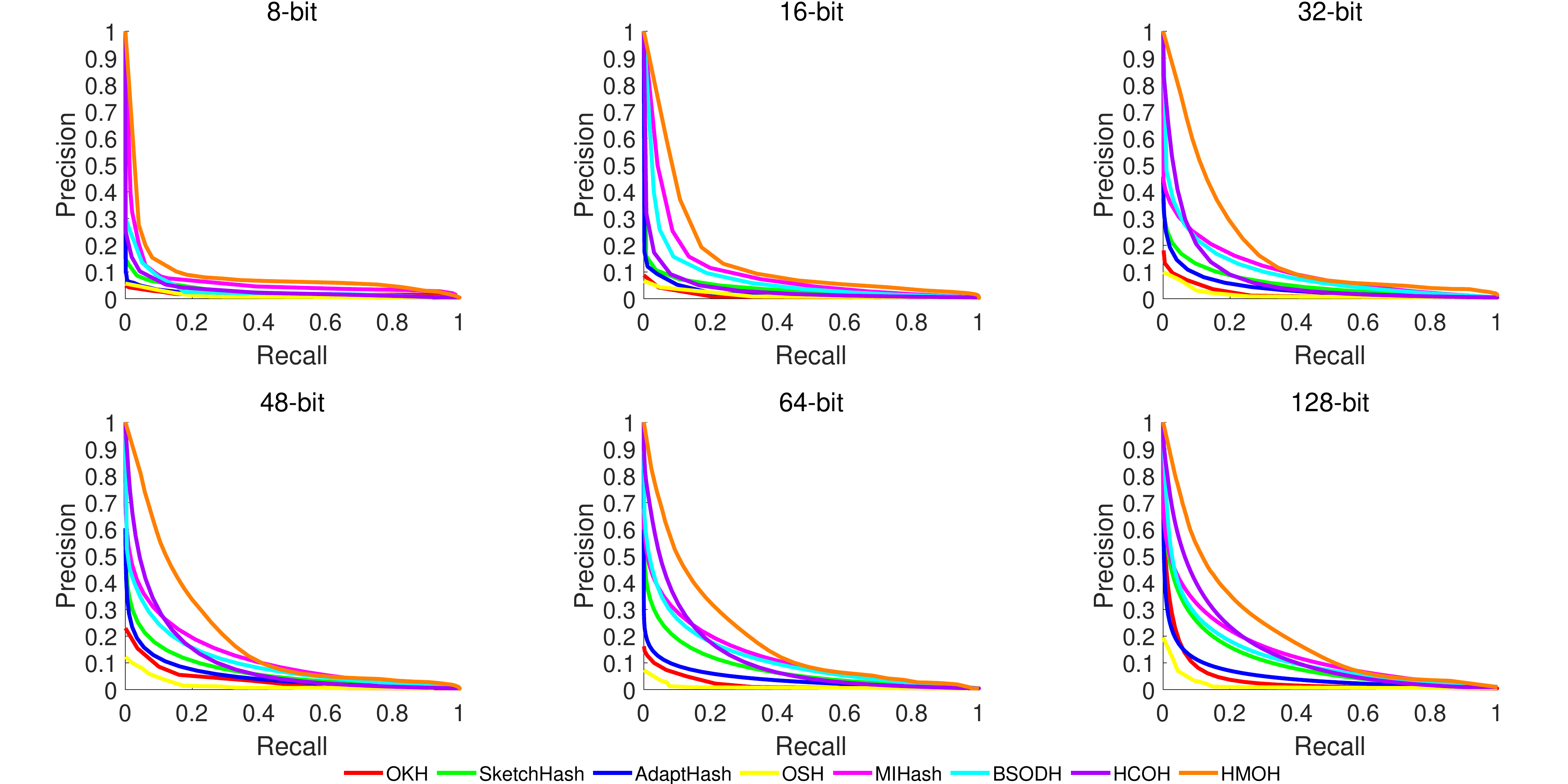}
\caption{\label{pr_places} Precision-Recall curves of compared algorithms on Places$205$.}
\end{center}
\vspace{-2em}
\end{figure*}

\subsubsection{Results on Places$205$}
Tab. \ref{map_precision_places} displays the \emph{m}AP@$1,000$ and Precision@H$2$ results.
Also, Fig.\,\ref{map_instance_places} illustrates the \emph{m}AP@$1,000$ vs. different sizes of training instances comparisons and their AUC results are plotted in Fig.\,\ref{map_auc_places}.
We show the Precision@K curves and their AUC curves in Fig.\,\ref{precision_places} and Fig.\,\ref{precision_auc_places}, respectively.
Besides, the Precision-Recall performance can be found in Fig.\,\ref{pr_places}.

We start with an analysis of the \emph{m}AP@$1,000$ performance.
From Tab.\,\ref{map_precision_places}, two observations can be derived.
First, the proposed HMOH keeps substantially best \emph{m}AP performance.
In particular, HMOH consistently outperforms the best baselines, \emph{i.e.}, MIHash or HCOH, by an average of $13.090\%$,
and transcends our previous version of HCOH by an average increase of $29.662\%$, respectively.
The second observation comes that the proposed HMOH overcomes the drawback of the previous HCOH, \emph{i.e.}, suffering poor performance in low code length (\emph{e.g.}, $8$ or $16$).
As mentioned in \citep{lin2018supervised}, HCOH is only suitable for learning models with high-dimensional features in low hash bit ($4096$-D for CIFAR-$10$, $128$-D for Places$205$ and $784$-D for MNIST).
Particularly, when the hashing bits are $8$ and $16$, the proposed HMOH not only ranks first but also obtains an increase of $108.163\%$ and $34.104\%$ compared with the previous HCOH, respectively.
Therefore, HMOH can well address the obstacles HCOH suffers, which further demonstrates the effectiveness of the proposed HMOH.

When it comes to the results of Precision@H$2$ in Tab. \ref{map_precision_places}, we can observe that with hashing bit being $8$, the proposed HMOH ranks second and MIHash holds the first position.
However, as the code length increases, the proposed HMOH still consistently shows the best.
Concretely speaking, in low bit of $8$, compared with HMOH, MIHash acquires $57.143\%$ gains.
When the code length is more than $8$, compared with the state-of-the-art method, \emph{i.e.}, MIHash or BSODH, HMOH shows a relative increase of $18.496\%$ which verifies the superiority of the proposed HMOH.
Notably, when the hashing bit is $128$, like other methods (MIHash: $0.202 \rightarrow 0.069$, BSODH: $0.212 \rightarrow 0.101$, HCOH: $0.114 \rightarrow 0.036$), the proposed HMOH also drops ($0.223 \rightarrow 0.137$), which contradicts with the observation on CIFAR-$10$.
We argue that this is owing to the large scale of Places$205$ which is in millions.
Searching for similar items within a Hamming ball of radius $2$ in large code length on a large dataset is tough.
Nevertheless, the proposed HMOH drops least and shows best results.

Further, we analyze the results of \emph{m}AP over time and the corresponding AUC curves in Fig.\,\ref{map_instance_places} and Fig.\,\ref{map_auc_places}, respectively.
Generally, the two key observations on CIFAR-$10$ can also be found in Places$205$, \emph{i.e.}, superior \emph{m}AP results over time and good generalization ability.
For the superior \emph{m}AP results over time, we analyze the AUC curves in Fig.\,\ref{map_auc_places}.
To be specific, the proposed HMOH surpasses the best baselines, \emph{i.e.}, MIHash or BSODH by an averaged improvement of $13.006\%$.
And comparing HMOH with its previous version, HCOH \citep{lin2018supervised}, the proposed HMOH gets an average boost of $24.577\%$. It can be concluded that, in low hash bit, the proposed HMOH improves quite a lot compared to HCOH.
As for the generalization ability, we can find that in most hash bits (except $128$), HMOH still obtains relatively high results with only a small number of training data.
To take the hash bit of $48$ as an example,
when the size of training data is $5K$, the proposed HMOH gets an \emph{m}AP of $0.277$.
However, the state-of-the-art methods suffer lower performance.
For example, it is $0.241$ \emph{m}AP for MIHash, $0.228$ \emph{m}AP for BSODH and $0.242$ \emph{m}AP for HCOH.
To achieve similar performance, it takes $40K$ training instances for MIHash, $100K$ training instances for BSODH, and $80K$ training instances for our previous version of HCOH.
It can be concluded that the proposed HMOH holds good generalization ability.

The Precision@K curves are presented in Fig.\,\ref{precision_places} and we plot their AUC results in Fig.\,\ref{precision_auc_places}.
Though in the case of $128$-bit, MIHash and HCOH perform best, the proposed HMOH outranks other methods in other hashing bits.
When the hashing bit is $128$, MIHash achieves an AUC improvement of $2.579\%$ and HCOH obtains a $1.719\%$ improvement over the proposed HMOH.
On the contrary, in other cases,
HMOH surpasses MIHash by an average of $6.775\%$ improvements, respectively.
Meanwhile, it increases the previous version of HCOH by an average of $29.355\%$.
Especially, it is clear that the proposed HMOH boosts its previous version, \emph{i.e.}, HMOH, by a large margin in low hashing bits.
Fig.\,\ref{pr_places} shows the Precision-Recall curves on Places$205$.
Generally, the proposed HMOH outperforms all baselines in all cases, which well demonstrates its effectiveness.
One common observation for all methods in Fig.\,\ref{pr_places} is that the areas covered by the Precision-Recall curves all relatively lower when compared with these in Fig.\,\ref{pr_cifar}.
To analyze, the Places$205$ is a large-scale benchmark, on which it is quite challenging to obtain a high performance.
In $8$-bit, the improvements for HMOH are not significant because $8$ bits can not encode well the abundant information contained in the large-scale Places$205$.
Nevertheless, with the code length increasing, the proposed HMOH still shows a clear better performance.

\subsubsection{Results on MNIST}

\begin{figure*}[!t]
\begin{center}
\includegraphics[height=0.48\linewidth]{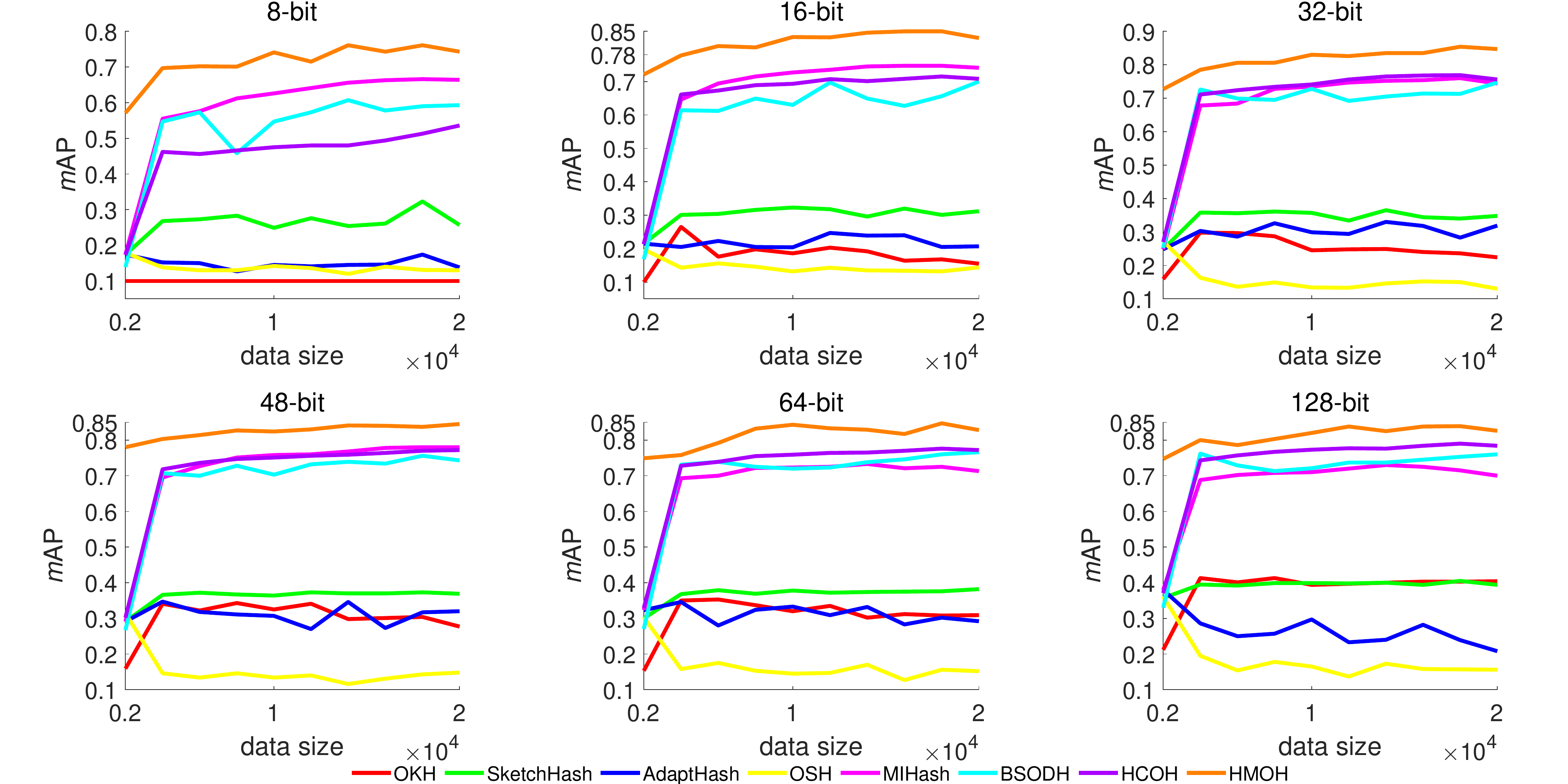}
\caption{\label{map_instance_mnist} \textit{m}AP performance with respect to different sizes of training instances on MNIST.}
\end{center}
\vspace{-1em}
\end{figure*}

\begin{figure}[!t]
\begin{center}
\includegraphics[height=0.53\linewidth]{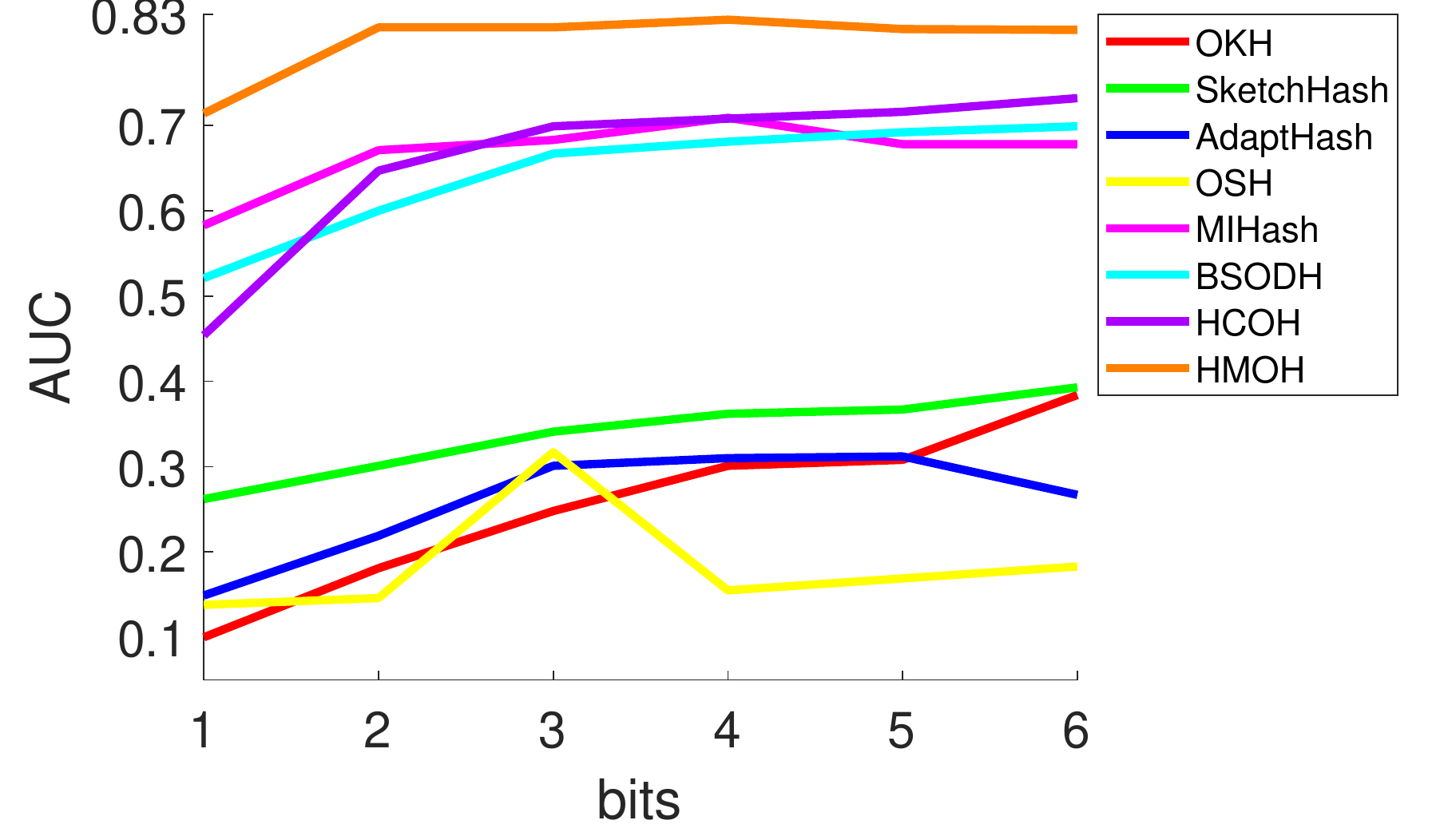}
\caption{\label{map_auc_mnist} AUC curves for mAP on MNIST.}
\end{center}
\vspace{-3.5em}
\end{figure}

\begin{figure*}[!t]
\begin{center}
\includegraphics[height=0.48\linewidth]{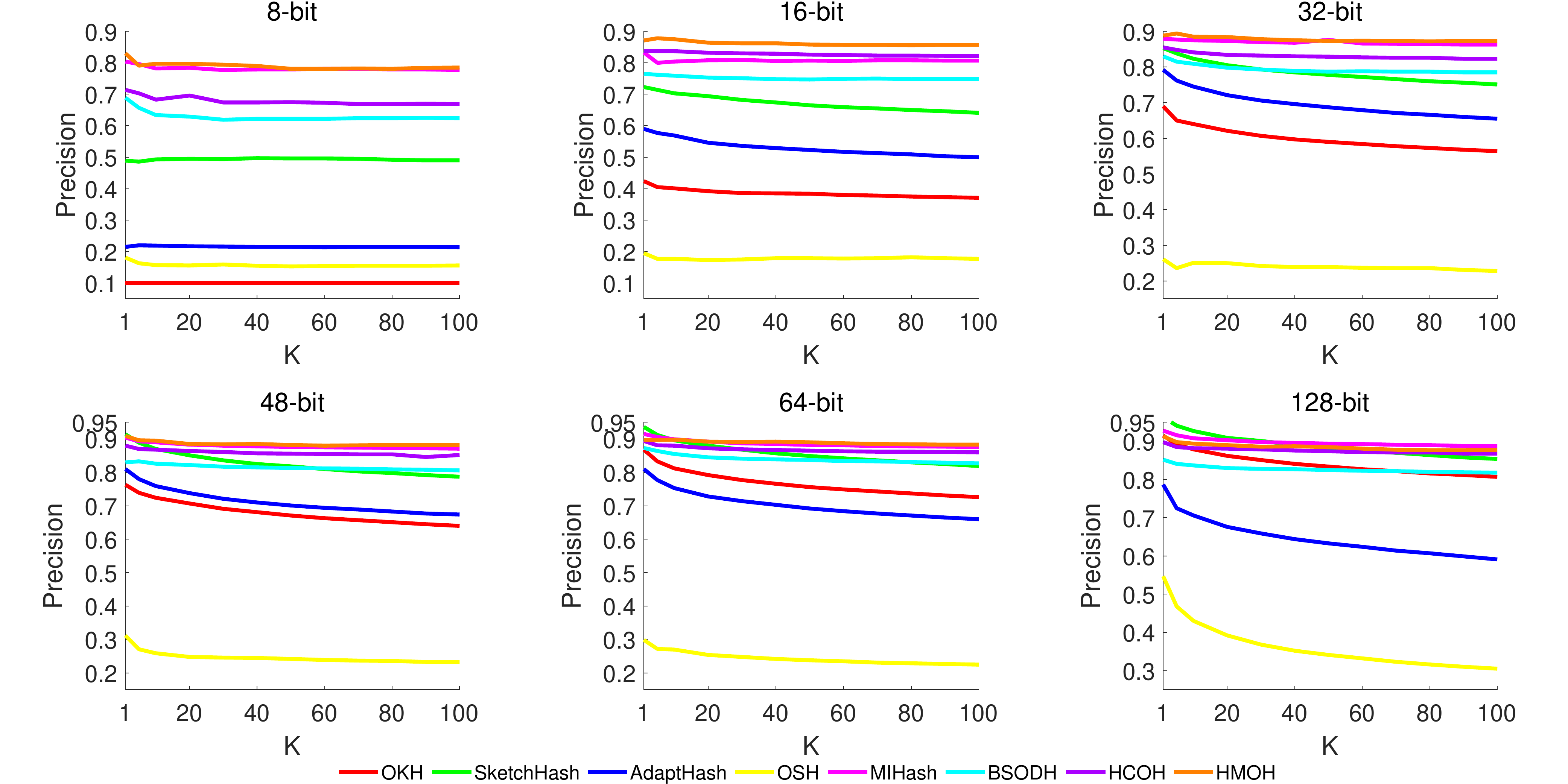}
\caption{\label{precision_mnist} Precision@K curves of compared algorithms on MNIST.}
\end{center}
\vspace{-1em}
\end{figure*}

\begin{figure}[!t]
\begin{center}
\includegraphics[height=0.53\linewidth]{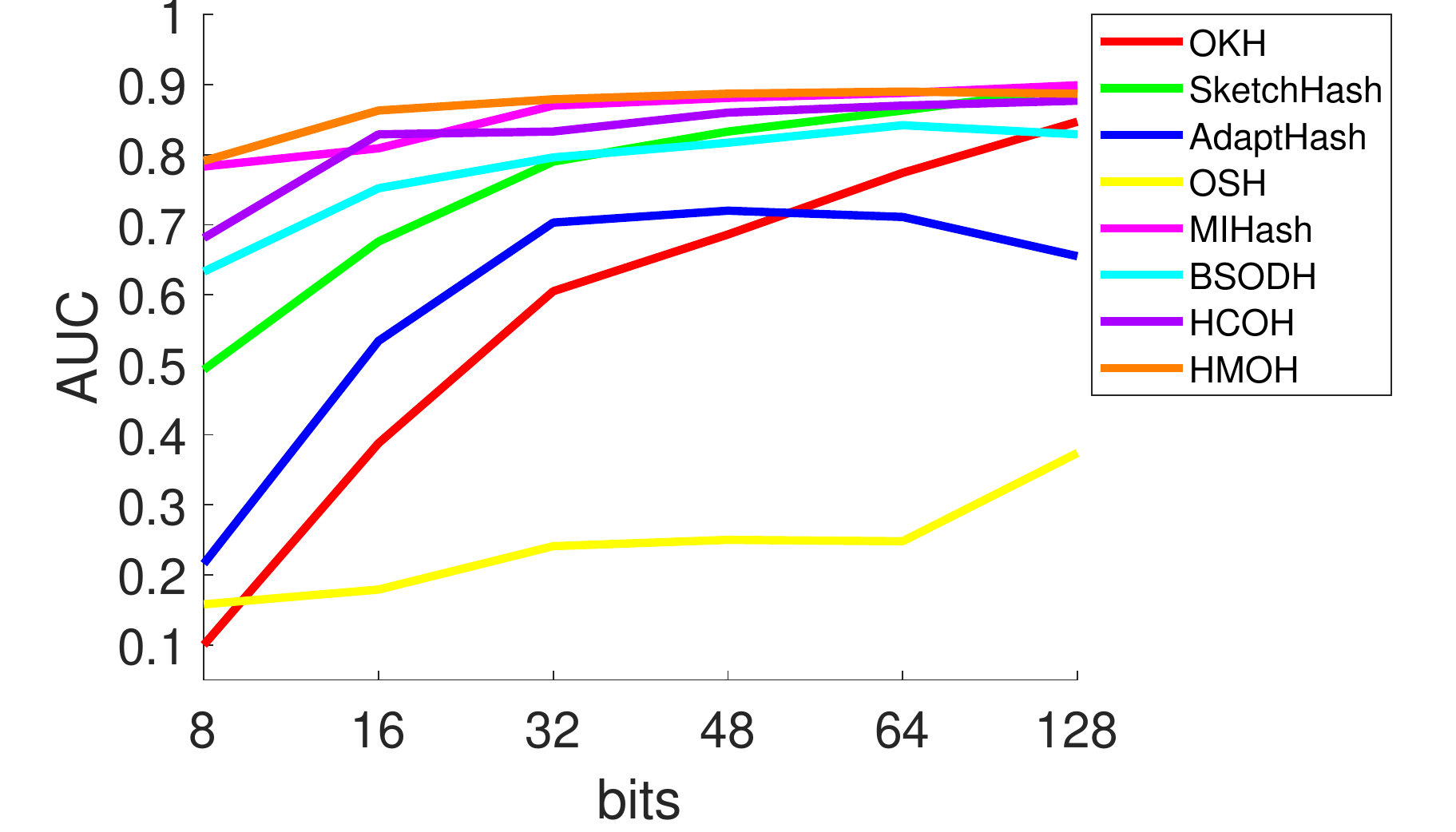}
\caption{\label{precision_auc_mnist} AUC curves for Precision@K on MNIST.}
\end{center}
\vspace{-2.5em}
\end{figure}

\begin{figure*}[!t]
\begin{center}
\includegraphics[height=0.48\linewidth]{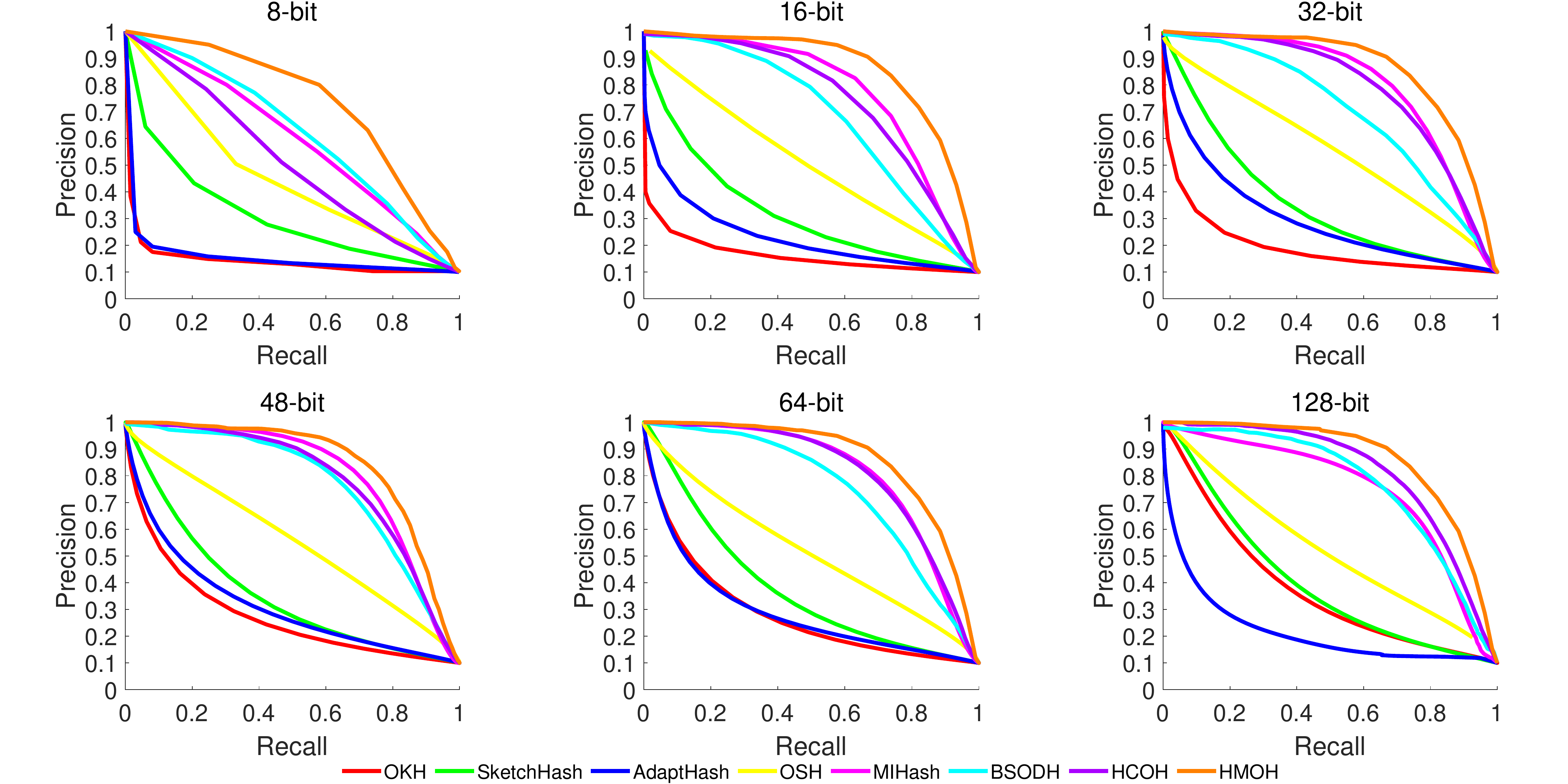}
\caption{\label{pr_mnist} Precision-Recall curves of compared algorithms on MNIST.}
\end{center}
\vspace{-2em}
\end{figure*}

Besides the quantitative evaluation on the above two datasets, we also apply our techniques on MNIST with features of pixel level.
The concrete values of \emph{m}AP and Preccision@H$2$ are filled in Tab.\,\ref{map_precision_mnist}.
Fig.\,\ref{map_instance_mnist} illustrates the \emph{m}AP curves under different training instances,
and Fig.\,\ref{map_auc_mnist} depicts their AUC performance.
The results for Precision@K and their AUC curves can be observed in Fig.\,\ref{precision_mnist} and Fig.\,\ref{precision_auc_mnist}, respectively.
Finally, Fig.\,\ref{pr_mnist} demonstrates the Precision-Recall curves.

With regards to \emph{m}AP in Tab. \ref{map_precision_mnist}, we have the following findings:
First, the proposed HMOH is competitive and far better than other methods.
More specifically, the state-of-the-art online hashing method, MIHash or BSODH, is surpassed by the proposed HMOH by large gaps, \emph{i.e.}, an average improvement of $10.401\%$.
Also, the previous version of HCOH is transcended by HMOH by $15.595\%$.
As the second finding, there is a great improvement of the proposed HMOH in low hashing bits (\emph{e.g.}, $8$ or $16$), which can also be found in Places$205$ as aforementioned.
With code length being $8$ and $16$, the previous version of HCOH falls behind MIHash and BSODH, which shows its inferiority in low hashing bits.
However, the proposed HCOH not only ranks first but also outranks HMOH by an improvement of $38.619\%$ in $8$-bit and $17.232\%$ in $16$-bit.

With regards to Precision@H$2$ in Tab.\,\ref{map_precision_mnist}, we analyze as follows:
First, similar to the performance on Places$205$, the proposed method ranks second, which is slightly worse than MIHash in the case of $8$-bit.
And as the code length increases, HMOH keeps substantially best performance.
Concretely, when the code length is $8$, MIHash gets an improvement of $3.397\%$.
Under other circumstances, compared with MIHash or BSODH, the performance of HMOH increases by an average of $8.924\%$.
What's more, when compared with the previous version, \emph{i.e.}, HCOH, the proposed HMOH obtains a continually averaged growth of $32.330\%$.
Second, we observe that the robustness of the proposed method can be also found on MNIST, which is similar to that on CIFAR-$10$.
As the code length increases to $128$-bit, other state-of-the-art methods degrade a lot (MIHash: $0.720 \rightarrow 0.471$, BSODH: $0.814 \rightarrow 0.643$, HCOH: $0.643 \rightarrow 0.370$).
%

%

\begin{table*}[]
\centering
\caption{\textit{m}AP and Precision@H$2$ Comparisons on MNIST with $8$, $16$, $32$, $48$, $64$ and $128$ bits. %
The best result is labeled with boldface and the second best is with an underline.}
\label{map_precision_mnist}
\begin{tabular}{c|cccccc|cccccc}
\hline
\multirow{2}{*}{Method} & \multicolumn{6}{c|}{\textit{m}AP}                   & \multicolumn{5}{c}{Precision@H$2$}          \\
\cline{2-13}
                        & 8-bit & 16-bit & 32-bit &48-bit & 64-it & 128-bit & 8-bit & 16-bit & 32-bit &48-bit & 64-bit
                        & 128-bit \\
\hline
OKH                     &0.100  &0.155   &0.224   &0.273  &0.301  &0.404    &0.100  &0.220   &0.457   &0.724   &0.522
                        &0.124 \\
\hline
SketchHash              &0.257  &0.312   &0.348   &0.369  & 0.376 &0.399     &0.261  &0.596   &0.691   &0.251  &0.091
                        &0.004  \\
\hline
AdaptHash               & 0.138 &0.207   &0.319   &0.318  &0.292  &0.208    &0.153  &0.442   &0.535   &0.335  &0.163
                        &0.168  \\
\hline
OSH                     &0.130  &0.144   &0.130   &0.148  &0.146  &0.143    &0.131  &0.146   &0.192   &0.134  &0.109
                        &0.019  \\
\hline
MIHash                  &\underline{0.664} &\underline{0.741}   &0.744   &\underline{0.780}  &0.713  &0.681 &\textbf{0.487} &\underline{0.803}   &0.814  &0.739 &0.720
                        &0.471 \\
\hline
BSODH                   &0.593  &0.700  &0.747 &0.743 &\underline{0.766}  &0.760   &0.308  &0.709   &\underline{0.826} &\underline{0.804}  &\underline{0.814}
                        &\underline{0.643}   \\
\hline
\hline
HCOH                    &0.536  &0.708 &\underline{0.756} &0.772  &0.759  &\underline{0.771}    &0.350  &0.800   &\underline{0.826}  &0.766  &0.643
                        &0.370   \\
\hline
HMOH                    &\textbf{0.743}  &\textbf{0.830}  &\textbf{0.847} &\textbf{0.845} &\textbf{0.828}  &\textbf{0.826}
                        &\underline{0.471}  &\textbf{0.838}   &\textbf{0.869}  &\textbf{0.854} &\textbf{0.855}  &\textbf{0.857}\\
\hline
\end{tabular}
\end{table*}

We then evaluate the \emph{m}AP \emph{vs}. different sizes of training instances in Fig.\,\ref{map_instance_mnist}. Obviously, the curve for the proposed HMOH is above other methods by a largin margin in all hashing bits.
To quantitatively evaluate the performance, we move to their corresponding AUC values in Fig.\,\ref{map_auc_mnist}, which clearly shows the high performance of the proposed HMOH.
Detailedly, compared with the best results between MIHash and BSODH, the proposed HMOH increases by $18.855\%$.
And it is $23.448\%$ compared with its previous version, \emph{i.e.}, HCOH.
It's clear that HMOH improves quite a lot especially in low hashing bits (\emph{e.g.}, $57.269\%$ in $8$-bit and $25.966\%$ in $16$-bit).
Besides, we can derive the generalization ability of the proposed HMOH as well from Fig.\,\ref{map_instance_mnist}.
That is, HMOH can achieve satisfactory performance with a much smaller batch of training instances.
To illustrate, we take the case with $48$-bit as an example.
When the number of instances is just $2K$, HMOH achieves a relatively high result of $0.780$, while it is only $0.291$ for MIHash, $0.267$ for BSODH and $0.302$ for HCOH.
As the number of training instances increases to $20K$, it is $0.780$ for MIHash, $0.743$ for BSODH and $0.772$ for HCOH, while HMOH achieves $0.845$.
Visibly, HMOH with only $2K$ training instances already shows competitive performance when compared with MIHash, BSODH and HCOH with training instances as many as $20K$.

We plot the Precision@K curves in Fig.\,\ref{precision_mnist} and the corresponding AUC results in Fig.\,\ref{precision_auc_mnist}.
Clearly, except the case of $128$-bit where our method ranks second, the proposed HMOH shows generally best performance.
To be concrete, in terms of the AUC curves for Precision@K, with hashing bit being $128$, MIHash gets a result of $0.899$ while it is $0.887$ for the proposed HMOH.
When hashing bit varies from $8$ to $64$, the proposed HMOH outperforms the best baseline, \emph{i.e.}, MIHash, $1.606\%$ on average.
While compared with the previous version of HCOH, it outperforms in all aspects by an average of $5.393\%$.
Again, the proposed HMOH boosts its previous version of HCOH by a large margin especially in low hashing bits.

Lastly, we plot the Precision-Recall curves on MNIST in Fig.\,\ref{pr_mnist}.
Similar to the \emph{m}AP results, the Precision-Recall curve for the proposed HMOH shows significant improvements in all code lengths.
Besides, it can be observed that, as the bit increases, the advantage of the proposed HMOH over the state-of-the-art methods decreases.
To analyze, on one hand, larger bits can encode more information in the dataset.
On the other hand, different with Places$205$, MNIST is a simple benchmark.
Hence, compared with Fig.\,\ref{pr_places}, the curve areas for most methods are larger.
Nevertheless, the proposed HMOH still yields a clearly better result.

\subsubsection{Results on NUS-WIDE}
\begin{figure*}[!t]
\begin{center}
\includegraphics[height=0.48\linewidth]{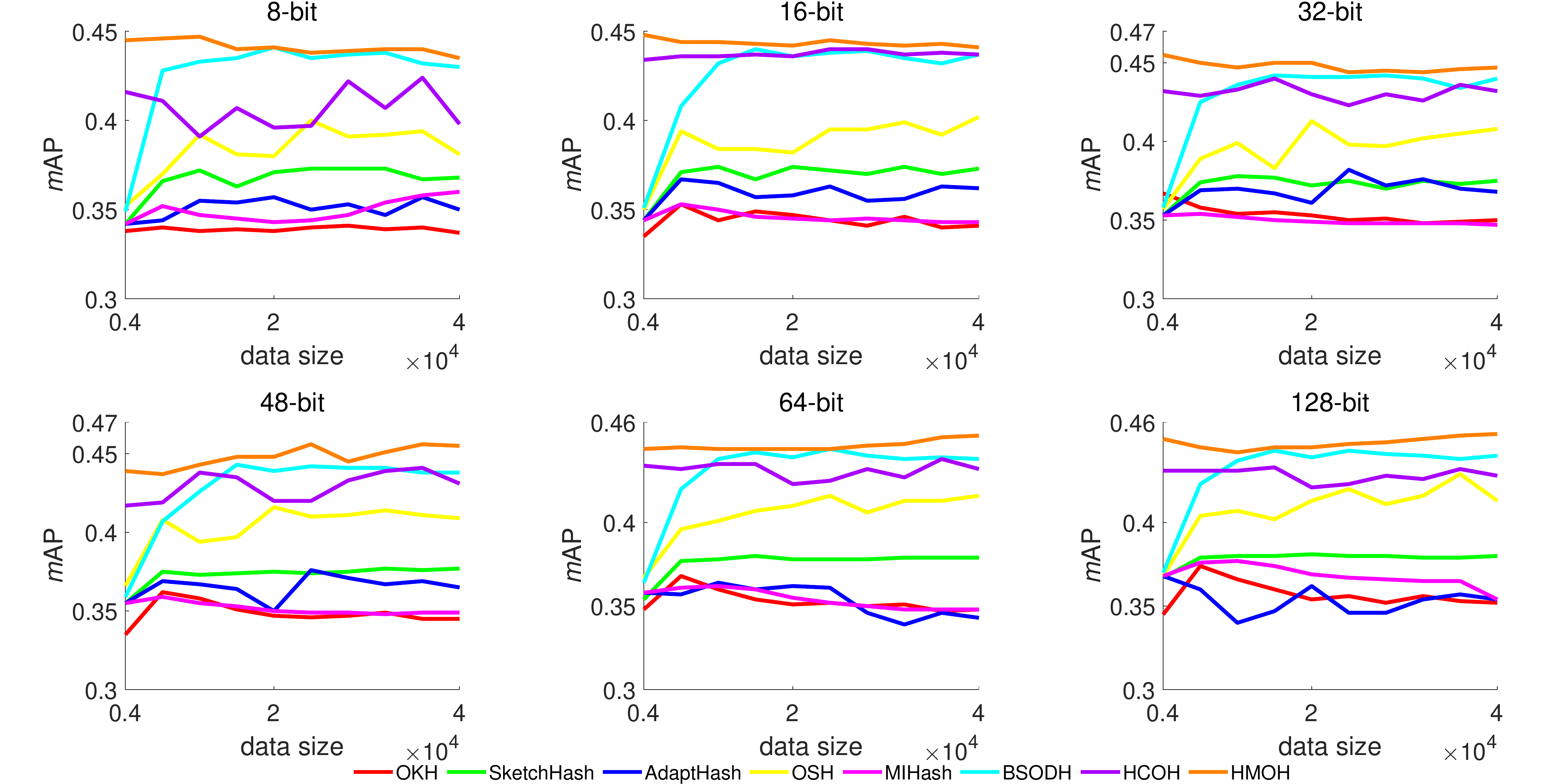}
\caption{\label{map_instance_nuswide} \textit{m}AP performance with respect to different sizes of training instances on NUS-WIDE.}
\end{center}
\vspace{-1em}
\end{figure*}

\begin{figure}[!t]
\begin{center}
\includegraphics[height=0.53\linewidth]{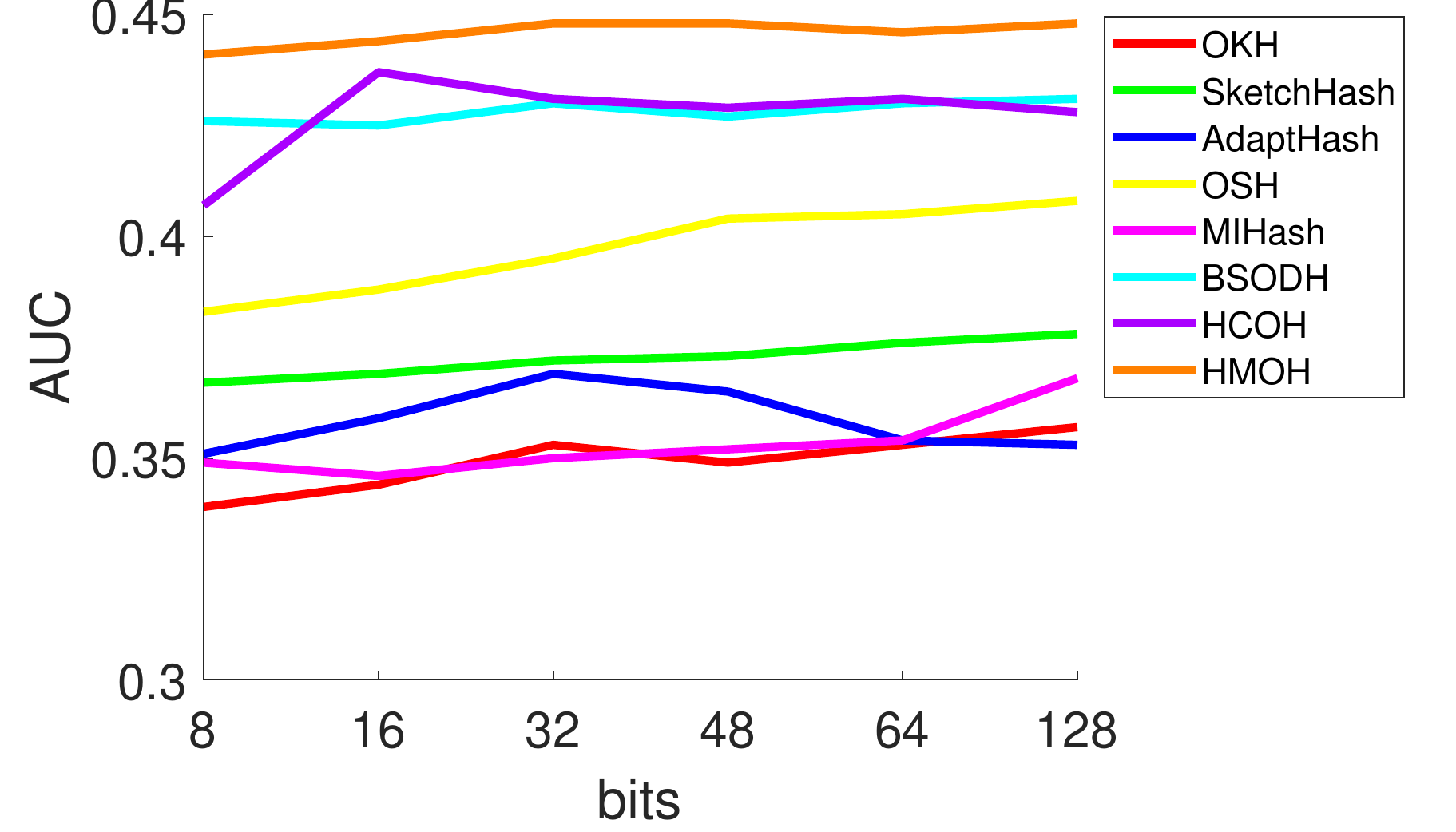}
\caption{\label{map_auc_wide} AUC curves for mAP on NUS-WIDE.}
\end{center}
\vspace{-2.5em}
\end{figure}

\begin{figure*}[!t]
\begin{center}
\includegraphics[height=0.48\linewidth]{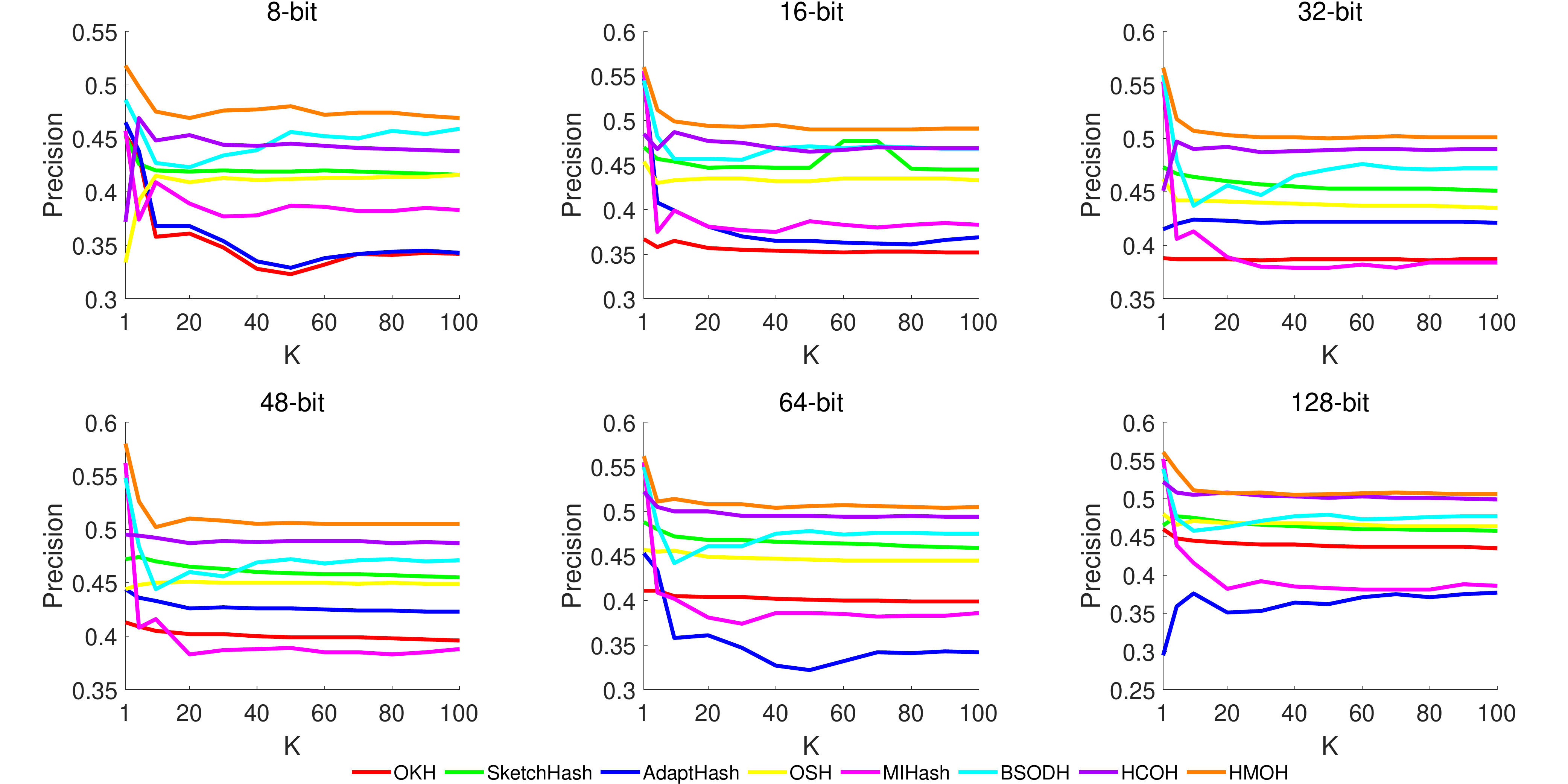}
\caption{\label{precision_nuswide} Precision@K curves of compared algorithms on NUS-WIDE.}
\end{center}
\vspace{-1em}
\end{figure*}

\begin{figure}[!t]
\begin{center}
\includegraphics[height=0.53\linewidth]{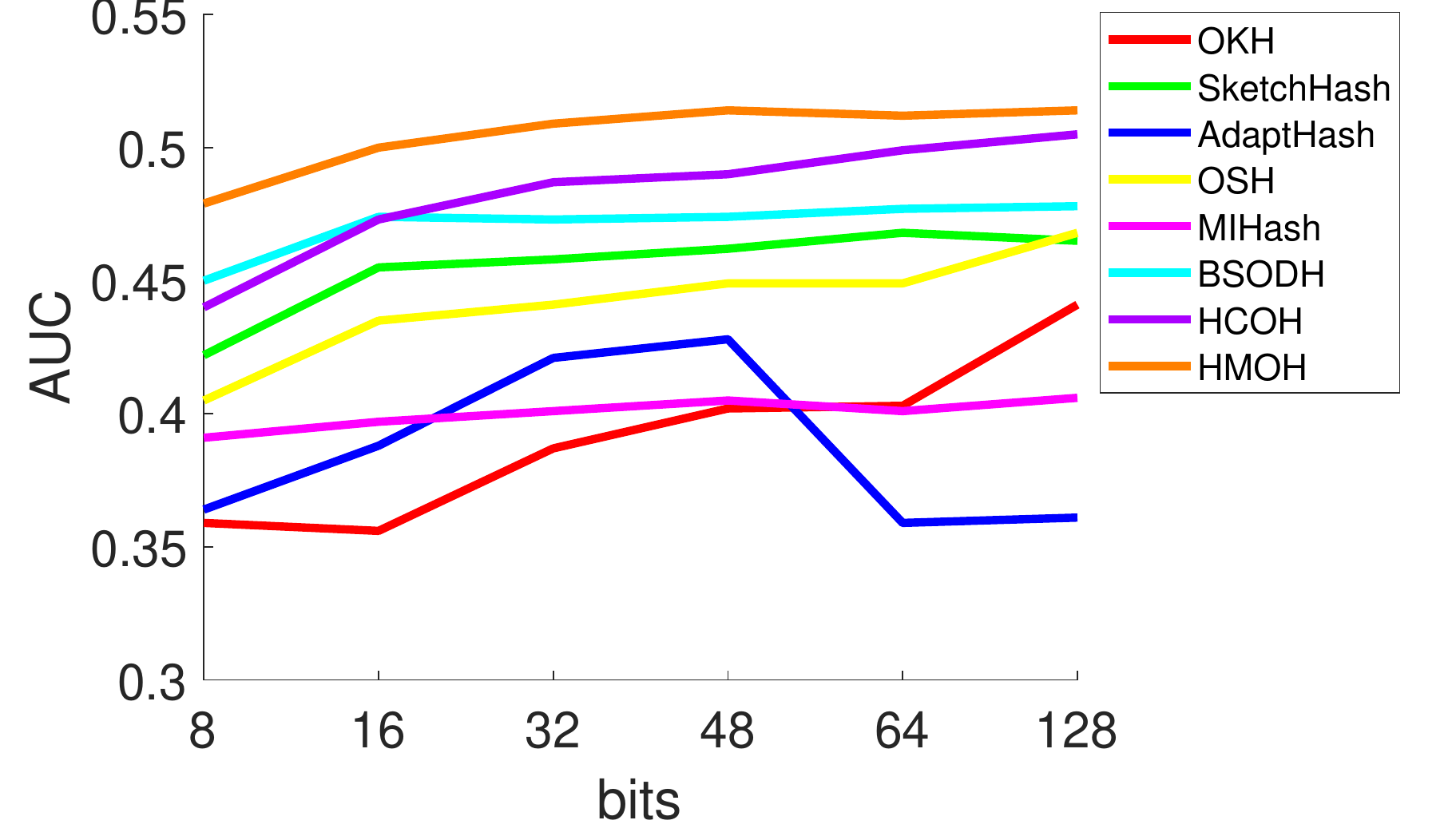}
\caption{\label{precision_auc_wide} AUC curves for Precision@K on NUS-WIDE.}
\end{center}
\vspace{-2.5em}
\end{figure}

\begin{figure*}[!t]
\begin{center}
\includegraphics[height=0.48\linewidth]{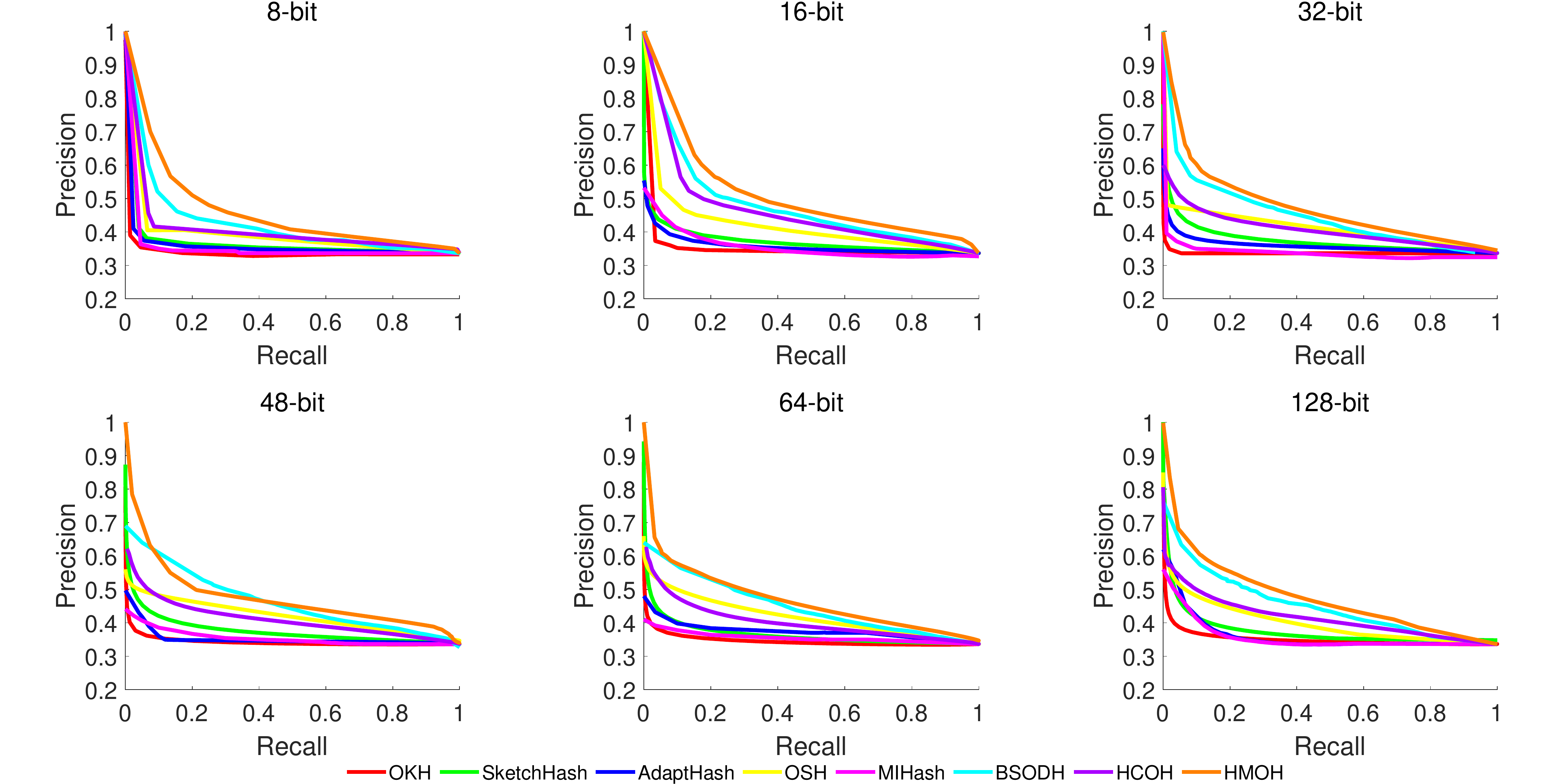}
\caption{\label{pr_nuswide} Precision-Recall curves of compared algorithms on NUS-WIDE.}
\end{center}
\end{figure*}

For NUS-WIDE, Tab.\,\ref{map_precision_nuswide} shows the $\emph{m}$AP and Precision@H$2$ results.
The \emph{m}AP \emph{vs}. different sizes of training instances curves and their AUC results are displayed in Fig.\,\ref{map_instance_nuswide} and Fig.\,\ref{map_auc_wide}, respectively.
Fig.\,\ref{precision_nuswide} plots the Precision@K curves and their AUC curves are illustrated in Fig.\,\ref{precision_auc_wide}.
Lastly, the Precision-Recall results are shown in Fig.\,\ref{pr_nuswide}.

We first analyze the \emph{m}AP performance in Tab.\,\ref{map_instance_nuswide}.
As can be observed, the proposed HMOH yields best results in all code lengths.
Different with the observations in Tab.\,\ref{map_precision_cifar} (CIFAR-$10$), Tab.\,\ref{map_precision_places} (Places$205$) and Tab.\,\ref{map_precision_mnist} (MNIST),
the BSODH obtains the second best in all bits, outperforming MIHash and HCOH while the performance of MIHash is not so good as on the other three benchmarks.
To explain:
BSODH adopts the inner-product scheme where the inner product of two hashing codes aims to approximate their similarity matrix \big(\{-1, +1\}\big).
However, the ``data-imbalance" issue in online learning disables the learning of BSODH.
On NUS-WIDE, the ``data-imbalance" issue can be relieved because any two data points are defined as similar if they share at least one same label and the quantitative difference between the number of similar pairs and dissimilar pairs is minor.
As for MIHash, given a query, it aims to separate the Hamming distance distributions between its neighbors and non-neighbors.
However, the low-level features (bag-of-visual-words) used on NUS-WIDE make it hard to learn the separable distance distributions in Hamming space.
Another observation is that the advantage of the proposed HMOH over BSODH is incremental gradually as code length increases ($0.001 \rightarrow 0.003 \rightarrow 0.006 \rightarrow 0.010 \rightarrow 0.013 \rightarrow 0.014$).
On one side, more bits can encode more information contained in the dataset.
One the other side, it well demonstrates the efficacy of the proposed ``majority principle" and ``balancedness principle" on multi-label benchmarks in Sec.\,\ref{multi_label_case}.

Then, we analyze the Precision@H$2$ in Tab.\,\ref{map_precision_nuswide}.
The performance of the proposed HMOH is similar to that on Places$205$ and MNIST.
In the case of $8$-bit, the proposed HMOH ranks second, slightly worse than BSODH and holds the first place in all other cases.
Quantitatively, in 8-bit, BSODH obtains $0.719\%$ improvements over the proposed method.
As the code length increases, the proposed method surpass the second best method by an improvement of $3.084\%$, $2.306\%$, $2.418\%$, $3.720\%$ and $1.961\%$, respectively.
One observation is that in 128-bit, the proposed HMOH shows a significant performance drop ($0.474 \rightarrow 0.468$).
Nevertheless, the proposed HMOH still gains the best performance compared with the second best BSODH with only $0.459$ Precision@H$2$ performance.
Hence, the superiority of the proposed HMOH on multi-label case is still undoubted.

In Fig.\,\ref{map_instance_nuswide}, we give a detailed analysis on performance of the \emph{m}AP \emph{vs}. different sizes of training instances.
Three observations can be obtained as follows:
First, the proposed HMOH consistently outperforms others and the second best is BSODH or HCOH.
To quantitatively analyze, we turn to the AUC curves in Fig.\,\ref{map_auc_wide}.
The proposed HMOH obtains $3.511\%$ ($8$-bit), $1.602\%$ ($16$-bit), $3.956\%$ ($32$-bit), $4.429\%$ ($48$-bit), $3.480\%$ ($64$-bit) and $3.944\%$ ($128$-bit) improvements over the second best.
Second, in $8$-bit, $16$-bit and $32$-bit, the proposed method shows a degenerated trend as the training data grows while in $48$-bit, $64$-bit and $128$-bit, the performance increases on the contrary.
As an analysis, NUS-WIDE is a relatively large benchmark and it is not enough to encode abundant information as the training data increases in low hashing bits.
Nevertheless, the proposed HMOH still shows best efficacy.
Third, the proposed HMOH can obtain fast adaptivity with less training instances.
As above, we take $48$-bit as an example.
When the number of training instance is $4K$, the proposed HMOH achieves \emph{m}AP of $0.439$.
While it is $0.417$ for the previous version, HCOH, and only $0.359$ for BSODH.
When the size of training data arrives at $40K$, HCOH obtains \emph{m}AP of $0.431$ while BSODH gains $0.438$, both of which still fall behind the performance of the proposed HMOH in $4K$.
Hence, the proposed HMOH can be well applied to the multi-label dataset in online learning.

\begin{table*}[]
\centering
\caption{\textit{m}AP and Precision@H$2$ Comparisons on NUS-WIDE with $8$, $16$, $32$, $48$, $64$ and $128$ bits. %
The best result is labeled with boldface and the second best is with an underline.}
\label{map_precision_nuswide}
\begin{tabular}{c|cccccc|cccccc}
\hline
\multirow{2}{*}{Method} & \multicolumn{6}{c|}{\textit{m}AP}                   & \multicolumn{5}{c}{Precision@H$2$}          \\
\cline{2-13}
                        & 8-bit & 16-bit & 32-bit &48-bit & 64-it & 128-bit
                        & 8-bit & 16-bit & 32-bit &48-bit & 64-bit & 128-bit \\
\hline
OKH                     &0.337  &0.341   &0.350   &0.345  &0.348  &0.352
                        &0.336  &0.340   &0.381   &0.319  &0.183   &0.004 \\
\hline
SketchHash              &0.368  &0.373   &0.375   &0.377  &0.375   &0.379
                        &0.381  &0.430   &0.375   &0.082  &0.027   &0.002 \\
\hline
AdaptHash               &0.350  &0.362   &0.368   &0.365  &0.343  &0.354
                        &0.341  &0.365   &0.407   &0.369  &0.336  &0.358  \\
\hline
OSH                     &0.381  &0.402   &0.408   &0.409  &0.416  &0.413
                        &0.395  &0.427   &0.421   &0.355  &0.208   &0.001  \\
\hline
MIHash                  &0.360  &0.343   &0.347   &0.349  &0.348  &0.354
                        &0.370  &0.372   &0.365   &0.361 &0.355   &0.345 \\
\hline
BSODH        &\underline{0.430}&\underline{0.437}&\underline{0.440}&\underline{0.438}&\underline{0.438}&\underline{0.440}
                        &\textbf{0.420}  &0.446   &0.454   &\underline{0.455}  &\underline{0.457}  &\underline{0.459}   \\
\hline
\hline
HCOH                    &0.398  &\underline{0.437}   &0.430   &0.431  &0.432  &0.428
                        &0.400  &\underline{0.454}   &\underline{0.477}   &0.409  &0.346  &0.153 \\
\hline
HMOH                    &\textbf{0.431}  &\textbf{0.440}   &\textbf{0.446}   &\textbf{0.448}  &\textbf{0.451}  &\textbf{0.454}
                        &\underline{0.417}  &\textbf{0.468}   &\textbf{0.488}   &\textbf{0.466}  &\textbf{0.474}  &\textbf{0.468}\\
\hline
\end{tabular}
\end{table*}


\begin{figure*}[!t]
\begin{center}
\includegraphics[height=0.22\linewidth]{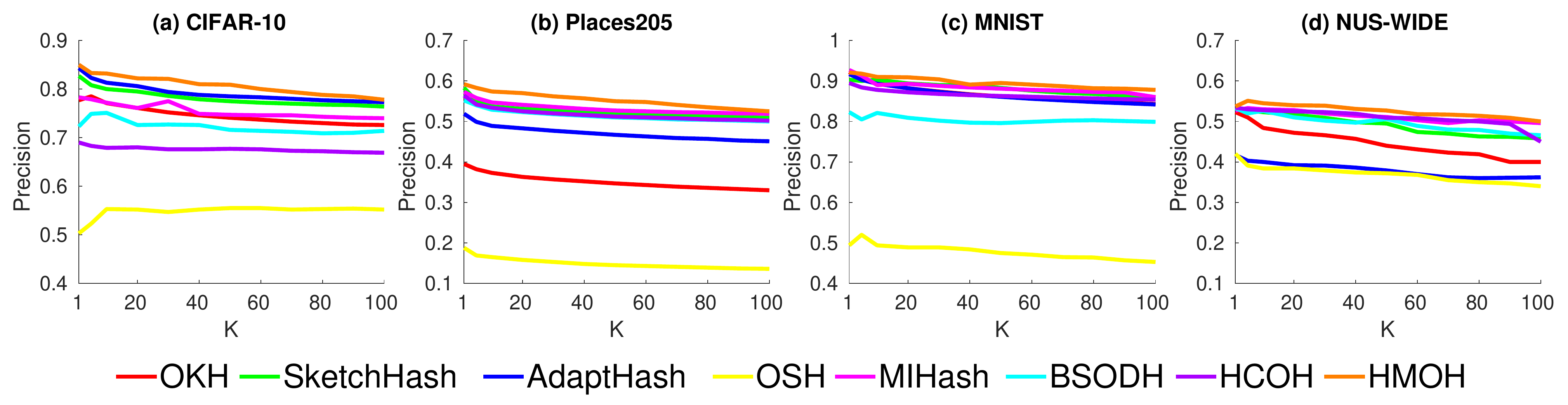}
\caption{\label{unseen} Precision@K curves on unseen class when the hash bit is 64.}
\end{center}
\vspace{-1.5em}
\end{figure*}
\begin{figure*}[!t]
\begin{center}
\includegraphics[height=0.22\linewidth]{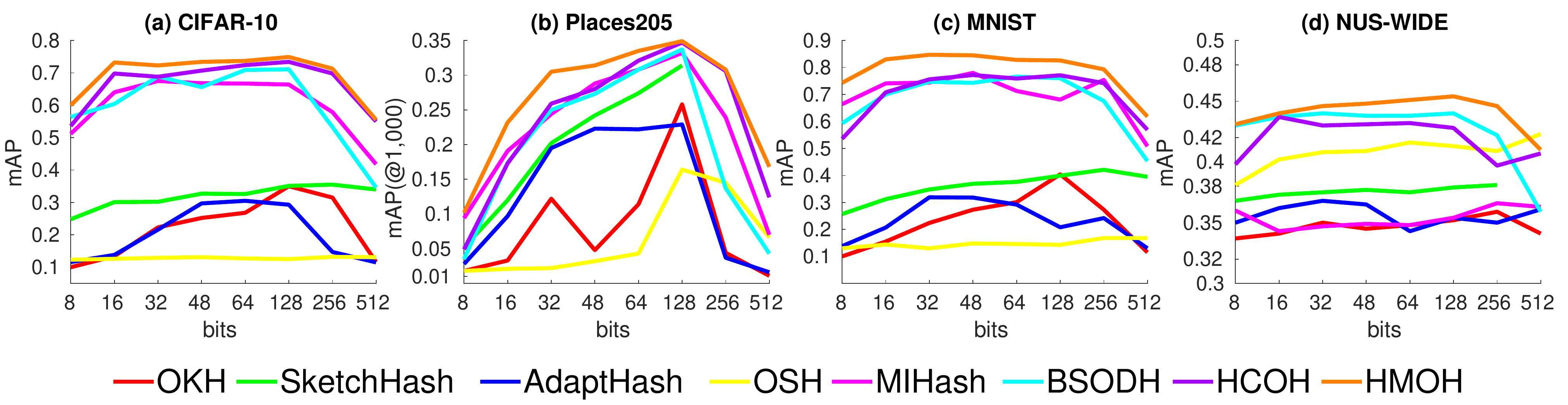}
\caption{\label{map_diff_bit} \textit{m}AP (\textit{m}AP@1,000) curves with different code lengths.}
\end{center}
\vspace{-1.5em}
\end{figure*}

Fig.\,\ref{precision_nuswide} plots the Precision@K results of all compared methods and Fig.\,\ref{precision_auc_wide} presents the AUC curves.
Generally, when the code length $\leq 64$, the proposed HMOH outperforms the second best, HCOH or BSODH by margins.
In the case of $128$-bit, HMOH yields a slightly better performance than its previous version, \emph{i.e.}, HCOH.
To explain, the Precision@K usually increases as the hash bit increases.
Hence, HCOH can also obtain a relatively good performance in $128$-bit.
Nevertheless, the superiority of the proposed HMOH is consistent.
Quantitatively, the proposed HMOH obtains an AUC improvement of $6.444\%$, $5.485\%$, $4.517\%$, $4.898\%$, $2.605\%$ and $1.782\%$ over the second best in $8$-bit, $16$-bit, $32$-bit, $48$-bit, $64$-bit and $128$-bit, respectively.

Lastly, we explore the impacts of different code lengths on the Precision-Recall performance on NUS-WIDE in Fig.\,\ref{pr_nuswide}.
We conclude that on one hand, the proposed HMOH obtains the best performance overall, which verifies the effectiveness of the proposed HMOH on multi-label benchmark and the usefulness of the proposed ``majority principle" and ``balancedness principle".
On the other hand, similar to the \emph{m}AP metric, the BSODH also shows the second best, which is even slightly better performance compared with the proposed HMOH in $48$-bit due to the advantage of inner-product scheme in multi-label datasets.
To sum, the above four benchmarks give a strong verification of the effectiveness of the proposed method.
It's worth noting that in terms of Precision@K performance under $128$-bit, the proposed HMOH ranks first, third, second and first on CIFAR-$10$ (Fig.\,\ref{precision_cifar}), Places$205$ (Fig.\,\ref{precision_places}), MNIST (Fig.\,\ref{precision_mnist}) and NUS-WIDE (Fig.\,\ref{precision_auc_wide}), respectively.
However, regarding \emph{m}AP (\emph{m}AP@$1,000$), HMOH holds a consistently first position, which means that not only the proposed HMOH retrieves the relevant instances to the query sample, but also ranks them at the top of the list.
It conforms with user experience in real-world applications.
Besides, the classification based HMOH shows significant improvements over the regression based HCOH, which demonstrates that using a classification setting is a better choice comparing to using a regression setting.
To explain, our method differs in the construction of Hadamard matrix, which serves as the target binary codes.
Under such a setting, the binary codes are known in advance and can be regarded as discrete labels.
And the hash functions aim to accurately predict the discrete label for each data.
On the other hand, HCOH adopts the regression learning which estimates the mapping from the input variables to numerical or continuous output variables.
Under such a setting, the gap between continuous space and discrete space makes it hard to fit well.
Consequently, it has to quantize the variables in continuous space to discrete space, which inevitably brings more quantization error
Instead, we consider the classification learning in this paper, which attempts to estimate the mapping function from the input variables to discrete/categorical output variables,
since the given binary label is discrete.
Hence, HMOH shows superior performance than HCOH.

\subsection{Retrieval on Unseen Classes}
We further conduct experiments with unseen classes by following the experimental settings in \citep{sablayrolles2017should} on all the four benchmarks.
Similar to \citep{sablayrolles2017should}, for each benchmark, 75\% of the categories are treated as seen classes to form the training set.
The remaining 25\% categories are regarded as unseen classes, which are further divided into a retrieval set and a test set to evaluate the hashing model.
For each query, we retrieve the nearest neighbors among the retrieval set and then compute the Precision@K.
The experiments are done as the hashing bit set as 64.
Fig.\,\ref{unseen} shows the experimental results with respect to different methods.
As can be seen from Fig.\,\ref{unseen}, the proposed HMOH shows consistently best performance on all four benchmarks, which demonstrates that HMOH can be well applied to scenarios with unseen classes compared with existing online hashing methods.

\subsection{Performance in Longer Code Length}
From Tab.\,\ref{map_precision_cifar}, Tab.\,\ref{map_precision_places}, Tab.\,\ref{map_precision_mnist} and Tab.\,\ref{map_precision_nuswide}, we can see that the map performance increases with the hash bit length.
However, the performance seems still not saturated even when the code lengths are 128-bit.
Following this, we further conduct experiments in the longer code length of 256-bit and 512-bit and show the experimental results in Fig.\,\ref{map_diff_bit}.
As can be observed from Fig.\,\ref{map_diff_bit}, most methods achieve optimal performance in the case of 128-bit.
As the code length continuously increases, the performance starts to decrease.
To analyze, the longer code lengths (256-bit and 512-bit) brings more redundant information, which inevitably introduces damage to the retrieval performance.
Hence, the 128-bit might be the optimal code length regarding the performance.
Longer code lengths are not encouraged based on our experimental observations.

\subsection{Ablation Study} \label{ablation_study}
In this section, we study the effects of the hyper-parameters including the bandwidth parameter $\eta$, the kernel size $m$, the learning rate $\lambda$, batch size $n_t$, and the usefulness of ensemble strategy.
For convenience, all the experiments are conducted on the four benchmarks in term of \emph{m}AP (\emph{m}AP@$1,000$) under the code length of $32$.
The experimental results can be generalized to other code lengths as well.

\begin{figure}[!t]
\begin{center}
\includegraphics[height=0.45\linewidth]{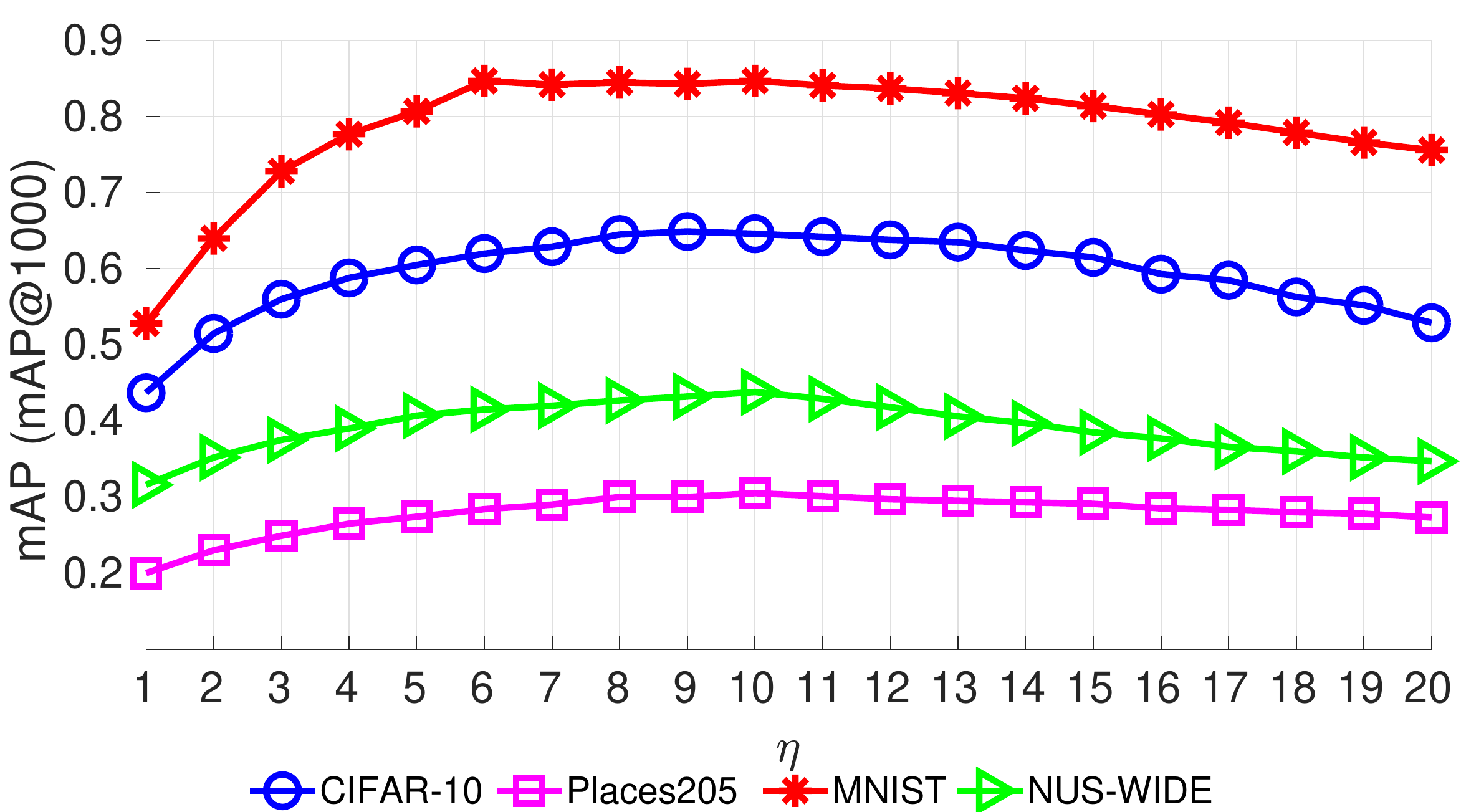}
\caption{\label{sigma_all} \emph{m}AP (\emph{m}AP@$1,000$) with varying temperature parameters.}
\end{center}
\vspace{-2.5em}
\end{figure}

\textbf{Effect of Bandwidth Parameter $\eta$.}
In this experiment, we evaluate the performance of the proposed HMOH \emph{w.r.t} different values of the bandwidth parameter $\eta$ applied in the kernelization process.
We report the experimental results in Fig.\,\ref{sigma_all}.
It can be observed that plotted data shows convex curves for all benchmarks.
Hence, it's easy to decide the best $\eta$ values.
Quantitatively, when set as $8$, $10$, $6$ and $10$, we obtain the best \emph{m}AP (\emph{m}AP@$1,000$) of $0.645$, $0.305$, $0.847$ and $0.438$ for CIFAR-$10$, Places$205$, MNIST and NUS-WIDE, respectively.

\begin{figure}[!t]
\begin{center}
\includegraphics[height=0.50\linewidth]{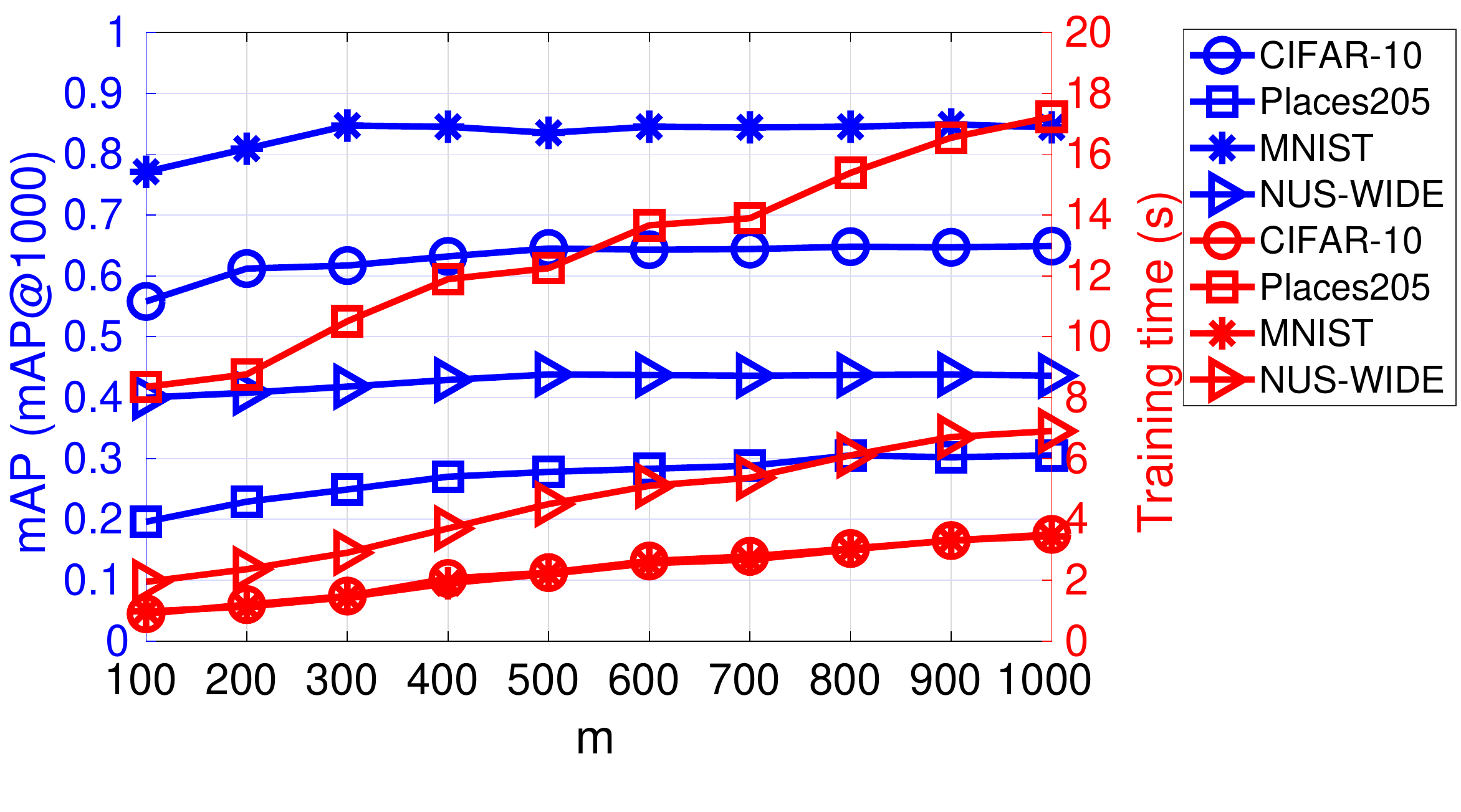}
\caption{\label{kernel_size_all} \emph{m}AP (\emph{m}AP@$1,000$) over different sizes of kernel.}
\end{center}
\vspace{-2.5em}
\end{figure}

\textbf{Effect of Kernel Size $m$.}
In this part, we aim to evaluate the performance of the proposed HMOH regarding different sizes of kernel.
The size of kernel not only affects the effectiveness but also the efficiency (Large kernel size brings more burdens on training time).
Hence, the choice of kernel size $m$ depends on the trade-off between effectiveness and efficiency.
We plot these two factors in Fig.\,\ref{kernel_size_all}.
Generally, as the kernel size increases, more training time is needed, while the performance of the proposed method first increases and then saturates.
We can observe that the time curves for CIFAR-$10$ and MNIST overlap and the time cost for Places$205$ and NUS-WIDE are almost five times and two times as much as that of CIFAR-$10$ and MNIST.
We analyze that this is because CIFAR-$10$ and MNIST have the same size of training instances ($20K$), while it is $100K$ for Places and $20K$ for NUS-WIDE.
To take care of both effectiveness and efficiency, we choose $m$ as $1000$, $800$, $300$ and $500$ for CIFAR-$10$, Places$205$, MNIST and NUS-WIDE, respectively.

During the experiments, we find that when applying the kernel trick on CIFAR-$10$ and NUS-WIDE, it doesn't do any benefit to the performance.
Taking the hashing bit as $32$ as an example, we obtain \emph{m}AP of $0.645$ and $0.446$ without kernelization while it is only $0.305$ and $0.438$ with kernelization.
We argue that this may be that CIFAR-$10$ and NUS-WIDE are linearly separable benchmarks in the original space.
Therefore, for all the experiments related to CIFAR-$10$ and NUS-WIDE, we do not apply the kernel trick.


\begin{figure}[!t]
\begin{center}
\includegraphics[height=0.45\linewidth]{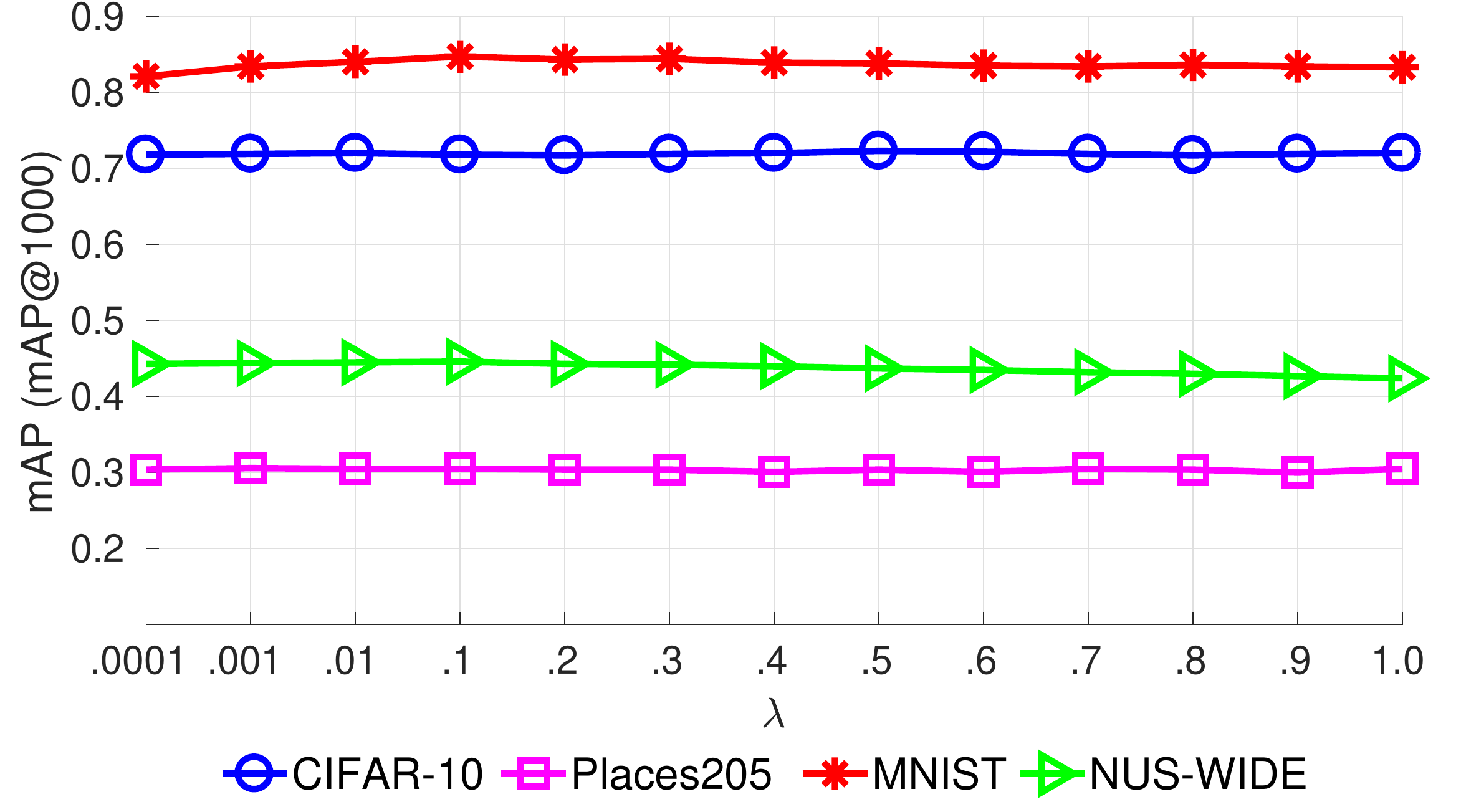}
\caption{\label{lr_all} \emph{m}AP (\emph{m}AP@$1,000$) with varying learning rates.}
\end{center}
\vspace{-1em}
\end{figure}

\textbf{Effect of Learning Rate $\lambda$.}
The obtained \emph{m}AP (\emph{m}AP@\\$1,000$) with learning rate $\lambda$ varying from $0.0001$ to $1$ are shown in Fig.\,\ref{lr_all}.
We can find that the proposed HMOH is not sensitive to $\lambda$ in a large range, as HMOH achieves almost constant performance on all the four datasets.
Nevertheless, in the experiments, we empirically set $\lambda$ as $0.5$, $0.01$, $0.1$ and $0.1$ on CIFAR-$10$, Places$205$, MNIST and NUS-WIDE, respectively, with which we get \emph{m}AP (\emph{m}AP@$1,000$) of $0.723$, $0.305$, $0.847$ and $0.446$ on each dataset.

\begin{figure}[!t]
\begin{center}
\includegraphics[height=0.45\linewidth]{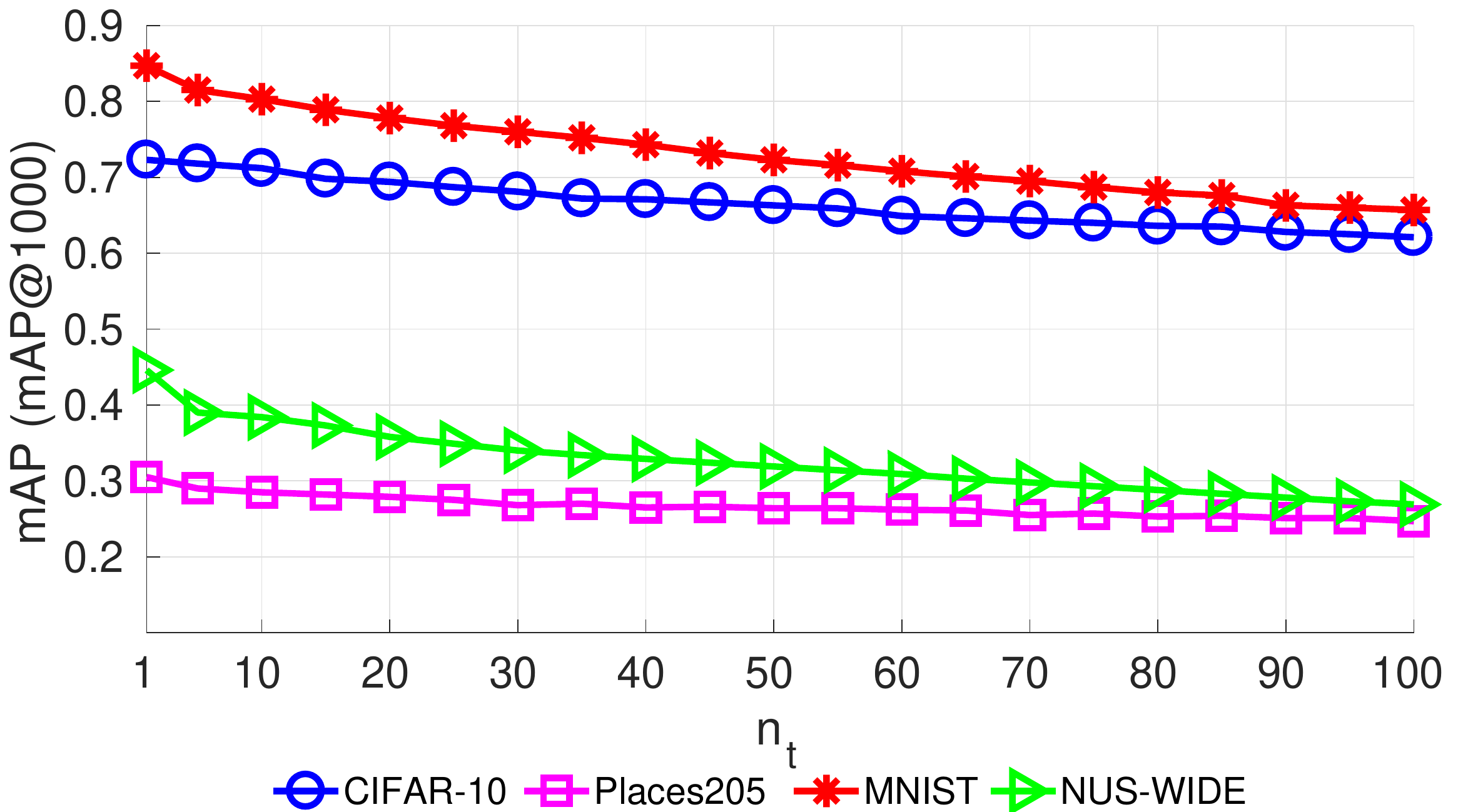}
\caption{\label{batch_size_all} \emph{m}AP (\emph{m}AP@$1,000$) over different sizes of batches.}
\end{center}
\vspace{-1em}
\end{figure}

\begin{figure}[!t]
\begin{center}
\includegraphics[height=0.50\linewidth]{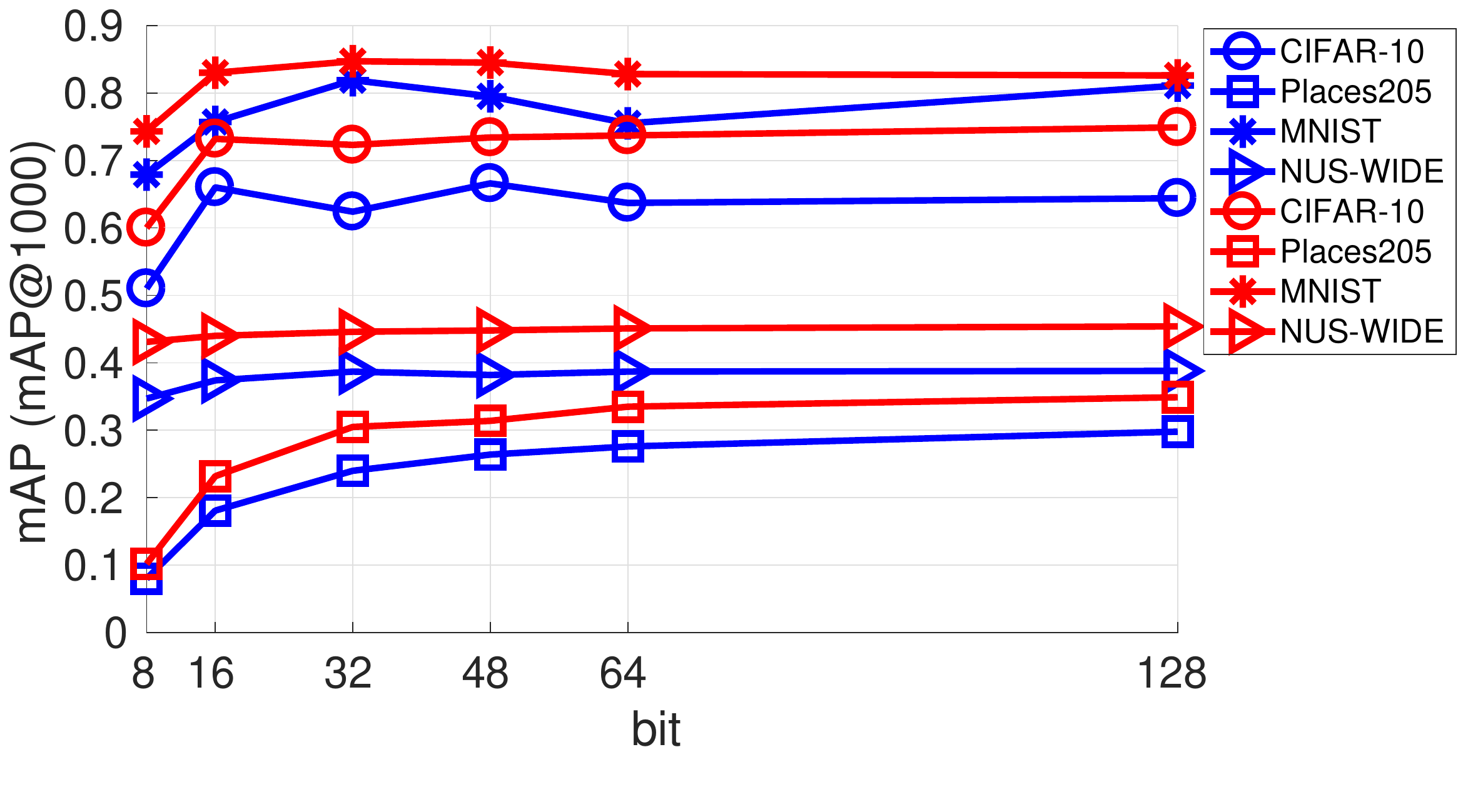}
\caption{\label{ensemble} Usefulness of the ensemble strategy (Blue lines denote the results without ensembling and red lines denote the results with ensembling).}
\end{center}
\end{figure}

%
\textbf{Effect of Batch Size $n_t$.}
This part of experiment mainly focuses on evaluating the effect of training size on the searching quality of the proposed HMOH.
For simplicity, we choose the \emph{m}AP as evaluation metric and vary the size of training data in the range of $\{1, 2, 3, ... ,49, 50, 51, ... 100\}$.
The experimental results are demonstrated in Fig.\,\ref{batch_size_all}.
As we can see, when the size increases from $1$ to $100$, we observe a slow decrease of the \emph{m}AP performance.
The precise values for $n_t = 1$ and $n_t = 2$ are $0.723$ and $0.721$, $0.305$ and $0.290$, $0.847$ and $0.833$, $0.446$ and $0.441$  on CIFAR-$10$, Places$205$, MNIST and NUS-WIDE, respectively.
Such experimental results conform with the observations in HCOH \citep{lin2018supervised}.
To explain,
firstly, the Perceptron algorithm for classification adopted by nature updates the classifier instance-wise.
%
%
%
%
Secondly, in online learning, the past stream data is not allowed to be reused to renew the models.
To gurantee the learning efficiency, for each arriving stream data, the optimizer usually updates the model only once instead of iteratively.
The number of training iterations significantly decreases as the batch size increases, which then degenerates the performance.
Therefore, we leave the training size $n_t$ set as $1$ for all four datasets.
\textbf{Effect of Ensemble Strategy.}
The quantitative evaluations for the effect of the ensemble strategy are shown in Fig.\,\ref{ensemble}.
The blue lines show the results with no ensembling while the red lines denote the results with ensembling.
As can be observed, the ensemble strategy takes effect on all four benchmarks.
Quantitatively, ensemble strategy obtains an average improvement of $14.439\%$, $23.365\%$, $6.716\%$ and $15.245\%$ on CIFAR-$10$, Places$205$, MNIST and NUS-WIDE, respectively.
These experiments validate the importance of considering the past learnt information, which can boost the performance of online hashing.

\subsection{Performance Variation on Places$205$}
As stated in Sec.\,\ref{experimental_setting}, in the Places$205$, $20$ images from each categories are randomly sampled, consisting of a total of $4,100$ query images.
However, the large scale of Places$205$ ($2.5$ million) may introduce large performance variation due to the random selection.
To test the performance variation, we analyze the ``mean $\pm$ std" of \emph{m}AP@$1,000$ in $32$-bit.
To that effect, the images sampled from each category range in $\{20, 30, 40, 50, 100, 200, 300, 400, 500\}$, composing of a test set of \{4100, 6150, 8200, 10250, 20500, 41000, 61500, 82000, 102500\} images, respectively.
The experimental results are shown in Tab.\,\ref{variation}.
As can be seen, the variations of most methods are small, and most methods are not sensitive to the size of test set, especially for state-of-the-art methods, \emph{i.e.}, MIHash, BSODH, HCOH and the proposed HMOH.
Besides, it can be observed that the proposed HMOH consistently show superioriities over the compared methods.
Especially, in the case of 500 images sampled from each category which composes of a large testing set with 102,500 images, HMOH still yields the best result, and the performance is similar to the case of 20 images sampled from each category that composes of a testing set with 4,100 images.
Hence, the proposed HMOH is robust to the large-scale benchmarks.

\begin{table*}[]
\centering
\caption{The performance variation analysis in the case of 32-bit on Places$205$.}
\label{variation}
\begin{tabular}{c|c|c|c|c|c|c|c|c}
\hline
Number         &OKH                 &SketchHash          &AdaptHash           &OSH
               &MIHash              &BSODH               &HCOH                &HMOH \\
\hline
20             &0.122$\pm$0.008     &0.202$\pm$0.002     &0.195$\pm$0.003     &0.022$\pm$0.002
               &0.244$\pm$0.004     &0.250$\pm$0.002     &\underline{0.259$\pm$0.001}     &\textbf{0.305$\pm$0.003}\\
\hline
30             &0.102$\pm$0.015     &0.213$\pm$0.001     &0.200$\pm$0.000     &0.023$\pm$0.001
               &0.245$\pm$0.002     &0.246$\pm$0.001     &\underline{0.256$\pm$0.002}     &\textbf{0.297$\pm$0.002}\\
\hline
40             &0.105$\pm$0.006     &0.214$\pm$0.001     &0.193$\pm$0.006     &0.023$\pm$0.001
               &0.246$\pm$0.003     &0.244$\pm$0.002     &\underline{0.254$\pm$0.002}     &\textbf{0.297$\pm$0.002}\\
\hline
50             &0.106$\pm$0.009     &0.213$\pm$0.003     &0.198$\pm$0.001     &0.022$\pm$0.001
               &0.246$\pm$0.001     &0.245$\pm$0.001     &\underline{0.255$\pm$0.003}     &\textbf{0.298$\pm$0.002}\\
\hline
100  &0.098$\pm$0.008&0.212$\pm$0.001&0.193$\pm$0.003&0.023$\pm$0.001&0.245$\pm$0.003&0.247$\pm$0.003&\underline{0.256$\pm$0.001}&\textbf{0.300$\pm$0.002} \\
\hline
200  &0.110$\pm$0.004&0.213$\pm$0.001&0.193$\pm$0.005&0.023$\pm$0.001&0.247$\pm$0.001&0.247$\pm$0.001&\underline{0.253$\pm$0.001}&\textbf{0.301$\pm$0.003} \\
\hline
300 &0.097$\pm$0.011&0.211$\pm$0.001&0.193$\pm$0.001&0.023$\pm$0.002&0.247$\pm$0.002&0.248$\pm$0.003&\underline{0.257$\pm$0.002}&\textbf{0.301$\pm$0.001} \\
\hline
400 &0.098$\pm$0.017&0.211$\pm$0.001&0.193$\pm$0.002&0.024$\pm$0.001&0.246$\pm$0.004&0.247$\pm$0.002&\underline{0.256$\pm$0.002}&\textbf{0.299$\pm$0.002} \\
\hline
500 &0.097$\pm$0.015&0.211$\pm$0.002&0.193$\pm$0.001&0.023$\pm$0.002&0.244$\pm$0.001&0.246$\pm$0.001&\underline{0.256$\pm$0.001}&\textbf{0.300$\pm$0.001} \\
\hline

\end{tabular}
\end{table*}

\subsection{Training Efficiency}\label{training_efficiency}
We quantitatively evaluate the efficiency of the proposed HMOH in Tab. \ref{training_time} when hashing bit is set as $32$.
The reported training time is the summarization of all training batches.
Generally speaking, SketchHash and OKH hold the best training efficiency, which however suffer poor effectiveness as analyzed before.
To stress, regarding state-of-the-art methods, \emph{i.e.}, MIHash and BSODH, both HMOH and HCOH are much more efficient.
To make a comparison between HMOH (classification based hashing) and HCOH (regression based hashing), the former performs more efficiently on CIFAR-$10$, MNIST and NUS-WIDE.
And, on Places$205$, HCOH is better.
We notice that on Places$205$, the required kernel size $m$ is $800$ as in Tab.\,\ref{setting}.
However, the original feature dimension on Places$205$ is $128$.
The increased dimension needs more training time.
For a fair comparison , we further test the training time of HMOH without kernelization.
And it takes only 6.34 seconds for HMOH compared with 10.54 seconds for HCOH.
Hence, the classification based HMOH is significantly more efficient than regression based HCOH.
%

\begin{table}[]
\centering
\caption{Training time (Seconds) on four benchmarks under $32$-bit hashing codes.}
\label{training_time}
\begin{tabular}{c|c|c|c|c}
\hline
Method     & CIFAR-$10$       &Places$205$      &MNIST      &NUS-WIDE \\
\hline
\hline
OKH        &4.53                 &15.66         &4.58      &15.50\\
\hline
SketchHash &4.98                 &3.52          &1.27      &23.35\\
\hline
AdaptHash  &20.73                &14.49         &6.26      &15.94\\
\hline
OSH        &93.45                &56.68         &24.07     &65.24\\
\hline
MIHash     &120.10               &468.77        &97.59     &504.33\\
\hline
BSODH      &36.12                &69.73         &4.83      &33.32\\
\hline
\hline
HCOH       &12.34                &10.54         &4.01      &6.23\\
\hline
HMOH       &9.29                 &28.57         &2.76      &5.21\\
\hline
\end{tabular}
\end{table}

\section{Conclusion}  \label{conclusion}
In this paper, we present an online hashing method which comes with the inspiration of Hadamard matrix.
To this end, the streaming data from the same class is assigned with a unique column of the Hadamard matrix as its target code.
And the hash functions aim to fit the assigned code.
To that effect, the assigned code is regarded as virtual binary categories.
The learning of hash functions is further transformed into learning a set of binary classification problem, which can be well solved by off-the-shelf kernelized perceptual algorithm.
Moreover, To guarantee the consistency between length of target code and the number of hashing bit, LSH algorithm is applied and theoretical analysis is given.
Lastly, we propose to ensemble the hashing models learned in every round by simply adding them to boost the performance.
Extensive experiments demonstrate the effectiveness and efficiency of the proposed method.

\begin{acknowledgements}
This work is supported by the National Key R\&D Program (No. 2017YFC0113000, and No. 2016YFB1001503), Nature Science Foundation of China (No. U1705262, No. 61772443, No. 61572410, and No.61702136).
\end{acknowledgements}

\bibliographystyle{spbasic}      
\bibliography{main}   


\end{document}